\documentclass[twocolumn]{aastex631}
\usepackage{amsmath}
\shorttitle{The EDGE-CALIFA survey: The resolved star formation efficiency and local physical conditions}
\shortauthors{Villanueva et al.}
\graphicspath{{./}{figures/}}

\begin{document}

\title{The EDGE-CALIFA survey: The resolved star formation efficiency and local physical conditions}

\author[0000-0002-5877-379X]{V. Villanueva}
\affiliation{Department of Astronomy, University of Maryland, College Park, MD 20742, USA}

\author[0000-0002-5480-5686]{A. Bolatto} 
\affiliation{Department of Astronomy, University of Maryland, College Park, MD 20742, USA}
\affiliation{Visiting Scholar at the Flatiron Institute, Center for Computational Astrophysics, NY 10010, USA}

\author{S. Vogel}
\affiliation{Department of Astronomy, University of Maryland, College Park, MD 20742, USA}

\author[0000-0003-2508-2586]{R. C. Levy}
\affiliation{Department of Astronomy, University of Maryland, College Park, MD 20742, USA}

\author[0000-0001-6444-9307]{S. F. S\'anchez}
\affiliation{Instituto de Astronom\'ia, Universidad Nacional Aut\'onoma de M\'exico, A.P. 70-264, 04510 M\'exico, D.F., M\'exico}

\author[0000-0003-2405-7258]{J. Barrera-Ballesteros}
\affiliation{Instituto de Astronom\'ia, Universidad Nacional Aut\'onoma de M\'exico, A.P. 70-264, 04510 M\'exico, D.F., M\'exico}

\author[0000-0002-7759-0585]{T. Wong}
\affiliation{Department of Astronomy, University of Illinois, Urbana, IL 61801, USA}

\author[0000-0002-5204-2259]{E. Rosolowsky}
\affiliation{Department of Physics, University of Alberta, 4-181 CCIS, Edmonton, AB T6G 2E1, Canada}

\author{D. Colombo}
\affiliation{Max Planck Institute for Radioastronomy, Auf dem H\"ugel 69, D-53121, Bonn, Germany}

\author{M. Rubio}
\affiliation{Departamento de Astronom\'ia, Universidad de Chile, Casilla 36-D Santiago, Chile}

\author[0000-0001-5301-1326]{Y. Cao}
\affiliation{Department of Astronomy, University of Illinois, Urbana, IL 61801, USA}
\affiliation{Aix Marseille Universit\'e, CNRS, LAM (Laboratoire d’Astrophysique de Marseille), F-13388 Marseille, France}

\author{V. Kalinova}
\affiliation{Max Planck Institute for Radioastronomy, Auf dem H\"ugel 69, D-53121, Bonn, Germany}

\author[0000-0002-2545-1700]{A. Leroy}
\affiliation{Department of Astronomy, The Ohio State University, Columbus, OH 43210, USA}
\affiliation{Center for Cosmology and Astroparticle Physics, 191 West Woodruff Avenue, Columbus, OH 43210, USA}

\author[0000-0003-4161-2639]{D. Utomo}
\affiliation{Department of Astronomy, The Ohio State University, Columbus, OH 43210, USA}

\author[0000-0002-2775-0595]{R. Herrera-Camus}
\affiliation{Departamento de Astronom{\'i}a, Universidad de Concepci{\'o}n, Barrio Universitario, Concepci{\'o}n, Chile}

\author{L. Blitz}
\affiliation{Department of Astronomy and Radio Astronomy Laboratory, University of California, Berkeley, CA 94720, USA}

\author[0000-0002-4623-0683]{Y. Luo}
\affiliation{Department of Astronomy, University of Illinois, Urbana, IL 61801, USA}

\correspondingauthor{Vicente Villanueva}
\email{vvillanu@umd.edu}



\begin{abstract}

We measure the star formation rate (SFR) per unit gas mass and the star formation efficiency (SFE$_{\rm gas}$ for total gas, SFE$_{\rm mol}$ for the molecular gas) in 81 nearby galaxies selected from the EDGE-CALIFA survey, using $^{12}$CO(J=1-0) and optical IFU data. For this analysis we stack CO spectra coherently by using the velocities of H$\alpha$ detections to detect fainter CO emission out to galactocentric radii $r_{\rm gal} \sim 1.2 r_{25}$ ($\sim 3 R_{\rm e}$), and include the effects of metallicity and high surface densities in the CO-to-H$_2$ conversion. We determine the scale lengths for the molecular and  stellar components, finding a close to 1:1 relation between them. This result indicates that CO emission and  star formation activity are closely related. We examine the radial dependence of SFE$_{\rm gas}$ on physical parameters such as galactocentric radius, stellar surface density $\Sigma_{\star}$, dynamical equilibrium pressure $P_{\rm DE}$, orbital timescale $\tau_{\rm orb}$, and the Toomre $Q$ stability parameter (including star and gas $Q_{\rm star+gas}$). We observe a generally smooth, continuous exponential decline in the SFE$_{\rm gas}$ with $r_{\rm gal}$. The SFE$_{\rm gas}$ dependence on most of the physical quantities appears to be well described by a power-law. Our results also show a flattening in the SFE$_{\rm gas}$-$\tau_{\rm orb}$ relation at $\log[\tau_{\rm orb}]\sim 7.9-8.1$ and a morphological dependence of the SFE$_{\rm gas}$ per orbital time, which may reflect star formation quenching due to the presence of a bulge component. We do not find a clear correlation between SFE$_{\rm gas}$ and $Q_{\rm star+gas}$. 

\end{abstract}

\date{September 2021. Accepted for publication in ApJ.}


\section{Introduction} \label{sec:intro}

Star formation is one of the most important evolutionary processes that shape galaxies over cosmic times. Either from the inter-galactic medium or through galaxy-galaxy interactions, the accretion of gas into a galaxy potential well provides the fuel for future star formation \citep[e.g.,][]{DiMatteo2007,Bournaud2009}. The mechanisms behind the conversion of gas into stars have been investigated in both distant and nearby galaxies \citep[][]{Kennicutt&Evans2012,Madau&Dickinson2014}. The \citet{Kennicutt1989,Kennicutt1998} seminal studies of the galaxy star formation scaling relations in terms of both the star formation rate and neutral gas surface densities ($\Sigma_{\rm SFR}$ and $\Sigma_{\rm gas}$, respectively), showed they are strongly correlated. More recent studies of the scaling laws between gas, stars, and star formation activity show that the latter is most closely related to molecular gas (H$_2$), and focus on the mechanisms that convert H$_2$ into stars, as the main gas reservoir for star formation \citep[][]{WongBlitz2002,Kennicutt2007,Bigiel2008,Leroy2008,Bigiel2011,Leroy2013}. 

Stars form in Giant Molecular Clouds (GMCs) in which the molecular gas is the main constituent \citep[e.g.,][]{Sanders1985}. We usually trace molecular gas through observations of the low-$J$ transitions of the carbon monoxide (CO) molecule which provide a good measure of the total molecular mass. The $^{12}$C$^{16}$O($J=1\,-\,0$) transition has been commonly used as a tracer of H$_2$ since it is the second most abundant molecule and it can be easily excited in the cold Interestellar Medium (ISM). The CO(1-0) emission line is usually optically thick, and the conversion of CO luminosity, ${L}'_{\rm CO 1-0}$, into molecular gas mass, $M_{\rm H_2}$, is done through a CO-to-H$_2$ conversion factor $\alpha_{\rm CO}$ \cite[e.g.,][]{Bolatto2013} which appears reasonably constant in the molecular regions of galactic disks but changes at low-metallicities and frequently in galaxy centers in response to environmental conditions \citep[e.g.,][]{Wolfire2010,Narayanan2012}.

In the last decades a sharp increase in optical data on galaxies has enabled the detailed study of structure assembly in the Universe, with the goal of understanding the mechanisms that drive the Universe from the very smooth state imprinted on the cosmic microwave background radiation to the galaxies we observe today. Optical spectroscopic surveys (e.g, zCOSMOS, \citealt{Lilly2007};  Sloan Digital Sky Survey III, \citealt{Shadab2015};  KMOS$^{\rm 3D}$, \citealt{Wisnioski2015}; SINS, \citealt{ForsterSchreiber2009}) have shown the relations between star formation, stellar population, nuclear activity, and metal enrichment for unresolved galaxies in a broad range of redshits. Meanwhile, gas surveys of nearby galaxies have enabled the exploration of the physics behind the star formation relations \citep[e.g.][]{Leroy2008,Leroy2013,Saintonge2011b,Saintonge2017}. These data have revealed that the star formation rate responds to two main factors: the molecular gas content and the stellar potential of the system. An important piece of information is the internal structure of the galaxies. The new generation of Integrated Field Unit (IFU) spectroscopy surveys (e.g., Calar Alto Legacy Integral Field Area, CALIFA, \citealt{Sanchez2012}; SAMI, \citealt{Croom2012}; MaNGA, \citealt{Bundy2015}) have provided detailed spectral imaging data with unprecedented spectral and spatial coverage and good resolution, giving the opportunity to map metallicities, dynamics, extinctions, SFRs, stellar mass density, and other quantities across galaxies. In addition, imaging spectroscopy of the molecular gas from millimeter-wave interferometers  \citep{Bolatto2017,Lin2019,Leroy2021} adds invaluable information to understand the baryon cycle in galaxies in the local Universe, where star formation has experienced a drastic decline since the peak of cosmic activity \citep[][]{Madau&Dickinson2014}.

The study of star formation in galaxies demands a holistic approach, since the phenomenon is controlled by multiple processes and it covers a broad range of scales and environments. The analysis of a broad range of galaxy types with multi-waveband datasets is therefore essential to understand the physical conditions that drive star formation activity. The Extragalactic Database for Galaxy Evolution (EDGE) survey is one of the legacy programs completed by the Combined Array for Millimeter-wave Astronomy (CARMA) interferometer \citep{Bock2006}, spanning imaging observations of CO emission in 126 local galaxies. The EDGE survey, combined with the IFU spectroscopy from the CALIFA survey \citep[][]{Sanchez2012}, constitute the EDGE-CALIFA survey \citep[][]{Bolatto2017}, which provides $^{12}$CO and $^{13}$CO ($J=1-0$) images at good sensitivity and angular resolution covering the CALIFA field-of-view. 

In this work, we investigate the star formation efficiency (SFE$_{\rm gas}$, where SFE$_{\rm gas}$ [yr$^{-1}$] $=\Sigma_{\rm SFR}$/$\Sigma_{\rm gas}$) in the EDGE-CALIFA survey taking advantage of its large multiwavelength data for 81 local galaxies with low inclinations. In particular, we investigate how the SFE$_{\rm gas}$ depends on physical quantities such as galactocentric radius, stellar surface density, mid-plane gas pressure, orbital timescale, and the stability of the gas disk to collapse. This paper is organized as follows: Section \ref{S2_Observations} explains the main characteristics of the EDGE-CALIFA survey and the sample selection. In section \ref{S4_Methods} we present the methods employed for data analysis, including the CO stacking procedure and the equations we used to derive the basic quantities. Finally, in sections \ref{S5_Results} and \ref{S6_Conclusion} we present our results, discussion, summary and conclusions of this work, respectively.

\section{DATA PRODUCTS}
\label{S2_Observations}

\subsection{The EDGE and CALIFA surveys}
\label{The_sample}

The EDGE-CALIFA survey \citep[][]{Bolatto2017} is based on the optical Integrated Field Spectroscopy (IFS) CALIFA and CO EDGE surveys. In the next paragraphs, we briefly summarize the main features of these two datasets.

The Calar Alto Legacy Integral Field Area survey, CALIFA \citep[][]{Sanchez2012}, comprises a sample of approximately 800 galaxies at $z\approx0$. The data were acquired by using the combination of the PMAS/PPAK IFU instrument \citep[][]{Roth2005} and the 3.5 m telescope from the Calar Alto Observatory. PMAS/PPAK uses 331 fibers each with a diameter of ${2}''.7$ sorted in an hexagonal shape which covers a field-of-view (FoV) of $\sim$ 1 arcmin$^2$. Its average resolution is $\lambda / \Delta \lambda \sim 850$ at ${\rm \sim 5000 \AA}$ with a wavelength range that spans from $3745$ to ${\rm 7300 \AA}$. CALIFA galaxies are selected such that their isophotal diameters, $D_{25}$, match well the PMA/PPAK FoV, and they range from $45$ to $80$ arcsec in the SDSS $r$-band \citep[][]{Walcher2014}. The CALIFA survey uses a data reduction pipeline designed to produce data cubes with more than $5000$ spectra and with a sampling of $1\times1$ arcsec$^2$ per spaxel. For more details, see \cite{Sanchez2012}.

The Extragalactic Database for Galaxy Evolution, EDGE, is a large intereferometric CO and $^{13}$CO $J = 1-0$ survey which comprises 126 galaxies selected from the CALIFA survey. The observations were taken using the Combined Array for Millimeterwave Astronomy \citep[CARMA,][]{Bock2006} in a combination of the E and D configurations for a total of roughly 4.3 hr per source, with a typical resolution of 8 and 4 arcsec, respectively. The observations used half-beam-spaced seven-point hexagonal mosaics giving a half-power power field-of-view of radius $\sim{50}\arcsec$. The data are primary-gain corrected and masked where the primary beam correction is greater than a factor of $2.5$. The final maps, resulting from the combination of E and D array data, have a velocity resolution of 20 km s$^{-1}$ and typical velocity coverage of 860 km s$^{-1}$, a typical angular resolution of $4.5\arcsec$, and a rms sensitivity of 30 mK at the velocity resolution. For more details, see \cite{Bolatto2017}.

\begin{figure*}
  \includegraphics[width=18.cm]{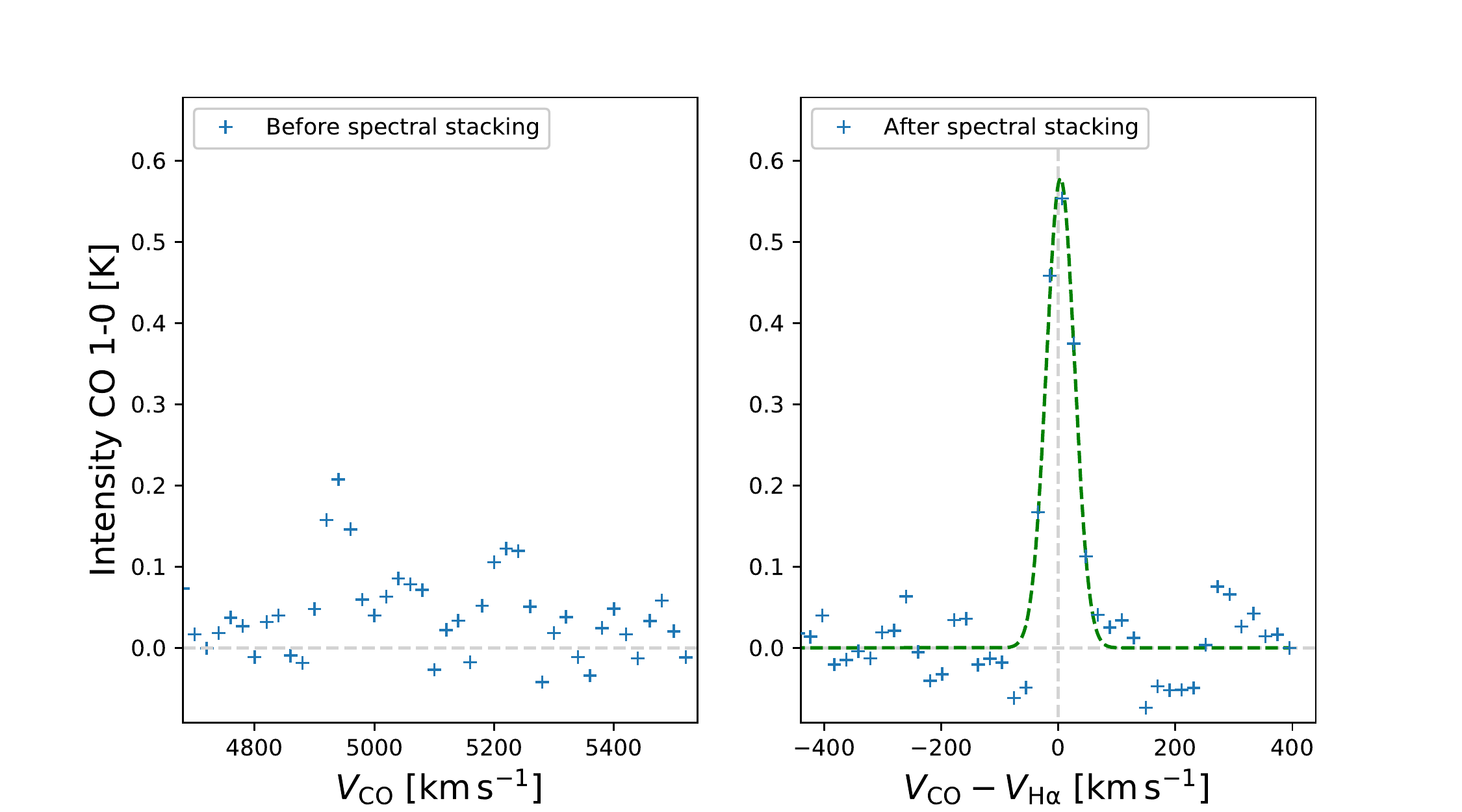}
  \caption{Example showing effects of spectral stacking. The average CO spectrum within an annulus that spans from $0.65$ to $0.75$ $r_{25}$ in NGC $0551$ is shown. The left panel shows the average of all spectra in the annulus in the observed velocity frame. The right panel shows the average in the velocity frame relative to H$\alpha$, along with the best Gaussian fit profile (green dashed line).}
  \label{Stacked_Example}
\end{figure*}

\vspace{1.cm}
\subsection{{\tt{edge}\_\tt{pydb}} database}
\label{Database}

The EDGE-CALIFA survey provides global (integrated) and spatially resolved information about the molecular/ionized gas and stellar components in 126 nearby galaxies, comprising $\sim15000$ individual lines-of-sight. In the context of this work, and to provide easy yet robust access to this large volume of data, we have used one main source of data to perform our analysis.

The {\tt{edge}\_\tt{pydb}} database (Wong et al. in prep.) is a versatile {\tt PYTHON} environment that allows  easy access and filtering of the EDGE-CALIFA data in the variety of analyses we aim to perform. {\tt{edge}\_\tt{pydb}} encompasses a combination of  global galaxy properties and  spatially resolved information, with a special emphasis on estimation of the CO moments from smoothed and masked versions of the CARMA CO datacubes. All data have been convolved to a common angular resolution of ${7}\arcsec$. By using the {\tt{PIPE3D}} data analysis pipeline \citep[see][ for more details]{Sanchez2015,Sanchez2016_2}, the convolved optical datacubes are reprocessed to generate two dimensional maps at 7\arcsec\ resolution. The pipeline fits the stellar continuum to the emission lines for each spaxel in each datacube (adopting a \citealt{Salpeter1955} Initial Mass Function, IMF), generating maps sampled on a square grid with a spacing of ${3}''$ in RA and DEC. To identify a given pixel in the grid, the data are organized by using a reference position (taken from HyperLEDA\footnote{http://leda.univ-lyon1.fr/}) and an offset indicating spatial position. The final database also contains ancillary data, including information from HyperLEDA, NED\footnote{https://ned.ipac.caltech.edu/}, among others.

\section{Methods}
\label{S4_Methods}

\subsection{Stacking of the CO spectra}
\label{Stacking_CO}

Although many EDGE-CALIFA galaxies have high signal-to-noise detections of CO emission in their central regions, emission is generally faint in their outer parts. Typically, the decrease in emission takes place from $r=0.5r_{25}$ outwards (around $1.1R_{\rm e}$, by assuming that $r_{25}\approx 2.7 R_{\rm e}$; \citealt{Sanchez2014}). \cite{Bolatto2017} published maps of velocity-integrated CO emission and discussed various masking techniques for recovering flux and producing maps with good signal to noise; even so, they tend to miss flux in regions of weak emission and to underestimate the CO flux (see Figure 9 in \citealt{Bolatto2017}). Since one of the main goals of this work is to find how the $\rm H_2$ content changes as a function of radius, it is essential to recover low-brightness CO emission line in the outermost parts of galaxies. 

Maps with both good spatial coverage and sensitivity are crucial to set thresholds and timescales for these dependencies. In order to cover a broad range of galactocentric radii, we perform spectral stacking of the $^{12}$CO ($J=1-0$) emission using the H$\alpha$ velocities to coherently align the spectra while integrating in rings. The CO spectral stacking helps recover CO flux in the outer parts of our galaxies, improving our ability to probe the SFE$_{\rm gas}$ in a variety of environments.  Many of the molecular gas surveys have measured some of these dependencies in a similar fashion (e.g., using the CO [$J=2-1$] spectral stacking; \citealt{Schruba2011}), although they mostly covered a small range of morphological types and/or stellar masses, or were limited to very local volumes that are subject to cosmic variance because they represent our particular local environment. Although the EDGE-CALIFA survey does not yet encompass resolved H$\rm I$ observations, we will explore the efficiency with respect to total gas and compare it to previous results by assuming a prescription for the atomic gas while keeping in mind the limitations of this methodology.

We perform a CO emission line stacking procedure following the methodology described by \cite{Schruba2011}. The method relies on using the IFU H$\alpha$ velocity data to define the velocity range for integrating CO emission. The key assumption of this method is that both the H$\alpha$ and the CO velocities are similar at any galaxy location. This assumption is consistent with results by \cite{Levy2018}, who found a median value for the difference between the CO and H$\alpha$ rotation curve of $\Delta V = V_{\rm rot}({\rm CO}) - V_{\rm rot}({\rm H \alpha})= 14$ km s$^{-1}$ (within our 20 km s$^{-1}$ channel width) when analyzing a sub-sample of 17 EDGE-CALIFA rotation-dominated galaxies. As we will discuss later, after shifting CO spectra to the H$\alpha$ velocity, we integrate over a window designed to minimize missing CO flux. The smaller the  velocity differences between CO and H$\alpha$, the better the signal-to-noise. Similarly, the smaller the velocity window we implement, the smaller the noise in the integrated flux estimate.

We constructed an algorithm coded in {\tt PYTHON} that implements this procedure. Since we are interested in radial variations in galactic properties, we stack in radial bins $0.1r_{25}$ wide. In practice, galactocentric radius is usually a well determined observable and it is covariant with other useful local parameters, which makes it a very useful ordinate \citep{Schruba2011}.

We recover the CO line emission by applying radial stacking based on the following steps:  We convert H${\alpha}$ velocity from the optical into the radio velocity convention. Then, for each spaxel in an annulus we shift the CO spectrum by the negative H${\alpha}$ velocity. This step aligns the CO spectrum for each line-of-sight at zero velocity if the intrinsic H$\alpha$ and CO velocities are identical. We then average all the velocity-shifted CO spaxels in an annulus, and integrate the resulting average spectrum over a given velocity window to produce the average intensity in the annulus.

Figure \ref{Stacked_Example} shows the usefulness of the stacking procedure in recovering CO emission. As an example, we show the average CO spectrum of NGC 0551 within an annulus that spans from 0.65 to 0.75 $r_{25}$ ($\sim 1.3-1.7 \, R_{\rm e}$). The left panel contains the average CO spectra within the given annulus using the observed velocity frame, while the right panel shows the average CO spectra after shifting by the observed H${\alpha}$ velocity. If the CO and H$\alpha$ velocities are identical for all spaxels, then the resulting CO emission would appear at zero velocity. This procedure allows us to co--add CO intensities coherently and reject noise. Figure \ref{Stacked_Example} also shows the best Gaussian fit for the averaged-stacked spectra. We expect that in an ideal case the total intensity integrated over the full velocity range ($\sim$860 km s$^{-1}$) is exactly the same in both cases, but the noise would be much larger without the spectral stacking. Without performing the stacking procedure the CO line emission is not evident, and the signal-to-noise ratio in the measurement of CO velocity-integrated intensity is lower. Interferometric deconvolution artifacts that produce negative intensities at some velocities, resulting from incomplete $uv$ sampling and spatial filtering, would also get into the integration more easily without stacking and artificially reduce the intensity.

\begin{figure}
\includegraphics[width=7.9cm]{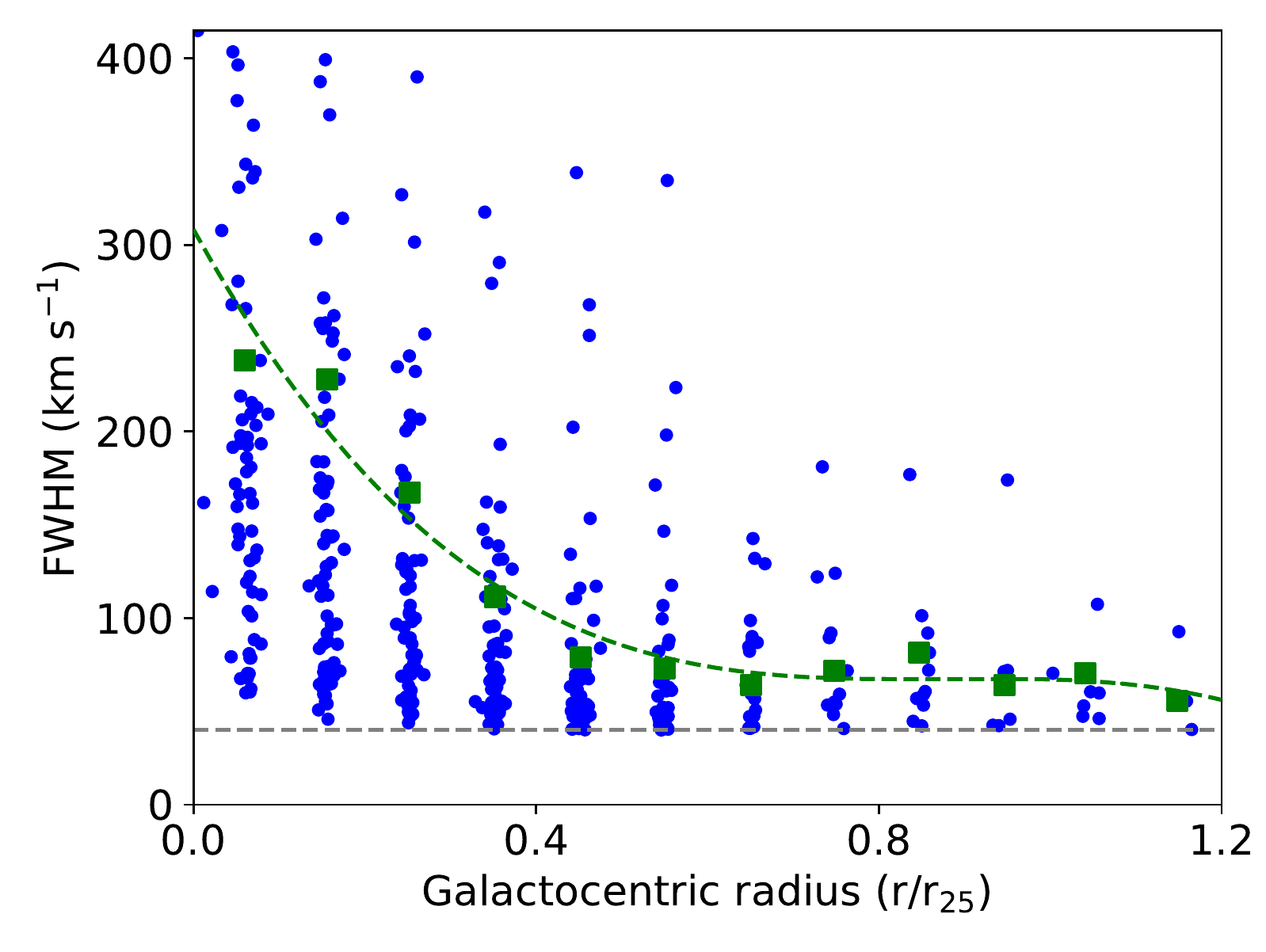}
\includegraphics[width=7.6cm]{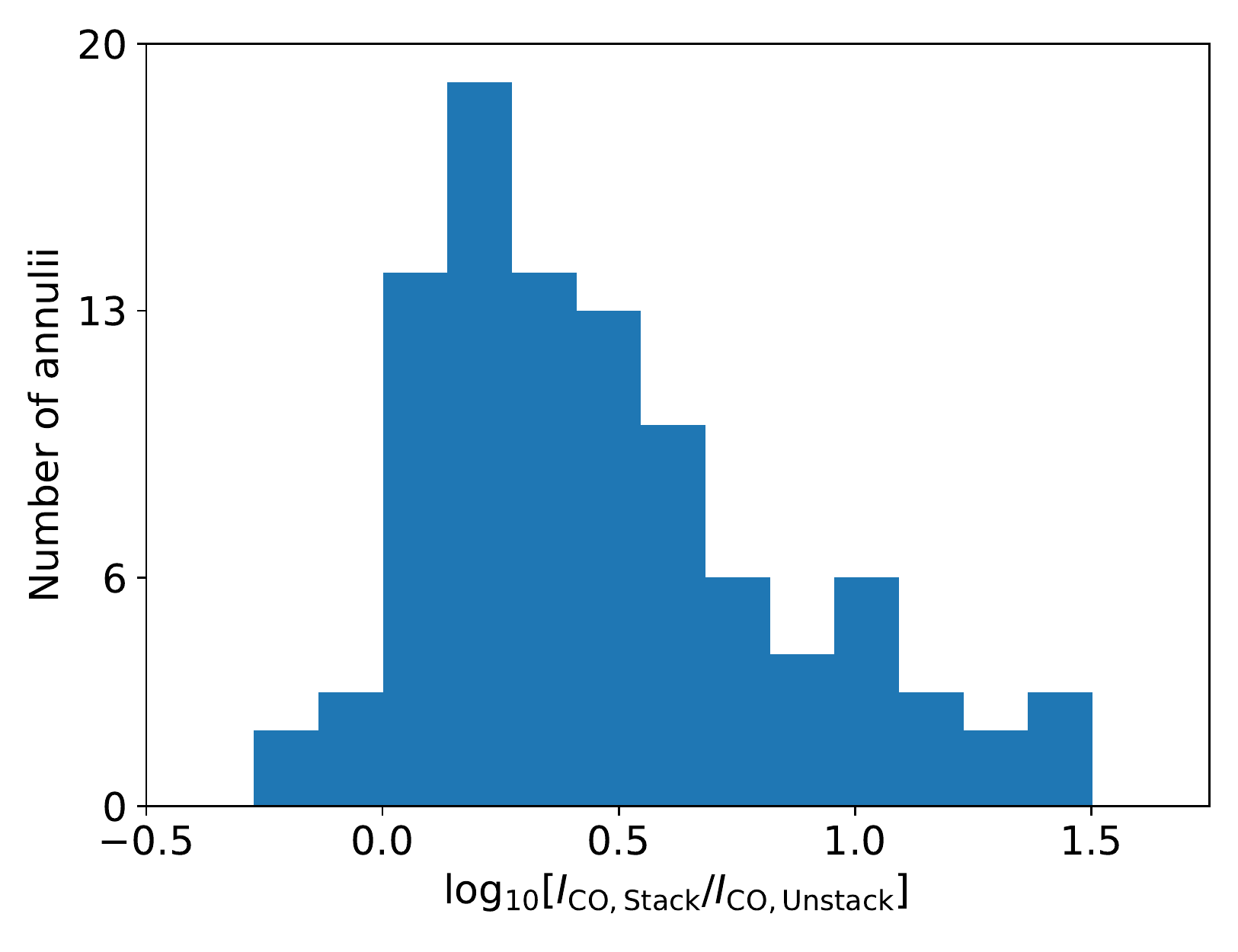}\\
\caption{ {\it Top:} FWHM of CO line as a function of galactocentric radius. Small colored dots show the FWHM of a Gaussian fit to the stacked spectrum in an annulus. Large green squares indicate the FWHM lying above 80\% of the points at that radius, and the dashed green line is the fit to the squares; we use this function to define the window of flux integration as a function of $r_{\rm gal}$. The gray-dashed line marks the limit at which we reject spectra with a Gaussian fit narrower than 40 km s$^{-1}$ (2 channels in the CO datacubes). {\it Bottom:} Ratio between the final stacked and unstacked integrated CO(1--0) line intensity per annulus for annuli located at $r/r_{25} > 0.5$, which include just $2\sigma$ detection spaxels.}
\label{FWHM_radius}
\end{figure}

\subsection{Extracting fluxes from stacked spectra}
\label{Extracting_fluxes}

After we compute the stacked spectra, we extract the total CO fluxes for each annulus as a function of galactocentric radius. To do this in a way that is likely to include all the CO flux but minimizes the noise, we want to select a matched velocity range that is just large enough to include all CO emission and exclude the baseline (which only adds noise). In order to investigate the ideal integration range we fit Gaussian profiles to each averaged-stacked CO spectrum with a detection. We reject fits which have central velocities more than $\pm 80$ km s$^{-1}$ from zero velocity. We also reject spectra with FWHMs narrower than 40 km s$^{-1}$ (2 channels). Results for valid stacked spectra fits are shown in the top panel of Figure \ref{FWHM_radius}, color coded by the reduced chi-squared of the fit and plotted against normalized galactocentric radius. 

We use these data to define a velocity window for the integrated CO line emission fluxes in the stacked spectra. For each radial bin, we define an integration range that guarantees that we integrate the CO line profile between $\pm {\rm FWHM}$ in at least 80\% of annuli. This is represented by the green dots in Figure \ref{FWHM_radius}. We assume that this window is sufficient to contain most of the CO flux, and we can use it to compute errors where no CO is detected. To obtain a prescription we fit the best third-order polynomial to the green squares (green dashed-line) as a function of galactocentric radius, ${\rm FWHM}$($r_{\rm gal}$). Finally, we recompute the CO line emission fluxes for the stacked spectra by integrating the CO stacked spectrum over $\pm {\rm FWHM}(r_{\rm gal})$. We extract the integrated flux uncertainties by taking the rms from the emission free part of the stacked CO spectra.

Using spectral stacking we reach a typical deprojected CO intensity $3\sigma$ uncertainty of $I_{\rm CO} \approx 0.25$ K km s$^{-1}$, or a $3\sigma$ surface density sensitivity of $\Sigma_{\rm mol} \approx 1.1\, M_{\odot}$ pc$^{-2}$, which represents the typical sensitivity in the outermost regions of galaxy disks. The bottom panel of Figure \ref{FWHM_radius} shows the ratio between the final stacked and unstacked integrated CO(1-0) line intensity, per annulus, located at $r>0.5 r_{25}$ (or $r>1.3 R_{\rm e}$), and includes just $2\sigma$ detection spaxels. The histogram shows that the distribution peaks at $\log[I_{\rm CO,Stack}/I_{\rm CO,Unstack}]\sim 0.47$, meaning that, in overall, we are recovering $\sim 3$ times more flux with the stacking procedure.

\subsection{Basic equations and assumptions}
\label{Basic_equations}

To compute the extinction-corrected SFRs, we estimate the extinction \citep[based on the Balmer decrement; see][]{Bolatto2017}  for each 7\arcsec\ spaxel  using

\begin{equation}
A_{\rm H{\alpha}} = 5.86 \log{\left (  \frac{F_{\rm H{\alpha}}}{2.86F_{\rm H{\beta}}}\right )}
\end{equation}
where $F_{\rm H \alpha}$ and $F_{\rm H \beta}$ are the fluxes of the respective Balmer lines, and the coefficients assumes a \cite{Cardelli1989} extinction curve and an unextincted flux ratio of 2.86 for case B recombination. Then, the corresponding SFR (in M$_{\odot}$ yr$^{-1}$) is obtained using \citep{RosaGonzalez2002}
\begin{equation}
{\rm SFR} = 1.61 \times 7.9 \times 10^{-42} F_{\rm H{\alpha}} 10^{\frac{A_{\rm H{\alpha}}}{2.5}},
\label{eq:sfr}
\end{equation}

\noindent which adopts a Salpeter Initial Mass Function (IMF) corrected by a factor of 1.61 to move it to a Kroupa IMF \citep{Speagle2014}. We use this to compute the Star Formation Rate surface density,  $\Sigma_{\rm SFR}$ in M$_\odot$\,yr$^{-1}$\,kpc$^{-2}$, by dividing by the face-on area corresponding to a 7\arcsec\ spaxel, given the angular diameter distance to the galaxy.

The gas surface density is computed as $\Sigma_{\rm gas} = \Sigma_{\rm mol} + \Sigma_{\rm atom}$, where $\Sigma_{\rm H_2}$ is derived from the integrated CO intensity, $I_{\rm CO}$, by adopting a Milky-Way constant CO-to-H$_2$ conversion factor, $X_{\rm CO} = 2 \times 10^{20}$ cm$^{-2}$ (K km s$^{-1}$)$^{-1}$ (or $\alpha_{\rm CO} = 4.3$ M$_\odot$ $[\rm K \, km \, s^{-1} \, pc^{-2}]^{-1}$). For the CO $J=1 - 0$ emission line, we use the following expression to obtain $\Sigma_{\rm mol}$ \citep[i.e., ][]{Leroy2008}
\begin{equation}
\Sigma_{\rm mol}  = 4.4 \cos i\, I_{\rm CO},
\end{equation}

\noindent where $I_{\rm CO}$ is in K\,km s$^{-1}$, $\Sigma_{\rm mol}$ is in M$_\odot$\,pc$^{-2}$, and $i$ is the inclination of the galaxy. This equation takes into account the mass correction due to the cosmic abundance of Helium.

To include in our calculations $\Sigma_{\rm atom}$ despite the fact that we do not have resolved H$\rm I$ data, we assume a constant $\Sigma_{\rm atom} = 6 $ M$_{\odot}$ pc$^{-2}$ for face-on disks. This is approximately correct (within a factor of 2) for spiral galaxies out to $r \sim r_{25}$ \citep{Walter2008,Leroy2008}. This value is also in agreement with Monte Carlo simulations performed by \cite{Barrera-Ballesteros2021} to test different values of $\Sigma_{\rm atom}$; they obtain a normal distribution of $\Sigma_{\rm atom}=7$ M$_\odot$ pc$^{-2}$, with a standard deviation of $2$ M$_\odot$ pc$^{-2}$. We also test the influence of metallicity in the CO-to-H$_2$ conversion factor, $\alpha_{\rm CO}$, by using the following equation \citep[from Equation 31 in][]{Bolatto2013}:
\begin{equation}
\alpha_{\rm CO}  = 2.9 \exp \left ( \frac{+0.4}{Z' \Sigma^{100}_{\rm GMC}} \right ) \left ( \frac{\Sigma_{\rm total}}{100 \, \rm M_\odot \, pc^{-2} } \right )^{-\gamma},
\label{Alpha_CO}
\end{equation}
in M$_\odot$ $\rm (K \, km s^{-1} \, pc^{-2} )^{-1}$,  $\gamma \approx 0.5$ for $\Sigma_{\rm total} > 100$ M$_\odot$ pc$^{-2}$ and $\gamma =0$ otherwise. We adopt the empirical calibrator based on the O3N2 ratio from \cite{Marino2013}, and then we use equation 2 from \cite{Marino2013} to obtain the oxygen abundances, $\rm 12 + \log(O/H)$. Finally, we derive the metallicity normalized to the solar value, ${Z}' = \rm [O/H]/[O/H]_\odot$, where $\rm [O/H]_\odot = 4.9\times 10^{-4}$ \citep[][]{Baumgartner&Mushotzky2006}.

Although there are many definitions for star formation efficiency (SFE$_{\rm gas}$), in this work we use SFR surface density per unit neutral gas surface density (atomic and molecular), $\Sigma_{\rm gas}=\Sigma_{\rm mol} +\Sigma_{\rm atom}$, in units of yr$^{-1}$ for each line-of-sight (LoS), 

\begin{equation}
{\rm SFE_{\rm gas}} = \frac{\Sigma_{\rm SFR}}{\Sigma_{\rm gas}}.
\label{eq:sfe}
\end{equation}

Midplane gas pressure, $P_{\rm h}$ is computed using the expression by \cite{Elmegreen1989},
\begin{equation}
P_{\rm h} \approx \frac{\pi}{2} \, G \, \Sigma^{2}_{\rm gas} + \frac{\pi}{2} \, G \, \frac{\sigma_{\rm g}}{\sigma_{\star , z} }\Sigma_{\star}\Sigma_{\rm gas}, 
\label{Midplane_gas_pressure}
\end{equation}
where $\sigma_{\rm g}$ and $\sigma_{\star , z}$ are the gas and stars dispersion velocities, respectively. We correct the $\Sigma_{\star }$ by the same 1.61 factor used for the SFR to translate them to a Kroupa IMF. We assume $ \sigma_{\rm g}=11$ km s$^{-1}$, which has been found to be a typical value in regions where HI is dominant (\citealt{Leroy2008}). This value is also in agreement with the second moments maps included in \cite{Tamburro2009}, and is also consistent with the CO velocity dispersion for a subsample of EDGE-CALIFA galaxies \citep[][]{Levy2018}. $\sigma_{\star , z}$ is the vertical velocity dispersion (in km\,s$^{-1}$) of stars. Although the EDGE-CALIFA database includes $\sigma_{z}$ measurements that could allow us to model $\sigma_{\star , z}$, the instrumental resolution of the survey constrains us to use them just in the central parts of the galaxies (for details, see \citealt{Sanchez2012}). Therefore, and following the assumptions and derivation included in \cite{Leroy2008}, we use the following expression for $\sigma_{\star , z} $:

\begin{equation}
\sigma_{\star , z} = \sqrt{\frac{2 \pi G l_{\star}}{7.3}} \Sigma^{0.5}_{\star},
\label{sigma_star}
\end{equation}
\noindent where $l_{\star}$ is the disk stellar exponential scale length obtained by fitting azimuthally averaged profiles to $\Sigma_{\star}$ in the SDSS $r$-band and $G=4.301\times10^{-3}$ pc \,M$_\odot^{-1}$\,km$^{2}$\,s$^{-2}$. In cases where we do not have $l_{\star}$ measurements, we use the relation $l_{\star} = [0.25\pm 0.01] r_{25}$ since it corresponds to the best linear fit for our data. See Section \ref{Exponential_scale} for more information about how both $l_{\star}$ and the $l_{\star}$-$r_{25}$ relation are derived.

The dynamical equilibrium pressure ($P_{\rm DE}$) is computed following a similar methodology as for $P_{\rm h}$ \citep[e.g][]{ElmegreenPavarro1994,HerreraCamus2017,Fisher2019,Schruba2019}. Assuming that the gas disk scale height is much smaller than the stellar scale height, and neglecting the gravity from dark matter, we write $P_{\rm DE}$ as \citep{Sun2020}:
\begin{equation}
P_{\rm DE} \approx \frac{\pi}{2} \, G \, \Sigma^{2}_{\rm gas} + \Sigma_{\rm gas} \sigma_{\rm gas , z} \sqrt{2\, G \, \rho_{\star}}. 
\label{Dynamical_equilibrium_pressure}
\end{equation}
Here, we assume that $\sigma_{\rm gas , z}=\sigma_{\rm g}=11\,{\rm km \, s^{-1}}$, and $\rho_{\star}$ is the mid-plane stellar volume density from the observed surface density in a kpc-size aperture,
\begin{equation}
\rho_\star = \frac{\Sigma_{\star}}{0.54 l_{\star}}.
\label{rho_star}
\end{equation}
\noindent This equation assumes that the exponential stellar scale height, $h_\star$, is related to the stellar scale length, $l_\star$, by $h_\star/l_\star = 7.3 \pm 2.2$ \citep[][]{Kregel2002}.

The orbital timescale, $\tau_{\rm orb}$, is usually used in the analysis of star formation law dependencies since it can be comparable to timescale of the star formation \citep[e.g.,][]{Silk1997,Elmegreen1997}. Following \cite{Kennicutt1998} and \cite{WongBlitz2002}, we compute $\tau_{\rm orb}$ using:

\begin{equation}
\tau^{-1}_{\rm orb} = \frac{v \left ( r_{\rm gal} \right )}{2 \pi r_{\rm gal}}
\label{Orbital_timescale}
\end{equation}
where $v \left ( r_{\rm gal} \right )$ is the rotational velocity at a galactocentric radius $r_{\rm gal}$. We obtain the H$\alpha$ rotation curves for EDGE-CALIFA galaxies from \cite{Levy2018}. We use them to adjust an Universal Rotation Curve \citep[URC, ][]{Persic1996} for each galaxy to avoid the noise in the inner and outer edges of the H$\alpha$ rotation curves.

\begin{figure*}
\begin{center}
  \includegraphics[width=8.8cm]{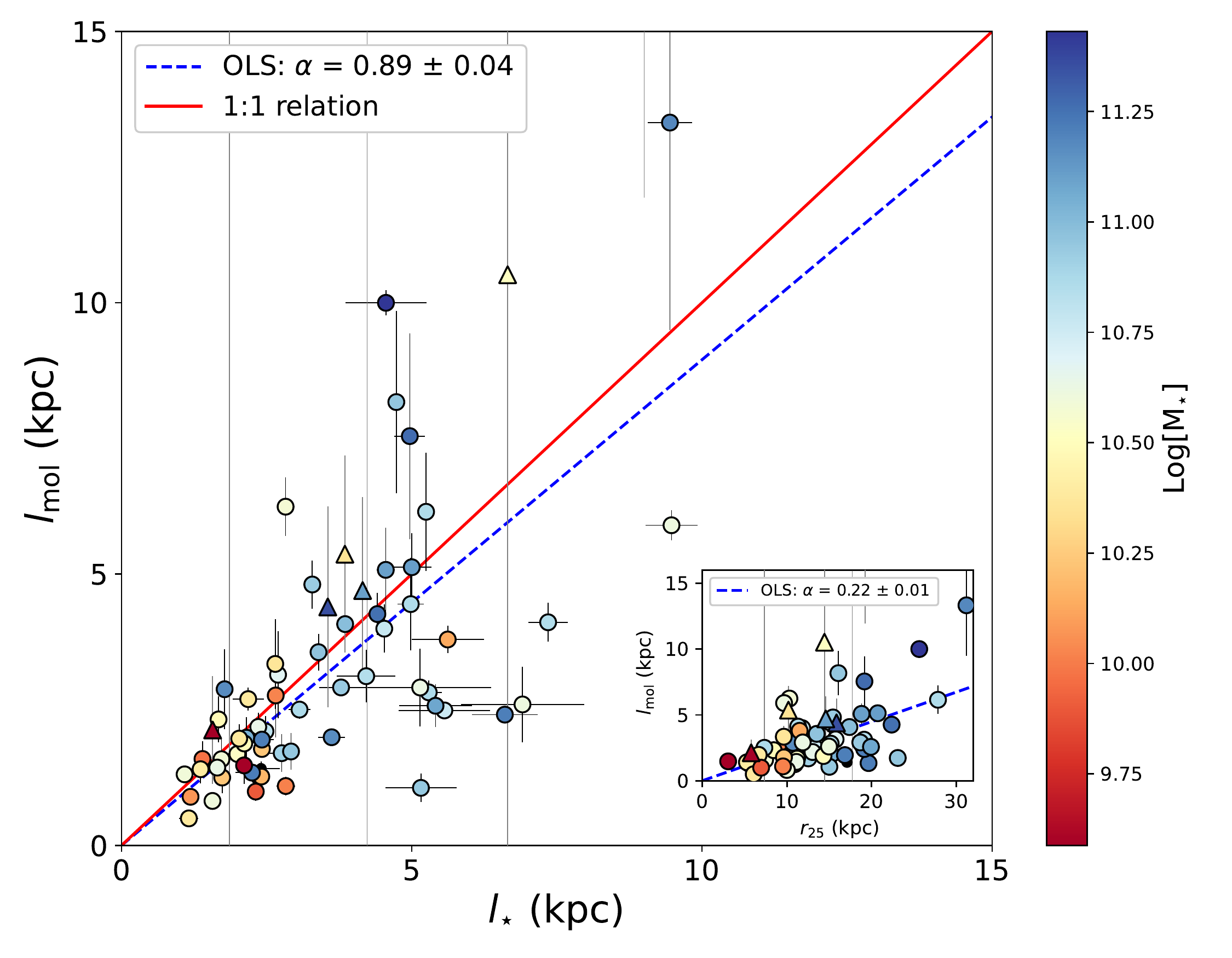}
  \includegraphics[width=8.8cm]{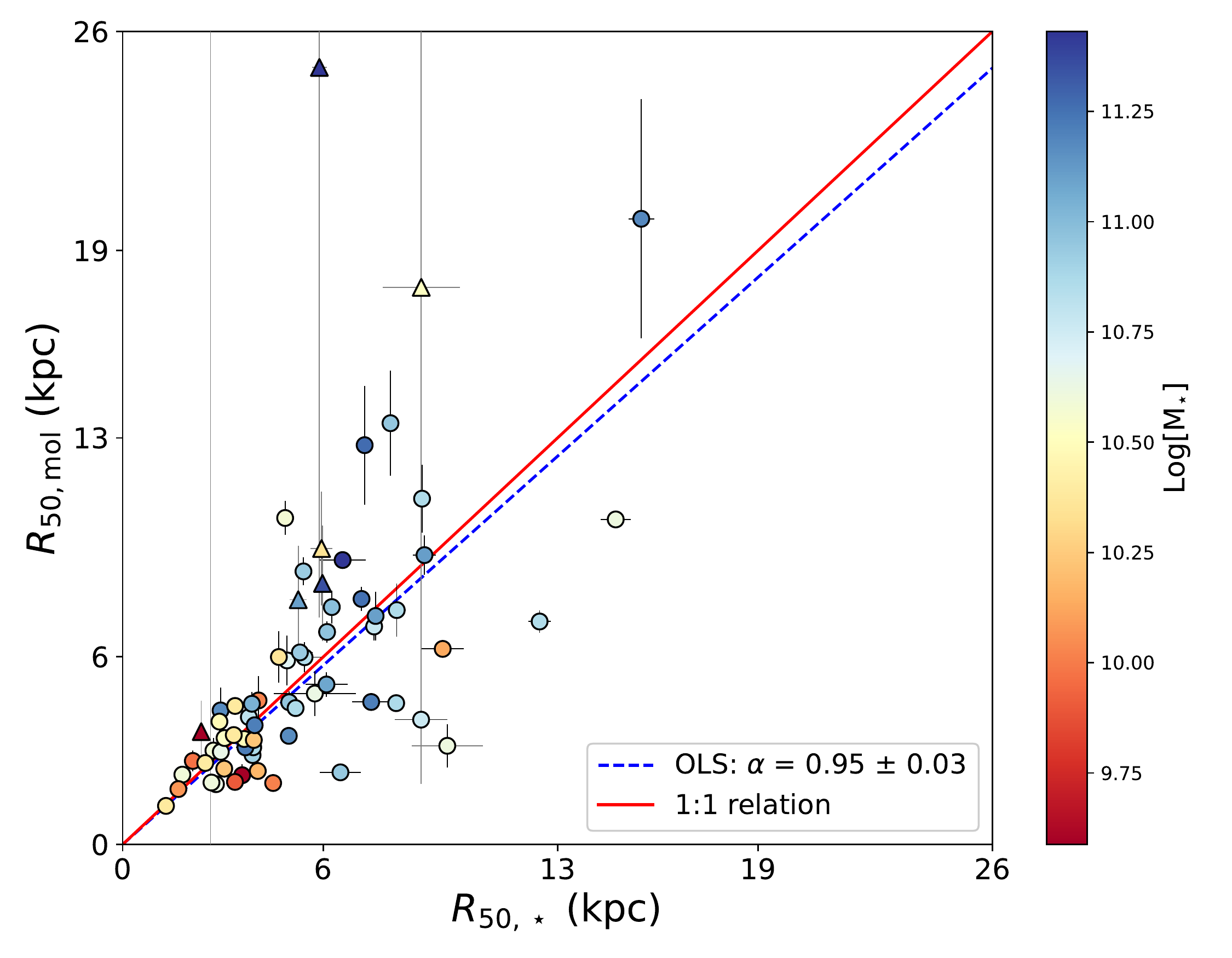}
  \end{center} 
  \caption{{\it Left:} Comparison between the stellar, ${\it l}_{\star}$, and molecular length scales, ${\it l}_{\rm mol}$, computed by fitting exponential profiles to the respective surface densities as a function of galactocentric radius. The colored circles correspond to 61 EDGE-CALIFA galaxies color-coded by stellar mass derived from SED fitting (see Section \ref{The_sample}). The inset panel shows the comparison between ${\it l}_{\rm mol}$ and the isophotal radius $r_{25}$. The triangles represent uncertain results for which measurements are smaller than $3\sigma$. The solid red and dashed blue lines illustrate the 1:1 scaling and the OLS linear bisector fit (forced through the origin) for all the sources, respectively. {\it Right:} Relationship between the radii that enclose 50\% of the molecular gas and the stellar mass, ${\it R}_{\rm 50, mol}$ and ${\it R}_{50, \star}$, respectively. Conventions and symbols are as in the left panel.}
 \label{FIG_14_LENGTH}
\end{figure*}

\begin{figure*}
\includegraphics[width=18cm]{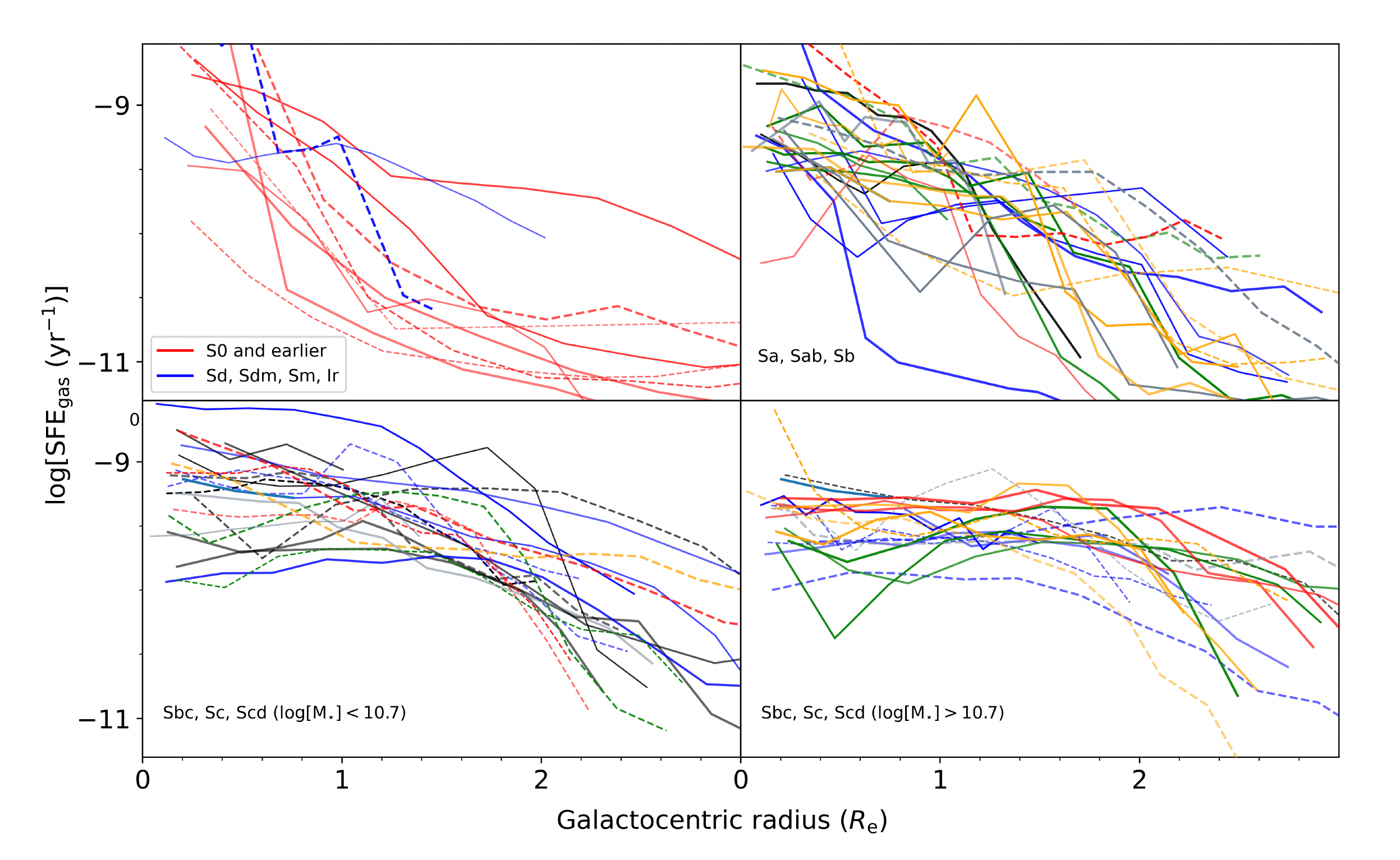}
\caption{SFE$_{\rm gas}$ vs galactocentric radius. Each line indicates the average SFE$_{\rm gas}$ for individual galaxies in $0.1r_{25}$-wide tilted annuli after stacking. The morphological group for the galaxies in each panel is indicated by the legend in that panel. The plot shows that the SFE$_{\rm gas}$ in individual galaxies generally decreases as a function of galactocentric radius and that the dispersion in SFE$_{\rm gas}$ at particular radii is due mostly to differences between galaxies.}
\label{FIG_1_Leroy_2008_1A}
\end{figure*}

\begin{figure}

\includegraphics[width=8cm]{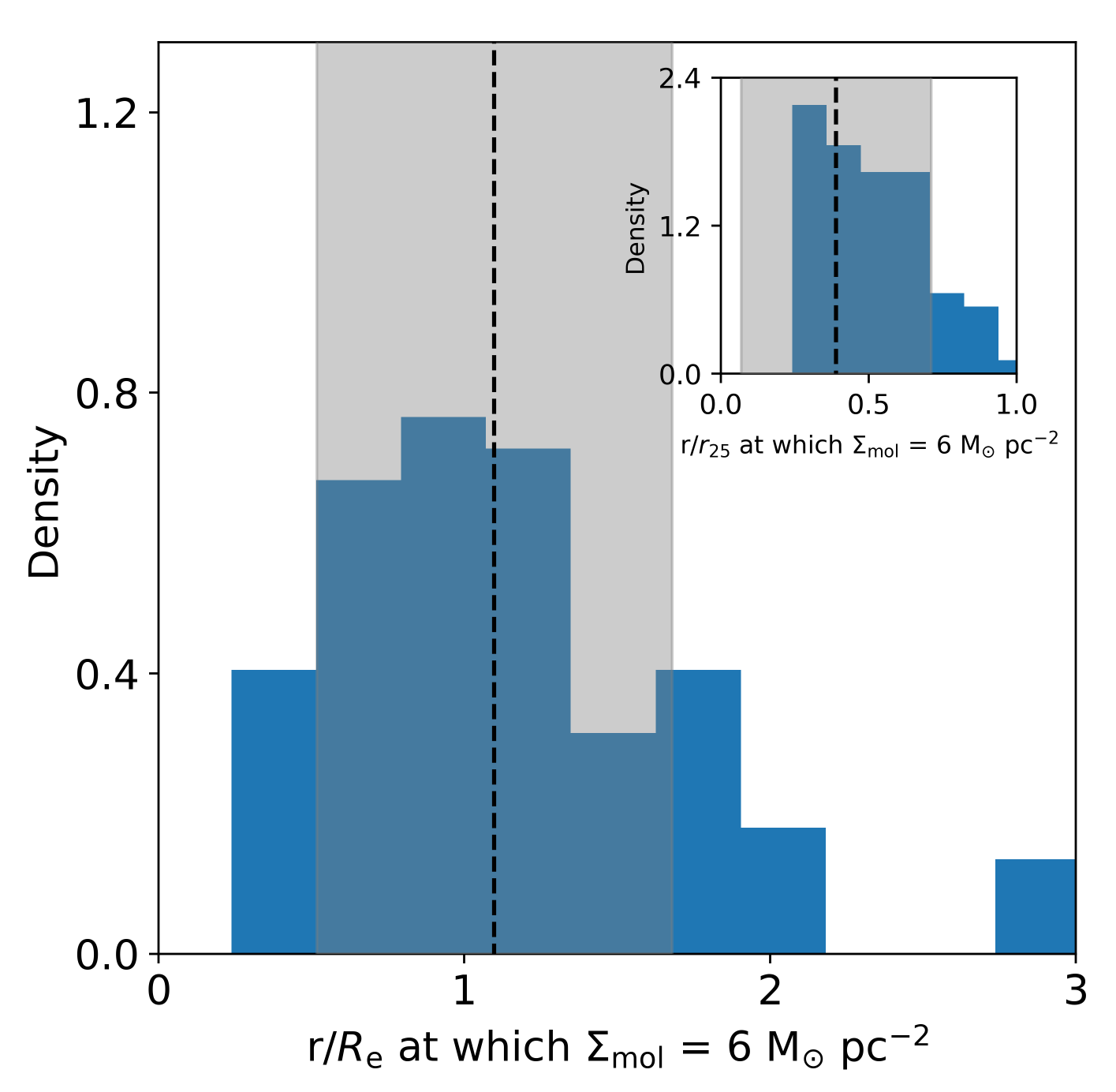}

\caption{Histogram of galactocentric radii at which $\Sigma_{\rm H_2}$ drops to 6 M$_\odot$ pc$^{-2}$, which is the value of $\Sigma_{\rm atom}$ assumed for EDGE-CALIFA galaxies in this work. The dashed vertical black line is the mean value of $r/R_{\rm e}$ at which this occurs, corresponding to $1.1$ ($0.4 r_{25}$; see inset panel). The gray area represents the uncertainty in mean value of $r/R_{\rm e}$. The inset shows a similar histogram for $r/r_{25}$.}
\label{FIG_1_Leroy2008_2}
\end{figure}

\begin{figure}

\includegraphics[width=8.75cm]{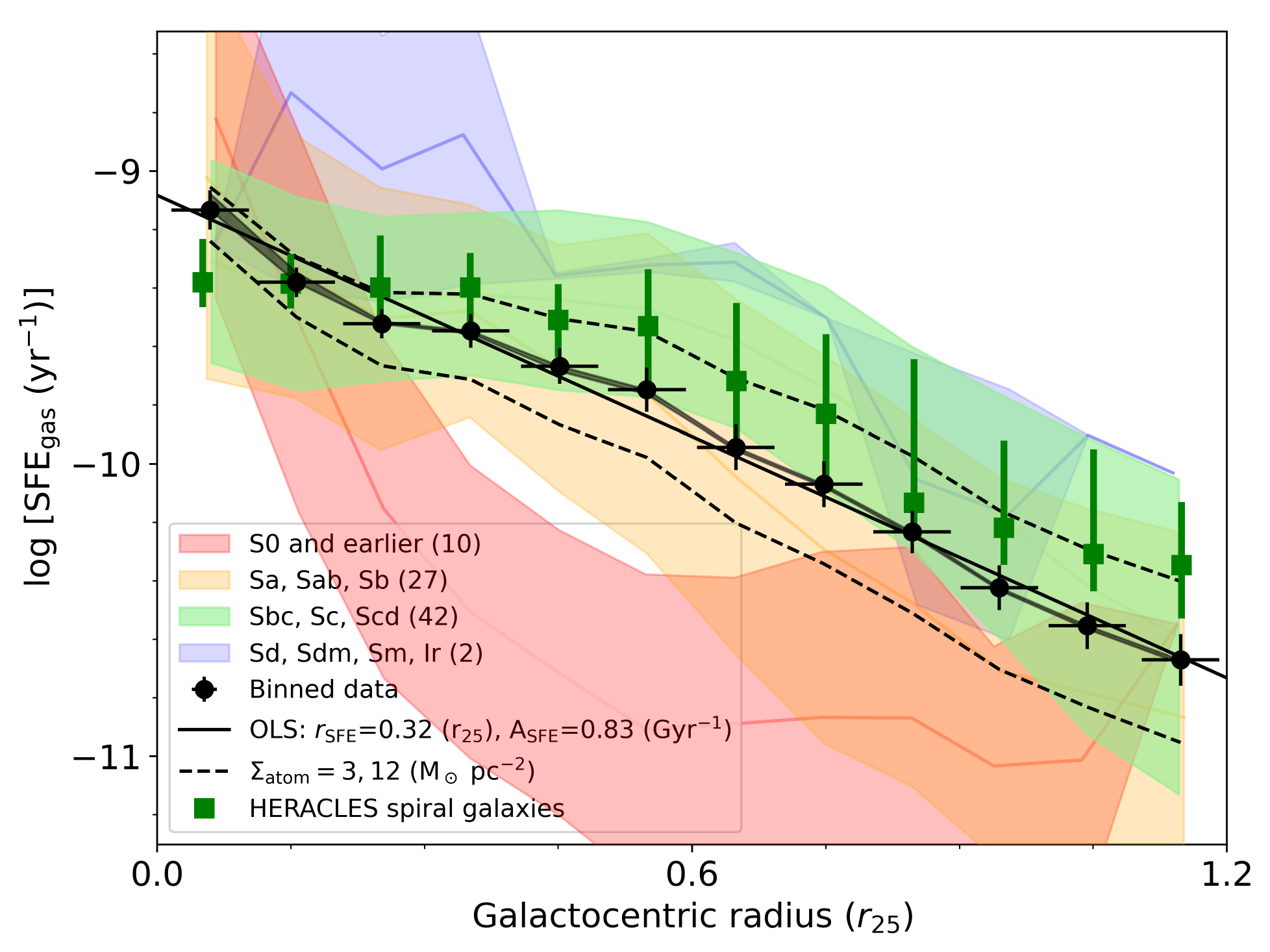}

\caption{{\it Top:}SFE$_{\rm gas}$ vs galactocentric radius for different morphological types of galaxies. SFE is averaged at each radius over all galaxies of the selected morphological type;  types  are indicated by shaded color as described in the legend. The vertical extent of the shaded area for each morphogical type is the 1$\sigma$ scatter distribution for that type (see Figure \ref{FIG_1_Leroy_2008_1A}). Circular dots indicate the average SFE$_{\rm gas}$ and galactocentric radius in stacked annuli for all EDGE-CALIFA galaxies; the black solid line is the OLS linear bisector fit to those points using the model $A_{\rm SFE} \times \exp{(-r/r_{\rm SFE})}$. The error bars are the uncertainties of the mean SFE values in each bin. The two dashed black lines show the effect of increasing and decreasing $\Sigma_{\rm atom}$ by a factor of two from its assumed value of 6 M$_\odot$ pc$^{-2}$. The shaded gray band indicates the amount by which the binned SFE$_{\rm gas}$ would increase if we use the metallicity dependent prescription for $\alpha_{\rm CO}$. The green squares are the HERACLES data for spiral galaxies. The figure shows that SFE$_{\rm gas}$ depends on radius, stellar mass and morphological type.}
\label{FIG_1_Leroy2001B}
\end{figure}

We compute the Toomre's instability parameter \citep[$Q$]{Toomre1964} including the effect of stars \citep{Rafikov2001}. The Toomre's instability parameter for the stellar component ($Q_{\rm star}$) is 
\begin{equation}
Q_{\rm stars} = \frac{\sigma_{\star, r} \kappa }{\pi G \Sigma_{\star}},
\label{Instability_parameter}
\end{equation}

\noindent where $\sigma_{\star, r}$ is the radial velocity dispersion of the stars. We compute it using $\sigma_{\star, r}=1.67\,\sigma_{\star, z}$, valid for most late-type galaxies \citep[][]{Shapiro2003}. The parameter $\kappa$ is the epicyclic frequency and can be computed as
\begin{equation}
\kappa = 1.41 \frac{v(r)}{r} \sqrt{1+\beta},
\end{equation}
where $\beta = \frac{d \log{\, v \left ( r \right )}}{d \log{r}}$. This derivative is computed based on the URC fit to the H$\alpha$ rotation curve. The Toomre's instability parameter for the gas ($Q_{\rm gas}$) is

\begin{equation}
Q_{\rm gas} = \frac{\sigma_{\rm gas} \kappa }{\pi G \Sigma_{\rm gas}} = \frac{(11\,{\rm km \, s^{-1}}) \kappa }{\pi G \Sigma_{\rm gas}}.
\end{equation}
Since $\Sigma_{\star}$ and $\Sigma_{\rm gas}$ are averaged and stacked by annuli, respectively, then both $Q_{\rm stars}$ and $Q_{\rm gas}$ are derived radially. The condition for instability in the gas+stars disk is then given by
\begin{equation}
\frac{1}{Q_{\rm star+gas}}=\frac{2}{Q_{\rm stars}}\frac{q}{1+q^2}+ \frac{2}{Q_{\rm gas}}R\frac{q}{1+q^2 R^2} > 1,
\label{Rafikov}
\end{equation}
where $q= k \sigma_{\star, r}/\kappa$. Here, $k=2\pi/\lambda$ is the wavenumber at maximum instability. Finally, $R=\sigma_{\rm g}$/$\sigma_{\star,r}$.

\section{Results and Discussion}
\label{S5_Results}

\subsection{Exponential scale lengths}
\label{Exponential_scale}

To investigate the spatial relationship between molecular and stellar components, we compute their exponential scale lengths, $l_{\rm mol}$ and $l_{\star}$, respectively, for a subsample of 68 galaxies. Out of the 81 EDGE-CALIFA galaxies with $i < 75^{\circ}$ , these galaxies are selected since their disks are well fitted by exponential profiles and they have at least three annuli available for the fitting. To avoid annuli within the bulge or with significant variations in $\alpha_{\rm CO}$ usually found in central regions of galaxies \citep[e.g.,][]{Sandstrom2013}, we do not include $\Sigma_{\rm mol}$ and $\Sigma_{\star}$ for $r_{\rm gal} \leq 1.5$ kpc.

It is well known that the CO distribution and star formation activity are closely related \citep[e.g., ][]{Leroy2013}. For instance, \cite{Leroy2009} showed that HERACLES spiral galaxies can be well described by exponential profiles for CO emission in the H$_2$-dominated regions of the disk, with similar CO scale lengths to those for old stars and star-forming tracers, and an early study on the EDGE sample found similar results \citep{Bolatto2017}. Here we use the stacking technique to extend the molecular radial profiles and obtain a better measurement of the distribution. 

Although molecular clouds have lifetimes spanning a few to several Myr \citep[similar to the stars that give rise to the H$\alpha$ emission used to compute SFR; e.g.,][]{Blitz&Shu1980,Kawamura2009,Gratier2012}, these are quite short compared with lifetimes of the stellar population in galaxies in the EDGE-CALIFA survey \citep[$0.4$ to $3.9$ Gyr;][]{Barrera-Ballesteros2021}. Consequently, it is not necessarily expected to have comparable distributions for the molecular and the stellar components. However, stellar and CO emission distributions can be similar when the process of converting atomic gas to molecular is driven by the stellar potential \citep[][]{Blitz&Rosolowsky2004,Ostriker2010}. For instance, \cite{Schruba2011} showed a clear correspondence between $l_{\rm CO}$ and $r_{25}$; this correlation is maintained even in the HI-dominated regions of the disk, supporting the role that molecular gas plays in a scenario when the stellar potential well is relevant in collecting material for star formation \citep{Blitz&Rosolowsky2006}. Thus, it is interesting to use the CO stacked data to verify if the exponential decay of $\Sigma_{\rm mol}$ holds in the outer parts of EDGE-CALIFA galaxies.

The left panel of Figure \ref{FIG_14_LENGTH} shows the relation between $l_{\rm mol}$ and $l_{\rm \star}$. The $l_{\rm \star}$  values were obtained by fitting exponential profiles to $\Sigma_{\rm *}(r_{\rm gal})$, after averaging it in annuli, while $l_{\rm mol}$ values were determined from $\Sigma_{\rm mol}(r_{\rm gal})$ derived from the CO stacking procedure. The left panel of Figure \ref{FIG_14_LENGTH} also shows the ordinary least-square (OLS) bisector fit weighted by the uncertainties for all scale lengths measured with better than $3\sigma$ significance (blue dashed line); we find that  $l_{\rm mol} = [0.89 \pm 0.04] \,l_{\star}$. This result is in agreement with the relation found by \cite{Bolatto2017} for 46 EDGE-CALIFA galaxies, who obtain $l_{\rm mol} = [ 1.05 \pm 0.06 ] \, l_{\star}$. Compared with \cite{Bolatto2017}, however, the CO radial stacking allows us to compute exponential length scales for a larger galaxy sample (68 in our case) and to constrain them better over a broader range of galactocentric radii. Our results are also in agreement with the exponential length scales for HERACLES ($l_{\rm mol} = [0.9 \pm 0.2] \,l_{\star}$; \citealt{Leroy2008}). The inset in the left panel of Figure \ref{FIG_14_LENGTH} shows the relation between $l_{\rm mol}$ and $r_{25}$. Using an OLS bisector fit, we find that  $l_{\rm mol} = [0.22 \pm 0.01]\times r_{25}$, which agrees reasonably with \cite{Young1995}, who find $l_{\rm mol} \approx 0.22 r_{25}$.

In general, resolved molecular gas surveys exhibit similarity between the stellar light and the CO distributions. \cite{Regan2001}, using the CO distribution from the BIMA SONG CO survey, showed that when comparing the scale lengths from exponential fits to the CO and the K-band galaxy profile data for 15 galaxies, the typical CO to stellar scale length ratio is $0.88\pm0.14$. Additionally, single-dish CO measurements plus 3.6$\mu$m data from the HERACLES galaxies show a correspondence between the stellar and molecular disk \citep[][]{Leroy2008,Schruba2011}, with an exponential scale length for CO that follows $l_{\rm mol} \approx 0.2\,r_{25}$. 

If the radial distributions for molecular gas and stars are similar, we would expect the radii containing 50\%\ of the CO emission and the star light to also be similar. The right panel of Figure \ref{FIG_14_LENGTH} demonstrates that our data confirm this expectation, as it shows the relation between the radii that enclose 50\% of the molecular gas and the stellar mass, $R_{\rm 50, mol}$ and $R_{50, \star}$, respectively. The dashed blue line represents an ordinary least-square bisector fit (weighted by the uncertainties) for all our 3$\sigma$ detections; we find that $R_{\rm 50, mol} = [0.95 \pm 0.03]\times R_{50, \star}$. 

Table \ref{Table} summarizes the properties of the 81 EDGE-CALIFA galaxies included in this work and together with the values for $l_{\rm mol}$, $l_{\star}$, $R_{\rm 50, mol}$, and $R_{50, \star}$ for the 68 galaxies analyzed in this section.

\subsection{SFE and Local Parameters}
In this section, we will look at how local physical parameters affect the star formation efficiency of the total gas, SFE$_{\rm gas}$ $=\Sigma_{\rm SFR}/[\Sigma_{\rm atom}+\Sigma_{\rm mol}]$, following methodologies similar to those used by HERACLES \citep{Leroy2008}, against which we will compare results. We compute efficiencies by dividing the star formation rate surface density obtained from H$\alpha$ corrected for extinction using the Balmer decrement (Eq.~\ref{eq:sfr}) by the total gas surface density (Eq.~\ref{eq:sfe}). As discussed in \S\ref{Basic_equations} we assume a constant $\Sigma_{\rm atom}=6$~M$_\odot$\,pc$^{-2}$.

The EDGE-CALIFA galaxies are generally at larger distances ($\sim$23 to 130 Mpc) than the much more local HERACLES sample (3 to 20 Mpc). Both samples have stellar masses spanning a similar range ($\log[M_{\star}/M_{\odot}] = 9.4$--$11.4$), but EDGE a larger representation of more massive disks and bulges as HERACLS includes mostly late Sb and Sc objects and lower mass galaxies. The parent sample CALIFA galaxies are selected in a large volume to allow adequate representation of the $z=0$ population and numbers that allow statistically significant conclusions for all classes of galaxies represented in the survey \citep[][]{Sanchez2012}. The EDGE follow-up selection is biased toward IR-bright objects, but otherwise tries to preserve the variety and volume of the mother sample. CALIFA does not include dwarf galaxies. EDGE otherwise spans a larger range of properties and has a larger sample size than HERACLES, although with lower spatial resolution ($\sim 1.5$~kpc versus $\sim200$~pc). 

We correct our calculations by the inclination of the galaxy (with a $\cos i$ factor, where $i$ is the inclination angle) to represent physical ```face-on'' deprojected surface densities (see \S\ref{Basic_equations}). Our typical $1\sigma$ uncertainty in the SFE$_{\rm gas}$ is $0.22$ dex, dominated by the CO line emission uncertainties derived from the stacking procedure after error propagation. 

\begin{figure}
\includegraphics[width=8.5cm]{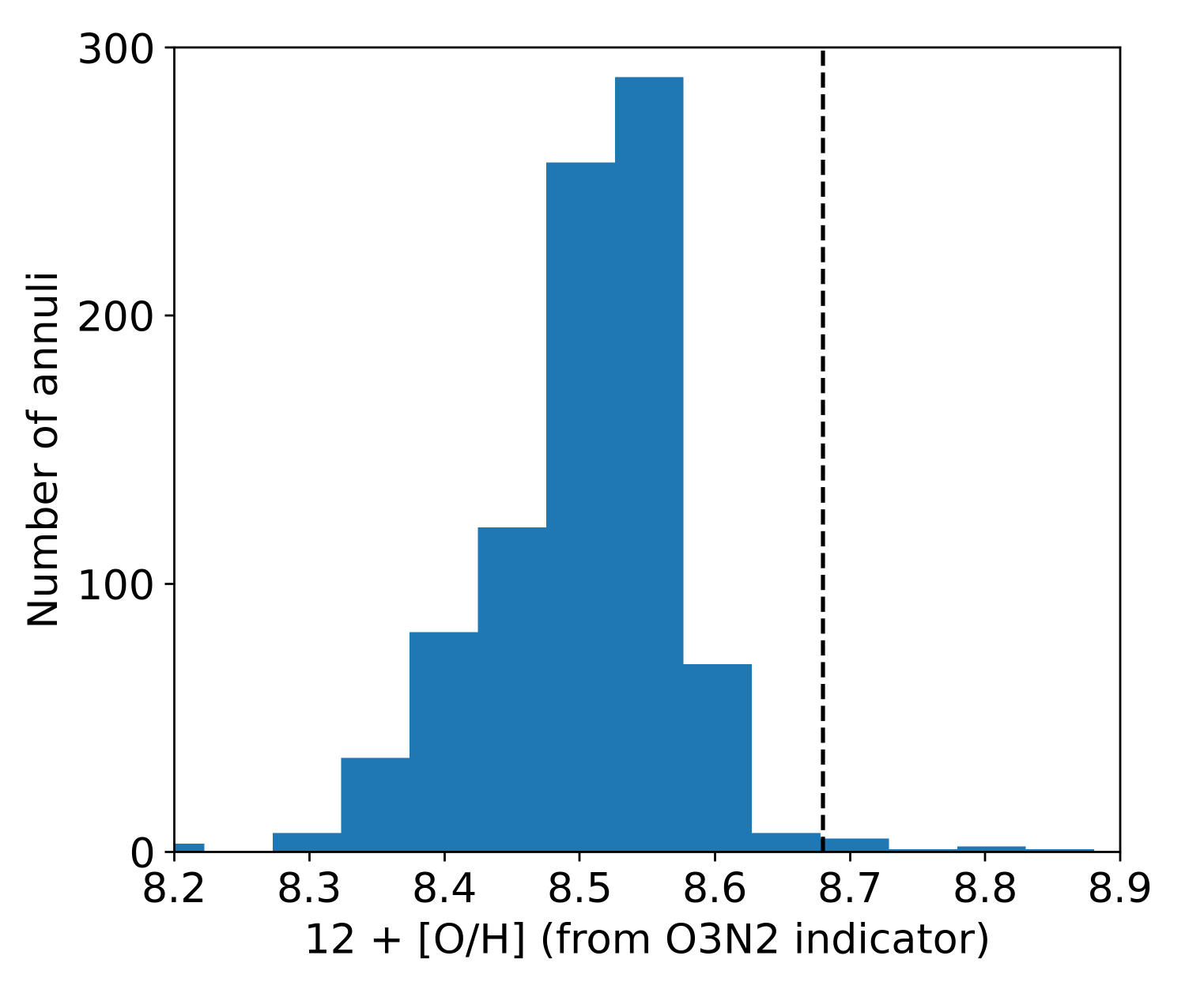}\\
\hspace{-.5cm}\includegraphics[width=9cm]{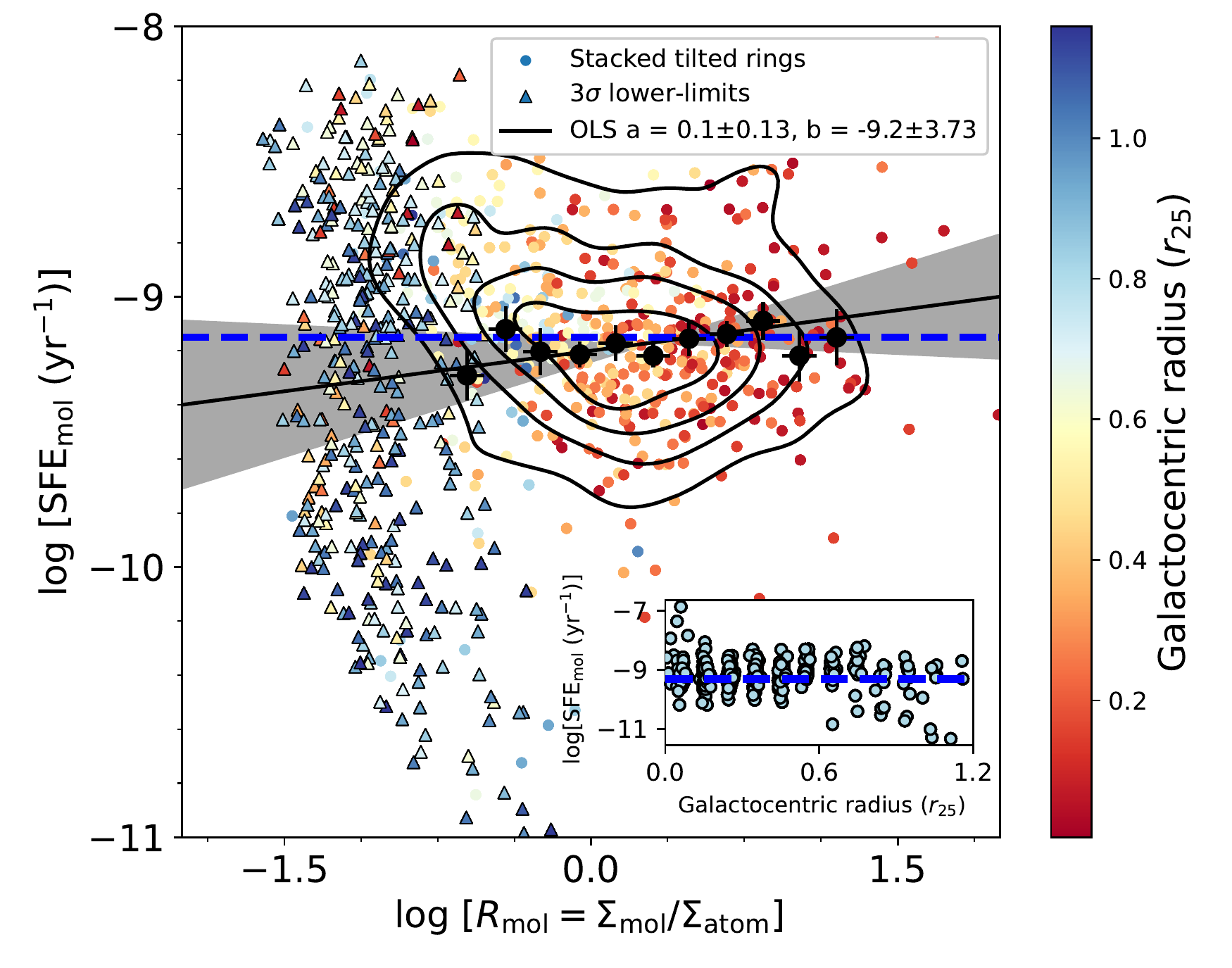}
\caption{{\it Top:} Sample distribution of the oxygen abundances, $\rm 12 + log(O/H)$, with the O3N2 as metallicity indicator. The dashed-black line is the assumed solar value, which corresponds to $\rm 12 + log(O/H)_{\odot}= 8.69$. {\it Bottom:} The star formation efficiency of the molecular gas, SFE$_{\rm mol}= \Sigma_{\rm SFR}/\Sigma_{\rm mol}$, vs the ratio between the molecular and the atomic gas surface densities, $R_{\rm mol}= \Sigma_{\rm mol}/\Sigma_{\rm atom}$. Colors code for galactocentric radius (in $R_{\rm e}$) are as indicated by the color bar. Black contours are 80\%, 60\%, 40\%, and 20\% of the points just for detections. Large black filled circles show the mean of EDGE-CALIFA data at each stellar surface density bin; the error bars are the uncertainties of the mean SFE$_{\rm mol}$ values in each bin. The black-solid line shows the OLS linear bisector fit for averaged points of SFE$_{\rm mol}$ over annuli by using the model $y = a x +b$. The shaded region represents uncertainty of the slope derived from the OLS linear bisector fit. The horizontal dashed-blue line is the average SFE$_{\rm mol}$, including the 3$\sigma$ detection, for the sample. The inset panel shows the SFE$_{\rm mol}$ for detections only as a function of galactocentric radius. The blue-dashed line is the average SFE$_{\rm mol}$.}
\label{FIG_1_Metallicity}
\end{figure}

\begin{table*}
\resizebox{.8\linewidth}{!}{ 
\begin{tabular}{cccccccccc}
\hline\hline
Name & Dist. (Mpc) & $\rm \log[M_{\star}/M_{\odot}]$  & Morph. Class  & $\rm \log[M_{\rm mol}/M_{\odot}]$  & Nuclear & $l_{\star}$ (kpc) & $l_{\rm mol}$ (kpc) & $R_{50, \star}$ (kpc) & $R_{50, \rm mol}$ (kpc) \\
\hline
ARP220 & 78.0 & 10.91$\pm$0.09 & Sm & 9.72$\pm$0.0 & LINER & 2.76$\pm$0.24 & 1.7$\pm$0.35 & 3.88$\pm$0.24 & 2.86$\pm$0.35 \\ 
IC0944 & 100.8 & 11.26$\pm$0.1 & Sa & 10.0$\pm$0.02 & SF & 4.41$\pm$0.11 & 4.26$\pm$0.39 & 7.14$\pm$0.11 & 7.85$\pm$0.39 \\ 
IC1151 & 30.8 & 10.02$\pm$0.1 & SBc & 7.93$\pm$0.14 & ... & 2.66$\pm$0.07 & 2.76$\pm$0.78 & 4.06$\pm$0.07 & 4.61$\pm$0.78 \\ 
IC1199 & 68.3 & 10.78$\pm$0.1 & Sbc & 9.35$\pm$0.04 & SF & 4.53$\pm$0.06 & 3.99$\pm$0.44 & 7.52$\pm$0.06 & 6.97$\pm$0.44 \\ 
IC1683 & 69.7 & 10.76$\pm$0.11 & Sb & 9.68$\pm$0.02 & SF & 5.56$\pm$0.79 & 2.49$\pm$0.13 & 8.92$\pm$0.79 & 3.99$\pm$0.13 \\ 
IC4566 & 80.7 & 10.76$\pm$0.11 & SABb & 9.68$\pm$0.02 & ... & $<$3.55 & $<$4.4 & $<$5.97 & $<$8.35 \\ 
NGC0447 & 79.7 & 11.43$\pm$0.1 & S0-a & 9.33$\pm$0.05 & ... & 4.56$\pm$0.7 & 10.0$\pm$0.23 & 6.58$\pm$0.7 & 9.09$\pm$0.23 \\ 
NGC0477 & 85.4 & 10.9$\pm$0.12 & Sc & 9.54$\pm$0.05 & SF & $<$9.01 & $<$21.78 & $<$14.43 & $<$35.98 \\ 
NGC0496 & 87.5 & 10.85$\pm$0.13 & Sbc & 9.48$\pm$0.04 & SF & 7.35$\pm$0.34 & 4.11$\pm$0.36 & 12.46$\pm$0.34 & 7.13$\pm$0.36 \\ 
NGC0528 & 68.8 & 11.06$\pm$0.1 & S0 & 8.36$\pm$0.13 & ... & ... & ... & ... & ... \\ 
NGC0551 & 74.5 & 10.95$\pm$0.11 & SBbc & 9.39$\pm$0.04 & ... & 4.73$\pm$0.07 & 8.17$\pm$1.68 & 8.01$\pm$0.07 & 13.47$\pm$1.68 \\ 
NGC1167 & 70.9 & 11.48$\pm$0.09 & S0 & 9.28$\pm$0.06 & LINER & ... & ... & ... & ... \\ 
NGC2253 & 51.2 & 10.81$\pm$0.11 & Sc & 9.62$\pm$0.02 & SF & 2.48$\pm$0.07 & 2.1$\pm$0.28 & 3.77$\pm$0.07 & 4.07$\pm$0.28 \\ 
NGC2347 & 63.7 & 11.04$\pm$0.1 & Sb & 9.56$\pm$0.02 & LINER & 2.15$\pm$0.06 & 1.99$\pm$0.38 & 3.86$\pm$0.06 & 4.5$\pm$0.38 \\ 
NGC2486 & 67.5 & 10.79$\pm$0.09 & Sa & $<$9.05 & ... & ... & ... & ... & ... \\ 
NGC2487 & 70.5 & 11.06$\pm$0.1 & Sb & 9.47$\pm$0.05 & ... & $<$4.23 & $<$16.66 & $<$5.88 & $<$24.85 \\ 
NGC2639 & 45.7 & 11.17$\pm$0.09 & Sa & 9.36$\pm$0.02 & LINER & 1.78$\pm$0.01 & 2.88$\pm$0.74 & 2.93$\pm$0.01 & 4.29$\pm$0.74 \\ 
NGC2730 & 54.8 & 10.13$\pm$0.09 & Sd & 9.0$\pm$0.06 & ... & 5.62$\pm$0.62 & 3.79$\pm$0.25 & 9.57$\pm$0.62 & 6.25$\pm$0.25 \\ 
NGC2880 & 22.7 & 10.56$\pm$0.08 & E-S0 & $<$7.93 & ... & ... & ... & ... & ... \\ 
NGC2906 & 37.7 & 10.59$\pm$0.09 & Sc & 9.11$\pm$0.03 & INDEF & 1.72$\pm$0.08 & 1.59$\pm$0.4 & 2.71$\pm$0.08 & 3.0$\pm$0.4 \\ 
NGC2916 & 53.2 & 10.96$\pm$0.08 & Sb & 9.05$\pm$0.06 & AGN & ... & ... & ... & ... \\ 
NGC3303 & 89.8 & 11.17$\pm$0.1 & Sa & 9.57$\pm$0.04 & LINER & 3.62$\pm$0.23 & 1.99$\pm$0.11 & 4.97$\pm$0.23 & 3.47$\pm$0.11 \\ 
NGC3381 & 23.4 & 9.88$\pm$0.09 & SBb & 8.11$\pm$0.08 & ... & ... & ... & ... & ... \\ 
NGC3687 & 36.0 & 10.51$\pm$0.11 & Sbc & $<$8.42 & ... & $<$1.86 & $<$39.56 & $<$2.63 & $<$66.35 \\ 
NGC3811 & 44.3 & 10.64$\pm$0.11 & SBc & 9.28$\pm$0.03 & ... & 2.36$\pm$0.09 & 2.18$\pm$0.26 & 2.93$\pm$0.09 & 2.96$\pm$0.26 \\ 
NGC3815 & 53.6 & 10.53$\pm$0.09 & Sab & 9.16$\pm$0.04 & ... & 2.0$\pm$0.16 & 1.68$\pm$0.27 & 3.05$\pm$0.16 & 3.4$\pm$0.27 \\ 
NGC3994 & 44.7 & 10.59$\pm$0.11 & Sc & 9.26$\pm$0.03 & ... & 1.09$\pm$0.04 & 1.31$\pm$0.08 & 1.78$\pm$0.04 & 2.23$\pm$0.08 \\ 
NGC4047 & 49.1 & 10.87$\pm$0.1 & Sb & 9.66$\pm$0.02 & SF & 2.37$\pm$0.02 & 1.26$\pm$0.25 & 3.9$\pm$0.02 & 3.11$\pm$0.25 \\ 
NGC4185 & 55.9 & 10.86$\pm$0.11 & SBbc & 9.08$\pm$0.07 & INDEF & 4.98$\pm$0.23 & 4.45$\pm$0.85 & 8.19$\pm$0.23 & 7.49$\pm$0.85 \\ 
NGC4210 & 38.8 & 10.51$\pm$0.1 & Sb & 8.86$\pm$0.05 & LINER & ... & ... & ... & ... \\ 
NGC4211NED02 & 96.9 & 10.53$\pm$0.13 & S0-a & 9.29$\pm$0.06 & ... & $<$6.65 & $<$10.52 & $<$8.93 & $<$17.82 \\ 
NGC4470 & 33.4 & 10.23$\pm$0.09 & Sa & 8.59$\pm$0.06 & SF & 1.73$\pm$0.05 & 1.25$\pm$0.29 & 3.04$\pm$0.05 & 2.42$\pm$0.29 \\ 
NGC4644 & 71.6 & 10.68$\pm$0.11 & Sb & 9.2$\pm$0.05 & ... & 2.7$\pm$0.05 & 3.15$\pm$0.8 & 4.91$\pm$0.05 & 5.88$\pm$0.8 \\ 
NGC4676A & 96.6 & 10.86$\pm$0.1 & S0-a & 9.88$\pm$0.02 & SF & ... & ... & ... & ... \\ 
NGC4711 & 58.8 & 10.58$\pm$0.09 & SBb & 9.18$\pm$0.05 & SF & 2.83$\pm$0.06 & 6.24$\pm$0.54 & 4.86$\pm$0.06 & 10.44$\pm$0.54 \\ 
NGC4961 & 36.6 & 9.98$\pm$0.1 & SBc & 8.41$\pm$0.08 & ... & 1.39$\pm$0.08 & 1.59$\pm$0.33 & 2.1$\pm$0.08 & 2.67$\pm$0.33 \\ 
NGC5000 & 80.8 & 10.94$\pm$0.1 & Sbc & 9.45$\pm$0.04 & SF & 5.16$\pm$0.61 & 1.06$\pm$0.26 & 6.51$\pm$0.61 & 2.31$\pm$0.26 \\ 
NGC5016 & 36.9 & 10.47$\pm$0.09 & SABb & 8.9$\pm$0.04 & ... & 1.67$\pm$0.02 & 2.32$\pm$0.42 & 2.89$\pm$0.02 & 3.93$\pm$0.42 \\ 
NGC5056 & 81.1 & 10.85$\pm$0.09 & Sc & 9.45$\pm$0.04 & ... & 4.22$\pm$0.51 & 3.12$\pm$0.48 & 5.44$\pm$0.51 & 5.99$\pm$0.48 \\ 
NGC5205 & 25.1 & 9.98$\pm$0.09 & Sbc & 8.37$\pm$0.07 & LINER & $<$1.57 & $<$2.13 & $<$2.35 & $<$3.61 \\ 
NGC5218 & 41.7 & 10.64$\pm$0.09 & SBb & 9.86$\pm$0.01 & ... & 1.65$\pm$0.08 & 1.43$\pm$0.18 & 2.79$\pm$0.08 & 1.93$\pm$0.18 \\ 
NGC5394 & 49.5 & 10.38$\pm$0.11 & SBb & 9.62$\pm$0.01 & SF & 2.18$\pm$0.27 & 2.7$\pm$0.21 & 3.36$\pm$0.27 & 4.43$\pm$0.21 \\ 
NGC5406 & 77.8 & 11.27$\pm$0.09 & Sbc & 9.69$\pm$0.04 & LINER & 4.97$\pm$0.26 & 7.54$\pm$1.9 & 7.23$\pm$0.26 & 12.77$\pm$1.9 \\ 
NGC5480 & 27.0 & 10.18$\pm$0.08 & Sc & 8.92$\pm$0.03 & LINER & 2.41$\pm$0.1 & 1.27$\pm$0.2 & 4.04$\pm$0.1 & 2.35$\pm$0.2 \\ 
NGC5485 & 26.9 & 10.75$\pm$0.08 & S0 & $<$8.09 & LINER & ... & ... & ... & ... \\ 
NGC5520 & 26.7 & 10.07$\pm$0.11 & Sb & 8.67$\pm$0.03 & ... & 1.19$\pm$0.07 & 0.9$\pm$0.11 & 1.67$\pm$0.07 & 1.77$\pm$0.11 \\ 
NGC5614 & 55.7 & 11.22$\pm$0.09 & Sab & 9.84$\pm$0.01 & ... & 2.25$\pm$0.28 & 1.34$\pm$0.16 & 3.67$\pm$0.28 & 3.1$\pm$0.16 \\ 
NGC5633 & 33.4 & 10.4$\pm$0.11 & Sb & 9.14$\pm$0.02 & SF & 1.36$\pm$0.03 & 1.4$\pm$0.26 & 2.47$\pm$0.03 & 2.61$\pm$0.26 \\ 
NGC5657 & 56.3 & 10.5$\pm$0.1 & Sb & 9.11$\pm$0.04 & ... & 2.11$\pm$0.07 & 1.88$\pm$0.13 & 3.63$\pm$0.07 & 3.37$\pm$0.13 \\ 
NGC5682 & 32.6 & 9.59$\pm$0.11 & Sb & $<$8.29 & SF & 2.11$\pm$0.05 & 1.47$\pm$0.34 & 3.57$\pm$0.05 & 2.21$\pm$0.34 \\ 
NGC5732 & 54.0 & 10.23$\pm$0.11 & Sbc & 8.82$\pm$0.07 & SF & 2.42$\pm$0.09 & 1.78$\pm$0.11 & 3.92$\pm$0.09 & 3.34$\pm$0.11 \\ 
NGC5784 & 79.4 & 0.0$\pm$0.0 & S0 & 9.4$\pm$0.04 & ... & 2.4$\pm$0.32 & 1.41$\pm$0.13 & 3.28$\pm$0.32 & 3.46$\pm$0.13 \\ 
NGC5876 & 46.9 & 10.78$\pm$0.1 & SBab & $<$8.56 & ... & ... & ... & ... & ... \\ 
NGC5908 & 47.1 & 10.95$\pm$0.1 & Sb & 9.94$\pm$0.01 & ... & 2.92$\pm$0.01 & 1.73$\pm$0.34 & 4.98$\pm$0.01 & 4.55$\pm$0.34 \\ 
NGC5930 & 37.2 & 10.61$\pm$0.11 & SABa & 9.33$\pm$0.02 & ... & 1.57$\pm$0.07 & 0.82$\pm$0.03 & 2.66$\pm$0.07 & 1.98$\pm$0.03 \\ 
NGC5934 & 82.7 & 10.87$\pm$0.09 & Sa & 9.81$\pm$0.02 & ... & 3.07$\pm$0.18 & 2.5$\pm$0.17 & 5.17$\pm$0.18 & 4.36$\pm$0.17 \\ 
NGC5947 & 86.1 & 10.87$\pm$0.1 & SBbc & 9.26$\pm$0.06 & AGN & $<$4.15 & $<$4.7 & $<$5.25 & $<$7.83 \\ 
NGC5953 & 28.4 & 10.38$\pm$0.11 & S0-a & 9.49$\pm$0.01 & ... & 1.16$\pm$0.17 & 0.5$\pm$0.07 & 1.3$\pm$0.17 & 1.23$\pm$0.07 \\ 
NGC6004 & 55.2 & 10.87$\pm$0.08 & Sc & 9.33$\pm$0.04 & ... & 5.29$\pm$0.23 & 2.82$\pm$0.21 & 8.18$\pm$0.23 & 4.52$\pm$0.21 \\ 
NGC6027 & 62.9 & 11.02$\pm$0.1 & S0-a0 & 8.01$\pm$0.22 & ... & ... & ... & ... & ... \\ 
NGC6060 & 63.2 & 10.99$\pm$0.09 & SABc & 9.68$\pm$0.03 & SF & 3.85$\pm$0.11 & 4.08$\pm$0.52 & 6.25$\pm$0.11 & 7.59$\pm$0.52 \\ 
NGC6063 & 40.7 & 10.36$\pm$0.12 & Sc & $<$8.53 & SF & 2.65$\pm$0.08 & 3.34$\pm$0.83 & 4.67$\pm$0.08 & 5.99$\pm$0.83 \\ 
NGC6125 & 68.0 & 11.36$\pm$0.09 & E & $<$8.83 & ... & ... & ... & ... & ... \\ 
NGC6146 & 128.7 & 11.72$\pm$0.09 & E & $<$9.36 & ... & ... & ... & ... & ... \\ 
NGC6155 & 34.6 & 10.38$\pm$0.1 & Sc & 8.94$\pm$0.03 & SF & 2.03$\pm$0.06 & 1.97$\pm$0.27 & 3.32$\pm$0.06 & 3.5$\pm$0.27 \\ 
NGC6186 & 42.4 & 10.62$\pm$0.09 & Sa & 9.46$\pm$0.02 & ... & 9.48$\pm$0.45 & 5.9$\pm$0.28 & 14.74$\pm$0.45 & 10.39$\pm$0.28 \\ 
NGC6301 & 121.4 & 11.18$\pm$0.12 & Sc & 9.96$\pm$0.03 & INDEF & 9.45$\pm$0.39 & 13.32$\pm$3.83 & 15.5$\pm$0.39 & 20.01$\pm$3.83 \\ 
NGC6314 & 95.9 & 11.21$\pm$0.09 & Sa & 9.57$\pm$0.03 & INDEF & 6.6$\pm$0.57 & 2.41$\pm$0.07 & 7.43$\pm$0.57 & 4.56$\pm$0.07 \\ 
NGC6394 & 124.3 & 11.11$\pm$0.1 & SBb & 9.86$\pm$0.04 & AGN & 5.0$\pm$0.34 & 5.13$\pm$0.63 & 9.02$\pm$0.34 & 9.25$\pm$0.63 \\ 
NGC7738 & 97.8 & 11.21$\pm$0.11 & Sb & 9.99$\pm$0.01 & LINER & 2.42$\pm$0.2 & 1.95$\pm$0.02 & 3.95$\pm$0.2 & 3.81$\pm$0.02 \\ 
NGC7819 & 71.6 & 10.61$\pm$0.09 & Sb & 9.27$\pm$0.04 & SF & 6.91$\pm$1.06 & 2.6$\pm$0.69 & 9.71$\pm$1.06 & 3.15$\pm$0.69 \\ 
UGC03253 & 59.5 & 10.63$\pm$0.11 & Sb & 8.88$\pm$0.06 & SF & 5.15$\pm$1.22 & 2.91$\pm$0.72 & 5.74$\pm$1.22 & 4.83$\pm$0.72 \\ 
UGC03973 & 95.9 & 10.94$\pm$0.08 & Sb & 9.51$\pm$0.05 & AGN & 3.78$\pm$0.38 & 2.91$\pm$0.11 & 5.3$\pm$0.38 & 6.14$\pm$0.11 \\ 
UGC05108 & 118.4 & 11.11$\pm$0.11 & SBab & 9.75$\pm$0.04 & ... & 4.55$\pm$0.16 & 5.08$\pm$0.78 & 7.56$\pm$0.16 & 7.31$\pm$0.78 \\ 
UGC05359 & 123.2 & 10.86$\pm$0.13 & SABb & 9.65$\pm$0.05 & SF & 5.25$\pm$0.15 & 6.15$\pm$1.09 & 8.95$\pm$0.15 & 11.06$\pm$1.09 \\ 
UGC06312 & 90.0 & 10.93$\pm$0.12 & Sa & $<$9.08 & ... & 3.29$\pm$0.07 & 4.81$\pm$0.44 & 5.41$\pm$0.07 & 8.73$\pm$0.44 \\ 
UGC07012 & 44.3 & 11.0$\pm$2.9 & SBc & 9.9$\pm$0.11 & SF & 2.31$\pm$0.1 & 0.99$\pm$0.17 & 3.36$\pm$0.1 & 2.0$\pm$0.17 \\ 
UGC09067 & 114.5 & 10.96$\pm$0.12 & Sab & 9.83$\pm$0.04 & SF & 3.39$\pm$0.06 & 3.56$\pm$0.34 & 6.11$\pm$0.06 & 6.8$\pm$0.34 \\ 
UGC09476 & 46.6 & 10.43$\pm$0.11 & SABc & 9.15$\pm$0.04 & SF & $<$3.85 & $<$5.37 & $<$5.95 & $<$9.46 \\ 
UGC09759 & 49.2 & 10.02$\pm$0.1 & Sb & 9.07$\pm$0.04 & ... & 2.83$\pm$0.16 & 1.09$\pm$0.17 & 4.5$\pm$0.16 & 1.96$\pm$0.17 \\ 
UGC10205 & 94.9 & 11.08$\pm$0.1 & Sa & 9.6$\pm$0.04 & SF & 5.41$\pm$0.63 & 2.57$\pm$0.39 & 6.09$\pm$0.63 & 5.12$\pm$0.39 \\
\hline
\end{tabular}}
\caption{Main properties of the 81 EDGE-CALIFA galaxies analyzed in this work. The columns Distance, $M_{\star}$, Morphological Class, and $M_{\rm mol}$ are taken from \cite{Bolatto2017}, where $M_{\rm mol}$ is computed using $\alpha_{\rm CO}=4.36$ M$_\odot$ (K km s$^{-1}$ pc$^{2}$)$^{-1}$. The column Nuclear corresponds to the emission-line diagnostic for the optical nucleus spectrum for CALIFA galaxies by \cite{Garcia-Lorenzo2015}, who classify the galaxies (with signal-to-noise larger than three) into star forming (SF), active galactic nuclei (AGN), and LINER-type galaxies. The columns $l_{\star}$, $l_{\rm mol}$, $R_{50,\star}$, and $R_{50,\rm mol}$ are the exponential scale lengths and the radii that enclose 50\% of the molecular gas and the stellar mass computed in Section \ref{Exponential_scale}, respectively.}
\label{Table}
\end{table*}

\subsubsection{SFE and Galactocentric Radius}  
\label{SFE_radius}
Figure \ref{FIG_1_Leroy_2008_1A} shows the relation between SFE$_{\rm gas}$ and galactocentric radius; the four different panels show grouping of the 81 galaxies. Following modern studies, we use $R_{\rm e}$ to normalize galactocentric distances, except when we need to compare to published data which use $r_{25}$. Note that for the EDGE galaxies in this sample, $r_{25} \approx 2.1 R_{\rm e}$. In this figure for clarity we split the Sbc, Sc, and Scd galaxies in two groups by choosing the median of stellar masses of the EDGE-CALIFA sample $\log_{10}$[$M_{\star}$]$=10.7$ \citep{Bolatto2017}. In general, there is a decreasing trend for SFE$_{\rm gas}$ with radius. It is important to note that SFE$_{\rm gas}$ is a fairly smooth function of radius for a given galaxy. In fact, variations between galaxies are frequently  larger than variations between most annuli in a galaxy, indicating that the radial decrease in SFE$_{\rm gas}$ within in a galaxy is often smooth and that galaxy to galaxy variations are significant. 

Figure \ref{FIG_1_Leroy2008_2} shows the radius at which our measured molecular surface density, averaged over an annulus, is the same as our assumed constant surface density in the atomic disk, $\Sigma_{\rm mol}=\Sigma_{\rm atom}=6$~M$_\odot$\,pc$^{-2}$. The typical radius at which this happens is $r/R_{\rm e} \sim [1.1\pm 0.5]$, or $r/r_{25}\sim[0.47\pm0.28]$ (see inset panel), which agrees with the value of $r/r_{25}\sim0.43 \pm 0.18$ found by \citet{Leroy2008}. Note that in Figure \ref{FIG_1_Leroy_2008_1A} the SFE$_{\rm gas}$ is generally smooth across that radius, suggesting that our assumption of a constant $\Sigma_{\rm atom}$ does not play a major role at determining the shape of the total gas SFE$_{\rm gas}$.

Figure \ref{FIG_1_Leroy2001B} shows the average SFE$_{\rm gas}$ as a function of the normalized galactocentric radius for each of the four different groups of morphological classification used in Figure \ref{FIG_1_Leroy_2008_1A}, with $\pm 1 \sigma$ variation indicated by the color bands. We note a systematic increase in the average SFE$_{\rm gas}$ from early type (red shaded area) to late type galaxies (blue shaded area). The SFE$_{\rm gas}$ tend to be lower for the early spirals (i.e., S0 and earlier; ten galaxies), which have a steeper profile when compared with the rest of the morphological groups, and therefore showing a significant anticorrelation between SFE$_{\rm gas}$ and $r_{\rm gal}$ (Pearson correlation coefficient of r$=-0.6$). This steepening may reflect the degree of central concentration seen in earlier-type galaxies. Sd-Ir galaxies show a SFE$_{\rm gas}$ flattening at $r_{\rm gal}<0.45\,r_{25}$; however, their small amount (only 2 galaxies in our sample) does not allow to conclude that this flattening is statistically significant. When looking at the average SFE$_{\rm gas}$ value, over $r_{\rm gal}$ for all the radial profiles (black-circular dots), we find that the SFE$_{\rm gas}$ decreases exponentially even in regions where the gas is mostly molecular. In EDGE we see an continuous exponential profile for the SFE$_{\rm gas}$ averaged over all galaxies (black line in Figure \ref{FIG_1_Leroy2001B}). Although still within the error bars, this is in contrast to HERACLES, which sees a leveling of the SFE$_{\rm gas}$ in the inner regions. The greater range of SFE$_{\rm gas}$s in our sample may be a reflection of the larger range of galaxy spiral types spanned by EDGE compared to HERACLES, which consisted mostly of late types. In fact the Sbc, Sc, and Scd galaxies in EDGE-CALIFA (green band) are very consistent with the measurements of HERACLES. Where the gas is dominated by the atomic component, $r \gtrsim 0.4 r_{25}$, the SFE$_{\rm gas}$ decreases rapidly to the galaxy edge. Because we assume a constant $\Sigma_{\rm atom}$, this is fundamentally a reflection of the rapid decrease of SFR in the atomic disks.  

We can describe the behaviour of the SFE$_{\rm gas}$ for our sample using an ordinary least-square (OLS) linear bisector method to fit a simple exponential decay:

\begin{equation}
{\rm SFE_{\rm gas}} = [0.83\pm 0.07\, {\rm (Gyr^{-1}})]  \exp \left (  \frac{-r_{\rm gal}}{[0.31\pm 0.02]\,r_{25}}\right ).
\label{SFE_radius_fit}
\end{equation}
We note that we do not see clear breaks in this trend; instead, we find a continuous smooth exponential decline of SFE$_{\rm gas}$ as a function of $r_{\rm gal}$. This is consistent with the rapid decline of star formation activity in the outer parts of galaxies \citep[e.g., ][]{Leroy2008,Kennicutt1989,MartinKennicutt2001}, and also is in agreement with previous results for low-redshift star-forming galaxies \citep[e.g., ][]{Sanchez_1_2020,Sanchez_2_2020}. In particular, our results agree with the inside out monotonic decrease of the SFE$_{\rm gas}$ shown by \cite{Sanchez_1_2020}. \cite{Sanchez_1_2020} also find that galaxies are segregated by morphology; for a given stellar mass, they show  that late-type galaxies present larger SFE$_{\rm gas}$ than earlier ones at any $r_{\rm gal}$, which is consistent with the trend we observe in Figure \ref{FIG_1_Leroy2001B}. In the outer parts, our steeper profiles may be influenced by our assumption of constant HI surface density. However, this does not explain our steeper profiles we also observe in the inner galaxy. 
The top and bottom dashed lines in Fig. \ref{FIG_1_Leroy2001B} show how SFE$_{\rm gas}$ changes if instead of 6~M$_\odot$\,pc$^{-2}$ we use $\Sigma_{\rm atom}=3$ and $12$~M$_\odot$\,pc$^{-2}$, which are the two extremes of $\Sigma_{\rm atom}$ values found in HERACLES \citep[][]{Leroy2008}. A better match between EDGE and HERACLES would require using $\Sigma_{\rm atom}=3$~M$_\odot$\,pc$^{-2}$, which appears extremely low. Note that these two studies use different SFR tracers: our extinction-corrected H$\alpha$, may behave differently from the GALEX FUV that dominates the SFR estimate in the outer disks of HERACLES \citep[e.g.,][]{Lee2009}. 

How sensitive is the SFE$_{\rm gas}$ determination to the CO-to-H$_2$ conversion factor? To test this we adopt a variable CO-to-H$_2$ conversion factor, $\alpha_{\rm CO}$, using equation \ref{Alpha_CO}. This includes changes in the central regions caused by high stellar surface densities, and changes due to metallicity. When comparing the effects of a constant and a variable prescription of $\alpha_{\rm CO}$ (shaded area in Figure \ref{FIG_1_Leroy2001B}) we observe that the central regions present larger SFE$_{\rm gas}$ variations than the outer disks within the range of galactocentric distances we study, as the latter do not exhibit $12+\log{\rm (O/H)}$ significantly below 8.4 according to the O3N2 indicator, as shown in the top panel of Figure \ref{FIG_1_Metallicity}. Therefore, the variations of the CO-to-H$_2$ conversion factor are generally small and consistent with the assumption of a constant $\alpha_{\rm CO}$.

So far, we have analyzed the SFE of the total gas, but it is also interesting to test whether the star formation efficiency responds to the phase of the ISM. The bottom panel of Figure \ref{FIG_1_Metallicity} shows the star formation efficiency of the molecular gas, SFE$_{\rm mol} = \Sigma_{\rm SFR}/\Sigma_{\rm mol}$ (in yr$^{-1}$), as a function of the ratio between the molecular and the atomic surface densities, $R_{\rm mol} = \Sigma_{\rm mol}/\Sigma_{\rm atom}$. Since we assume $\Sigma_{\rm atom}=6$ M$_\odot$ pc$^{-2}$, $R_{\rm mol}$ is a prescription for the $\Sigma_{\rm mol}$ normalized by a factor of 6. Although there is large scatter, the figure shows that the SFE$_{\rm mol}$, averaged by $R_{\rm mol}$ bins (filled-black dots), remains almost constant over the $R_{\rm mol}$ range, with an average log[SFE$_{\rm mol}$] $\sim -9.15$ (blue-dashed line in bottom panel of Fig. \ref{FIG_1_Metallicity}). The inset panel shows that the SFE$_{\rm mol}$ is also fairly constant over the range of galactocentric radii. These results are in agreement with \cite{Kazuyuki2019}, who find a similar flattening in SFE$_{\rm mol}$ for annuli at $r \lesssim0.6r_{25}$ when analyzing 80 nearby-spiral galaxies selected from the CO Multi-line Imaging of Nearby Galaxies survey \citep[COMING;][]{Sorai2019}. Using CO, FUV+24$\mu$m and H$_\alpha$+24$\mu$m data for 33 nearby-spiral galaxies selected from the IRAM HERACLES survey \citep{Leroy2009}, \cite{Schruba2011} found that H$_2$-dominated regions are well parameterized by a fixed SFE$_{\rm mol}$ equivalent to a molecular gas depletion time of $\tau_{\rm dep,mol}={\rm SFE^{-1}_{mol}}\sim 1.4$ Gyr, which is consistent with our average $\tau_{\rm dep,mol} \sim 1.45\pm 0.23$ Gyr. As for previous studies, these results support the idea that the vast majority of the star formation activity takes place in the molecular phase of the ISM instead of the atomic gas \citep[e.g.,][]{MartinKennicutt2001,Bigiel2008,Schruba2011}.
 
We explore possible trends between SFE$_{\rm gas}$, galactocentric radius, and nuclear activity. We adopt the nuclear activity classification performed by \cite{Garcia-Lorenzo2015}, who classify CALIFA galaxies (with signal-to-noise larger than three) into star forming (SF), active galactic nuclei (AGN), and LINER-type galaxies, and we apply it, when available, for the 81 galaxies analyzed in this work (see column Nuclear in Table \ref{Table}).  We do not identify significant trends as a function of galactocentric radius for any of these three categories.

\subsubsection{SFE versus Stellar and Gas Surface Density}
\label{SFE_SSD}

\begin{figure}
\begin{tabular}{c}
\hspace{-.5cm}\vspace{-.35cm}\includegraphics[width=8.7cm]{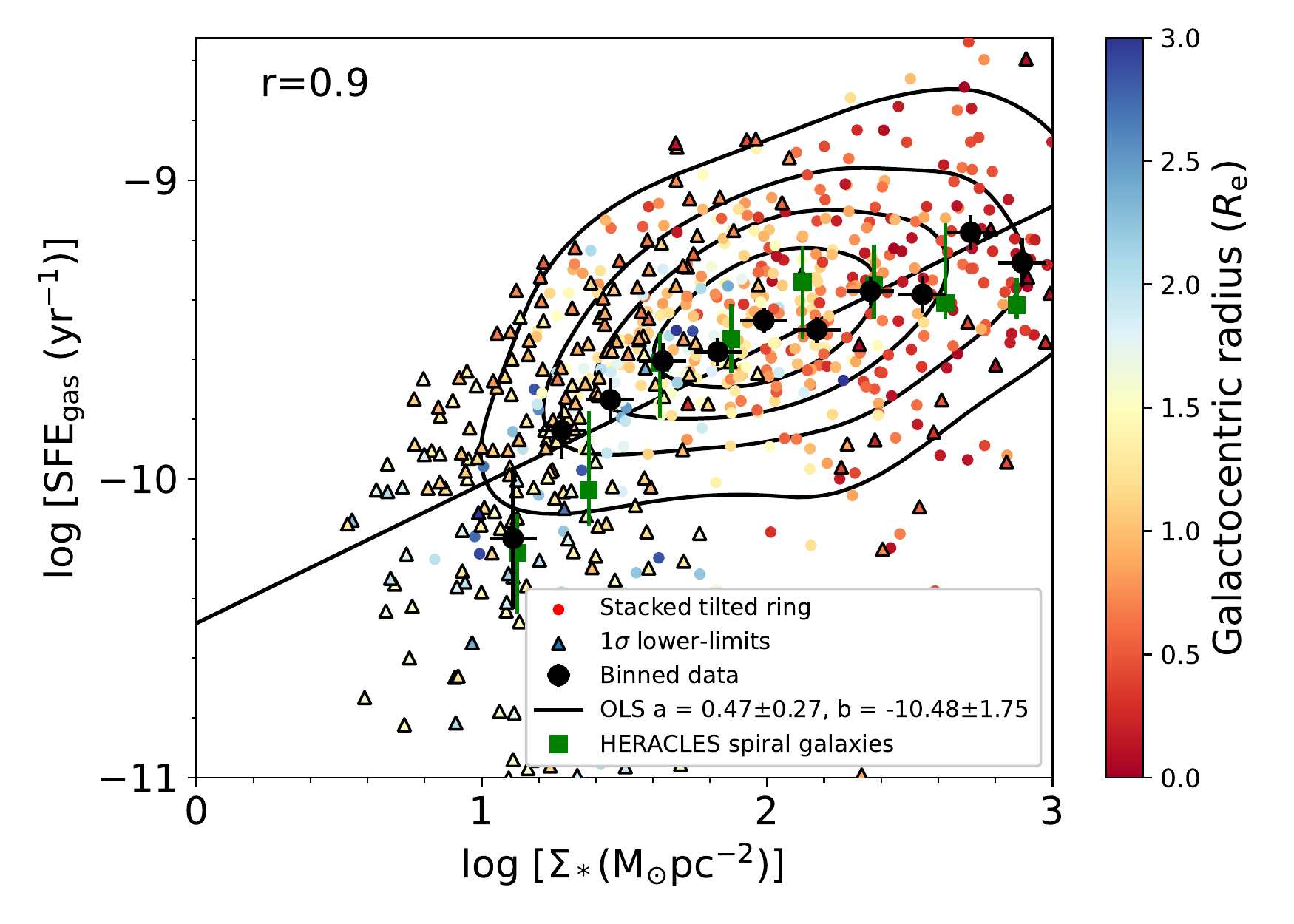} \\
\hspace{-.5cm}\vspace{-.35cm}\includegraphics[width=8.3cm]{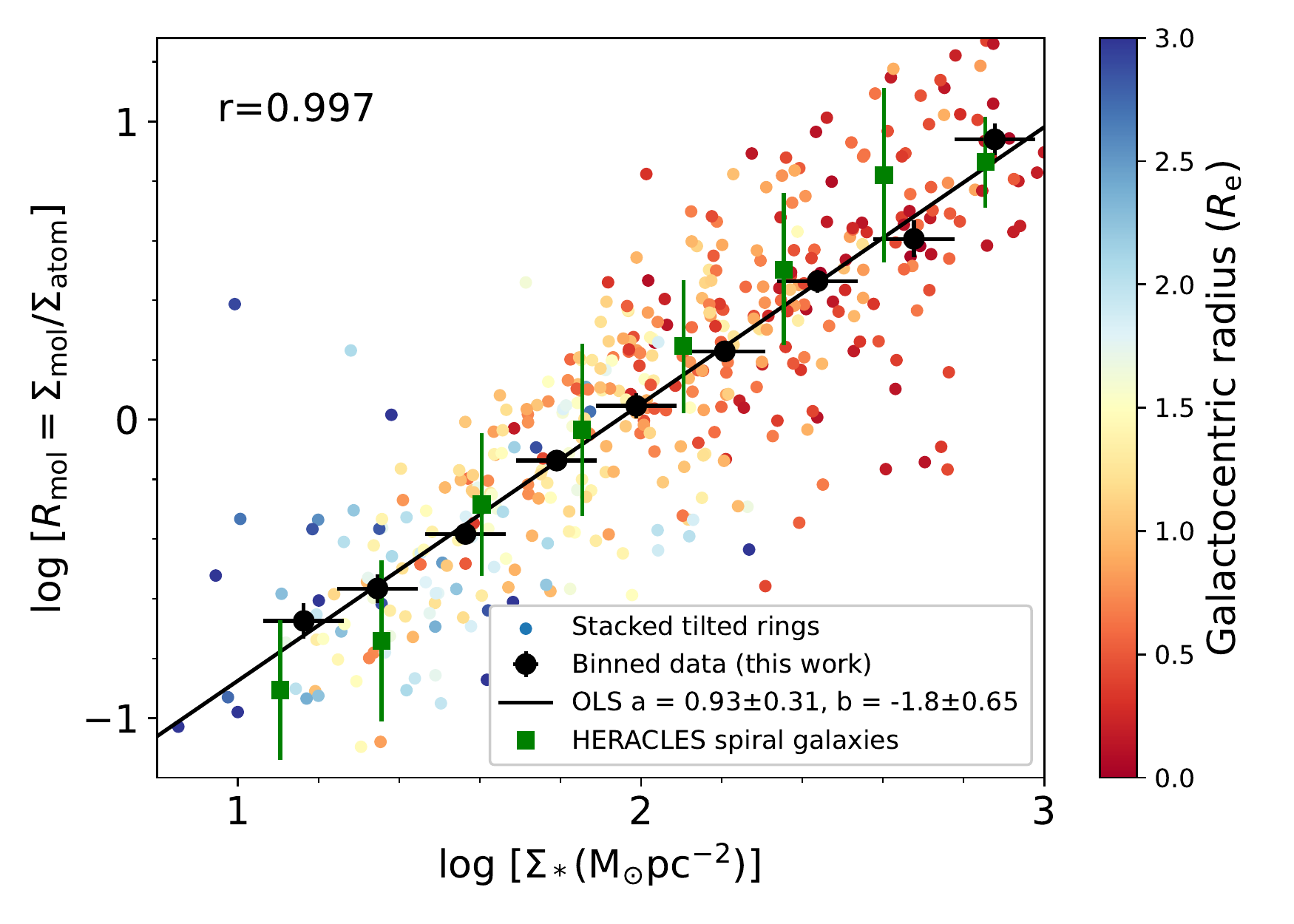} \\
\hspace{-.9cm}\includegraphics[width=8.7cm]{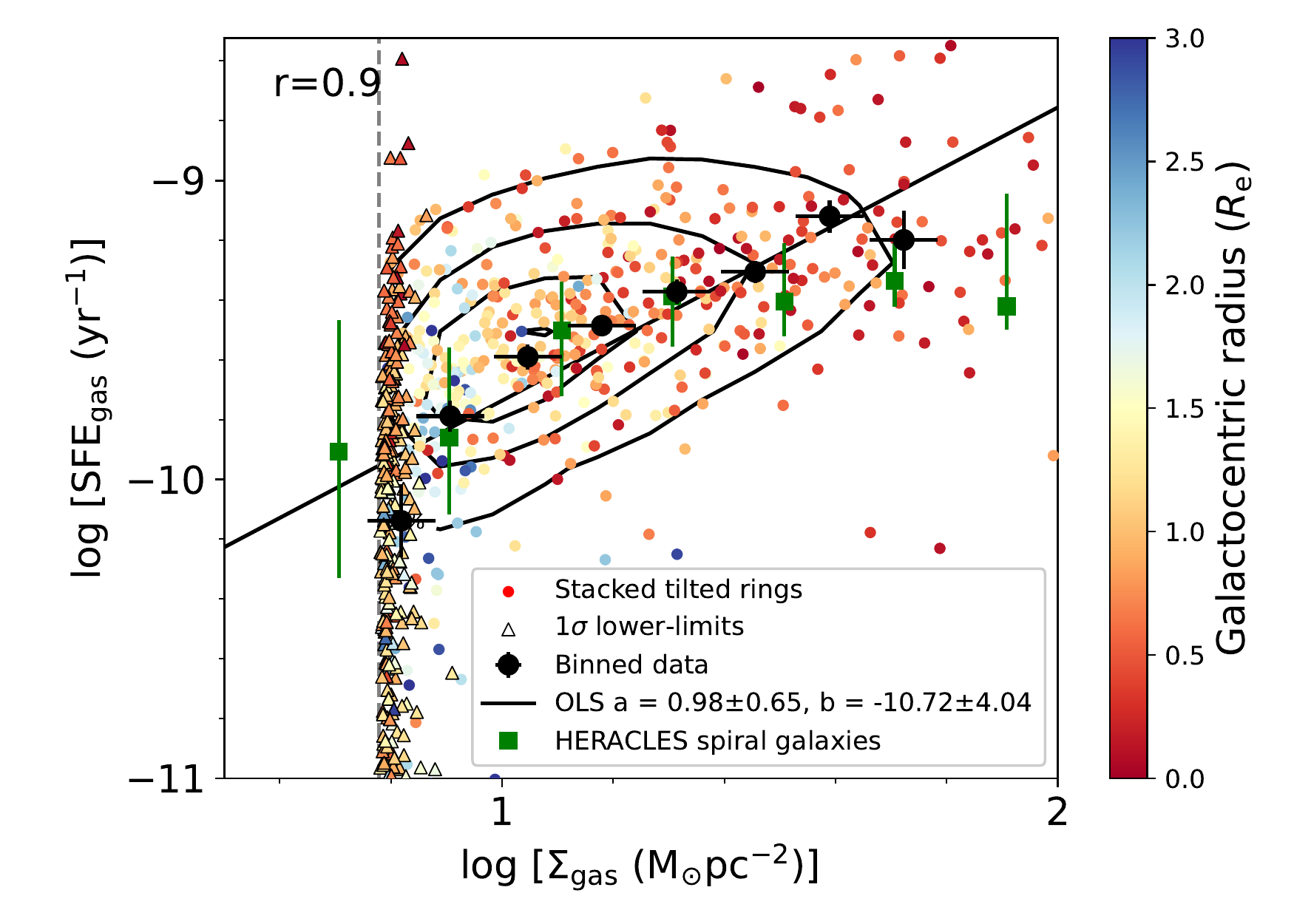}
\end{tabular}
\vspace{-0.5cm}
\caption{{\it Top:} SFE$_{\rm gas}$ as a function of stellar surface density, $\Sigma_{\star}$. Circular dots indicate the average SFE$_{\rm gas}$ and $\Sigma_{\star}$ in stacked annuli for the EDGE-CALIFA data. The `r' term represents the Pearson correlation coefficient, including the binned annuli, for the relation between the SFE$_{\rm gas}$ and $\Sigma_{\star}$. Conventions are as in bottom panel of Figure \ref{FIG_1_Metallicity}. {\it Middle:} The H$_2$-to-H$_{\rm I}$ ratio, $R_{\rm mol}$, as a function of $\Sigma_{\rm \star}$; we use $\Sigma_{\rm atom}= 6$~M$_\odot$\,pc$^{-2}$. Conventions are as in top panel. {\it Bottom:} SFE$_{\rm gas}$ as a function of gas surface density. The grey dashed line marks the point at which $\Sigma_{\rm gas}=\Sigma_{\rm atom}= 6$~M$_\odot$\,pc$^{-2}$. Conventions are as in top panel.}
\label{FIG_3_Leroy2008}
\end{figure}

Since in the previous section we show a clear dependence of SFE$_{\rm gas}$ on galactocentric distance, it is expected that SFE$_{\rm gas}$ will also depend on the stellar surface density, $\Sigma_\star$. Indeed, the top panel of Figure \ref{FIG_3_Leroy2008} shows an approximately power-law relationship between SFE$_{\rm gas}$ and $\Sigma_{\star}$. We quantify this relation by using an OLS linear bisector method in logarithmic space to estimate the best linear fit to our data (excluding upper-limits), obtaining
\begin{multline}
\log[{\rm SFE_{\rm gas}}\,({\rm yr^{-1}})]  = [0.32\pm 0.27]\times \log[\Sigma_\star \, ({\rm M_\odot \, pc^{-2}})]\\
-[10.13\pm1.75].
\label{OLS_Stellar_surface_density}  
\end{multline}

When comparing the EDGE average SFE$_{\rm gas}$, over $\Sigma_\star$ bins (black dots) with similar HERACLES bins (green squares), we find consistently slightly larger efficiencies at $\log[\Sigma_\star (\rm M_\odot\,pc^{-2})]\lesssim 1.4$ although the HERACLES points are still within the error bars of our data. Since these points are in the outer regions of the EDGE galaxies, this result may sensitive to the adoption of $\Sigma_{\rm atom}=6$~M$_\odot$\,pc$^{-2}$.  In the inner regions with $\log[\Sigma_\star (\rm M_\odot\,pc^{-2})]\geq 2.6$, our average efficiencies are also higher, although we do not expect these regions to be sensitive to the choice of $\Sigma_{\rm atom}$. Between these two extremes, however, there is good general agreement between the EDGE and HERACLES results.

The middle and bottom panels of Figure \ref{FIG_3_Leroy2008} show the relation between the H$_2$-to-H$_{\rm I}$ ratio ($R_{\rm mol}=\Sigma_{\rm mol}/\Sigma_{\rm atom} = \Sigma_{\rm mol}/6$ M$_{\odot}$ pc$^{-2}$), $\Sigma_{\rm *}$, and the gas surface density, $\Sigma_{\rm gas}=\Sigma_{\rm mol}+\Sigma_{\rm atom}$, respectively. In the middle panel, we observe a tight correlation between $R_{\rm mol}$ and $\Sigma_{\star}$. The relation is well described by a power-law and there is overall reasonable consistency between EDGE and HERACLES. Our measurements are also consistent with the resolved Molecular Gas Main Sequence relation (rMGMS, $\Sigma_{\rm gas}$-$\Sigma_{\star}$; \citealt{Lin2019}) found for EDGE-CALIFA galaxies by \cite{Barrera-Ballesteros2021}. The bottom panel shows very good agreement between the EDGE and HERACLES results in the range $0.9 \lesssim \log[\Sigma_{\rm gas}] \lesssim 1.5$; outside this range there are small differences, although there is still consistency within the error bars. Therefore, the discrepancies seen in the top panel are not the result of differences in efficiency at a given H$_2$-to-H$_{\rm I}$ ratio nor gas surface density, but likely reflect small systematic differences in the relation between gas and stellar surface density in HERACLES and EDGE. Since we have both a broader morphological and a more numerous sample selection than HERACLES (particularly in the H$_{\rm I}$-dominated regions), our results reflect on a more general power-law dependence of the SFE$_{\rm gas}$ on $\Sigma_{\star}$.  
Observations have shown that the fraction of gas in the molecular phase in which star formation takes place depends on the pressure in the medium \citep[][]{Elmegreen1993,Blitz&Rosolowsky2006}. These results suggest that high stellar densities in the inner regions of EDGE-CALIFA galaxies are helping self-gravity to compress the gas, resulting in H$_2$ dominated regions. Once the gas is predominantly molecular, our data suggests that a dependence of the SFE$_{\rm gas}$ on $\Sigma_\star$ persists even in high $\Sigma_\star$, predominantly molecular regions. 

Other studies have given different insights of the relation between star formation activity and the stellar surface density. For instance, analyzing 34 galaxies selected from the ALMA-MaNGA Quenching and STar formation \citep[ALMaQUEST;][]{Lin2019}, \cite{Ellison2020} find that $\Sigma_{\rm SFR}$ is mainly regulated by $\Sigma_{\rm mol}$, with a secondary dependence on $\Sigma_{\star}$. Conversely, analyzing 39 galaxies selected from EDGE-CALIFA, \cite{Dey2019}  find a strong correlation between $\Sigma_{\rm SFR}$ and $\Sigma_{\rm mol}$; they show that the $\Sigma_{\rm SFR}-\Sigma_{\star}$ relation is statistically more significant. \cite{Sanchez2021}, however, used the {\tt{edge}\_\tt{pydb}} database to show 
that secondary correlations can be driven purely by errors in correlated parameters, and it is necessary to be particularly careful when studying these effects. Errors in $\Sigma_{\rm gas}$, for example, will tend to flatten the relation between SFE$_{\rm gas}$ and $\Sigma_{\rm gas}$ because of the intrinsic correlations between the axes, and will have the same effect on the relation between SFE$_{\rm gas}$ and $\Sigma_*$ because of the positive correlation between $\Sigma_*$ and $\Sigma_{\rm gas}$.

\subsubsection{SFE, Pressure and SFR} 
\label{SFE_Midplane_Gas_Pressure}

We explore the dependency of SFE$_{\rm gas}$ on the dynamical equilibrium pressure, $P_{\rm DE}$. While the midplane gas pressure, $P_{\rm h}$ \citep{Elmegreen1989}, is a well studied pressure prescription in a range of previous works \citep[e.g., ][]{Elmegreen1993,Leroy2008}, $P_{\rm DE}$ has been extensively discussed recently (e.g., \citealt{Kim2013,HerreraCamus2017,Sun2020,Barrera-Ballesteros2021}). In both pressure prescriptions, it is assumed that the gas disk scale height is much smaller than the stellar disk scale height and the gravitational influence from dark matter is neglected. $P_{\rm h}$ and $P_{\rm DE}$ have an almost equivalent formulation, although they slightly differ in the term related to the gravitational influence from the stellar component (second term in equations \ref{Midplane_gas_pressure} and \ref{Dynamical_equilibrium_pressure}; see section \ref{Basic_equations}). We quantify this difference by computing the mean $P_{\rm DE}$-to-$P_{\rm h}$ ratio averaged in annuli for our sample, obtaining $P_{\rm DE}/P_{\rm h}\approx 1.51\pm 0.19$. We use this value to convert the $P_{\rm h}$ from HERACLES into $P_{\rm DE}$, since we perform our qualitative analysis using the dynamical equilibrium pressure.

\begin{figure}
 \begin{tabular}{c}
  \hspace{-1.2cm}\includegraphics[width=9.5cm]{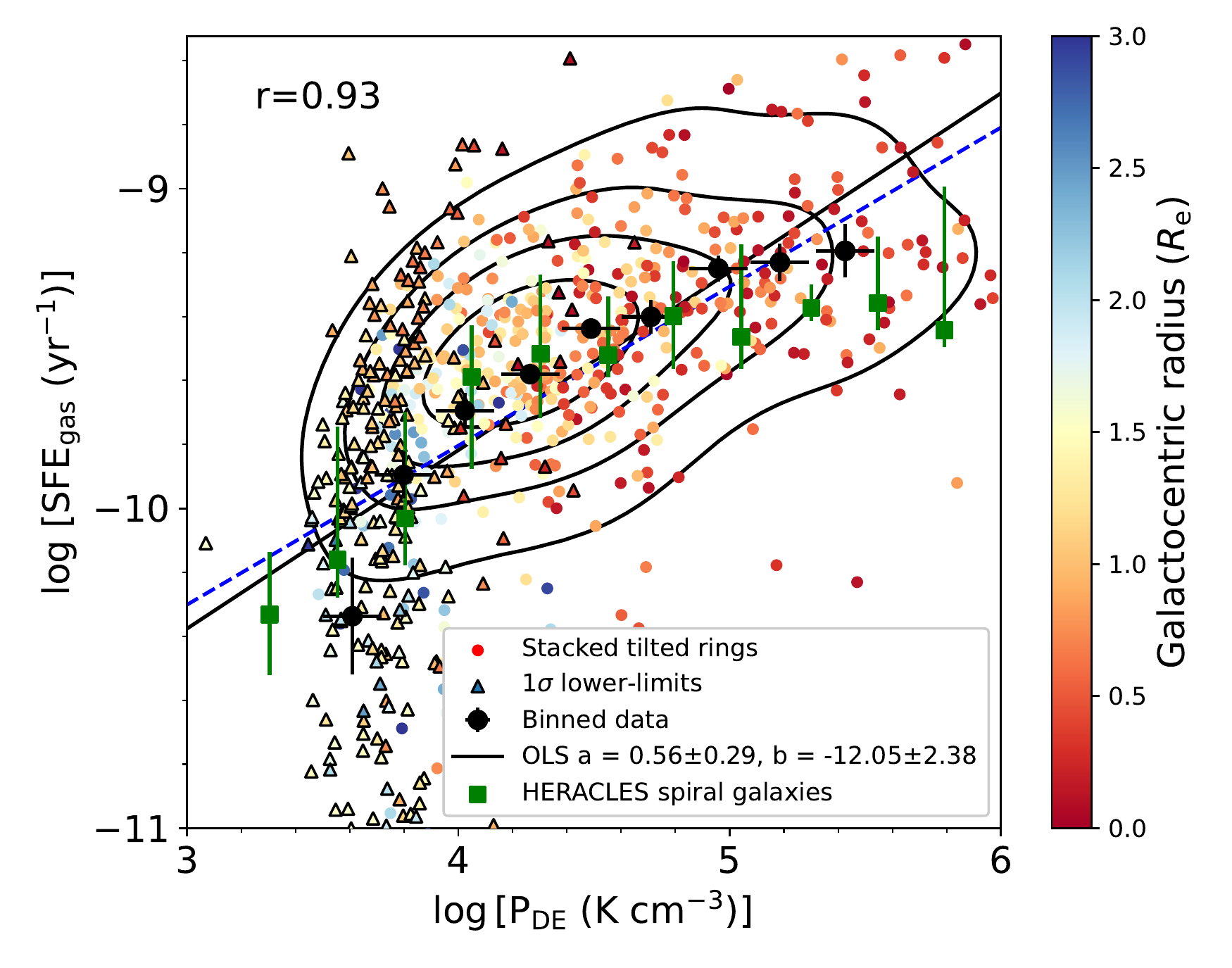} \\
  \hspace{-1.2cm}\includegraphics[width=8.8cm]{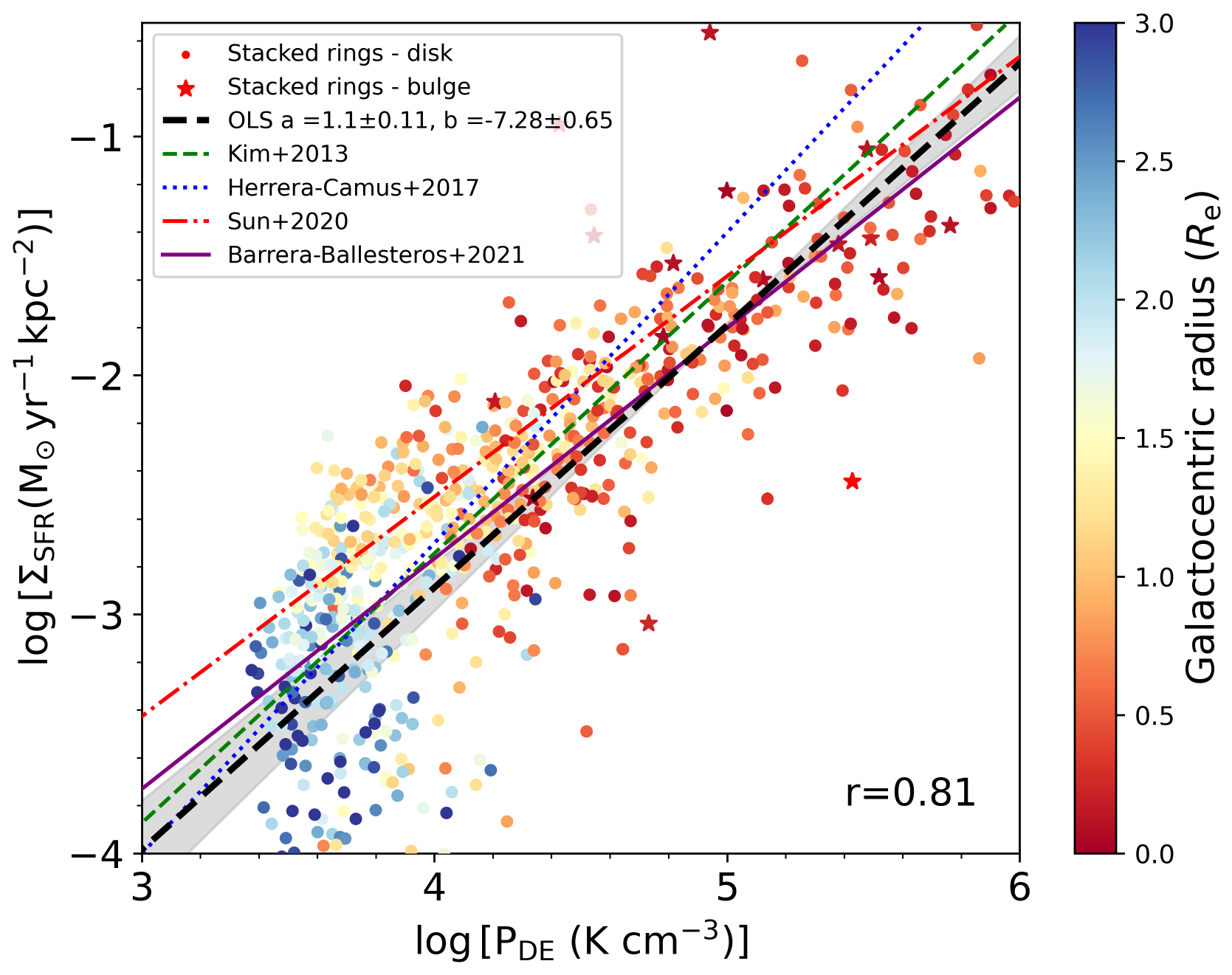}
  \end{tabular}
  \caption{{\it Top:} SFE$_{\rm gas}$ as a function of dynamical equilibrium pressure, $P_{\rm DE}$. The dashed-blue line corresponds to 1\% of gas converted to stars per disk free-fall time. {\it Bottom:} Star formation rate surface density, $\Sigma_{\rm SFR}$, as a function of $P_{\rm DE}$. The black dashed line is the OLS linear bisector fit for all points. The 'r' term is the Pearson correlation coefficient of this subsample. The shaded region represents uncertainty of the slope derived from the OLS linear bisector fit. Conventions are as in Figure \ref{FIG_3_Leroy2008}.}
  \label{FIG_6_Leroy2008}
\end{figure}

The top panel of Figure \ref{FIG_6_Leroy2008} shows the SFE$_{\rm gas}$ as a function of $P_{\rm DE}$ (in units of ${\rm K\, cm^{-3}})$. The slope of the SFE$_{\rm gas}$ vs $P_{\rm DE}$ relation  (averaged over $P_{\rm DE}$ bins (black dots) has a break at $\log[P_{\rm DE}]\sim 3.7$. Below $\log[P_{\rm DE}]\lesssim 3.7$ (i.e., where the ISM is H$\rm I$-dominated) we do not see a clear correlation between SFE$_{\rm gas}$ and $P_{\rm DE}$. This is at the sensitivity limit existing data for EDGE,  but it  is also consistent with the overall behaviour seen in HERACLES corresponding to a steepening of their mean relation. Above this pressure we find a clear linear trend in log-log space. For higher $P_{\rm DE}$ values (e.g., H$_{\rm 2}$-dominated regions) the EDGE average efficiencies are somewhat higher than observed in HERACLES, which flatten out at high $P_{\rm DE}$) although with a scatter that is within the respective 1$\sigma$ error bars. For $\log[P_{\rm DE}]\gtrsim 3.7$ the EDGE average efficiencies are well described by the blue-dashed line, which corresponds to $1\%$ of the gas converted to stars per disk free-fall time, $\tau_{\rm ff} = (G\rho)^{-1/2}$. To quantify this relation, we use an OLS linear bisector method to estimate the best linear fit to our data, obtaining
\begin{multline}
\log[{\rm SFE_{\rm gas}} \,({\rm yr^{-1}})] = [0.41\pm 0.29] \times \log[P_{\rm DE}/k \,({\rm K cm^{-3}})]\\ 
- [11.32 \pm 2.24].
\label{OLS_SFE_P_DE}
\end{multline}
The bottom panel of Figure \ref{FIG_6_Leroy2008} shows the $\Sigma_{\rm SFR}$ versus $P_{\rm DE}$, color-coded by galactocentric radius. When compared with other recent measurements (e.g., KINGFISH, \citealt{HerreraCamus2017}; PHANGS, \citealt{Sun2020}), our annuli have the advantage of covering a somewhat wider dynamic range in both $\Sigma_{\rm SFR}$ and $P_{\rm DE}$. We find a strong correlation between $\Sigma_{\rm SFR}$ and $P_{\rm DE}$ that is approximately linear for annuli at $\log[P_{\rm DE}/k] \gtrsim 3.7$, although below this limit we observe a break in the trend. As shown by the color coding of the  symbols, indicating $r_{\rm gal}$ in Figure \ref{FIG_6_Leroy2008}, this limit is apparently related to the $r_{\rm gal}$ at which the transition from H$_2$-dominated to HI-dominated annuli happens. This transition may be due to the large range of physical properties covered by our sample, which span from molecular dominated to atomic dominated regimes. Where the ISM weight is higher (e.g., H$_2$-dominated regions), the SFR is stabilized by the increasing feedback from star formation to maintain the pressure that counteracts the $P_{\rm DE}$ \citep{Sun2020}. The lack of correlation we observe at $\log[P_{\rm DE}/k] \lesssim 3.7$ ($r\gtrsim 0.7$) is mainly because we are reaching our CO sensitivity in the HI-dominated regions. To quantify the correlation, we estimate the best linear fit by using an OLS linear bisector method in logarithmic space for annuli at $r\gtrsim 0.7$,  

\begin{multline}
\log[\Sigma_{\rm SFR} \,({\rm M_\odot \,yr^{-1}})] = [1.10\pm 0.11] \times \log[P_{\rm DE}/k \, ({\rm K\, cm^{-3}})] \\ 
- [7.28 \pm 0.65].
\label{OLS_P_DE}
\end{multline}

\noindent Note that these results are potentially sensitive to the method we employ for the fitting. Nonetheless, using an orthogonal distance regression (ODR) to fit the same subsample, we obtain very comparable values $\log[\Sigma_{\rm SFR} \,({\rm M_\odot \,yr^{-1}})] = [1.09\pm 0.05] \times \log[P_{\rm DE}/k \, ({\rm K cm^{-3}})] - [7.25 \pm 0.25]$. \cite{Barrera-Ballesteros2021} analyze 4260 resolved star-forming regions of kpc size located in 96 galaxies from the EDGE-CALIFA survey, using a similar sample selection (e.g., inclination, $\sigma_{\rm gas}$ and $\Sigma_{\rm atom}$ constant values, among others) but they just consider Equivalent Widths for the H$\alpha$ line emission $\rm EW(H\alpha)>20 \AA$. Using an ODR fitting technique, they obtain $\log[\Sigma_{\rm SFR} \,({\rm M_\odot \,yr^{-1}})] = [0.97\pm 0.05] \times \log[P_{\rm DE}/k \, ({\rm K cm^{-3}})] - [7.88 \pm 0.48]$, which is in agreement with the distribution shown in the bottom panel of Figure \ref{FIG_6_Leroy2008}. The figure also shows that the correlation agrees with hydrodynamical simulations performed by \cite{Kim2013} (green dashed line), in which they obtain a slope of $1.13$. These results are also consistent with measurements obtained in other galaxy samples. \cite{Sun2020} obtain a slope of $0.84\pm0.01$ for 28 well-resolved CO galaxies ($\sim {1}''.5$, corresponding to $\sim100$ pc) selected from the ALMA-PHANGS sample by using a methodology very similar to ours. Smaller slopes have been referenced in local very actively star-forming galaxies (e.g., local ultra luminous infra-red galaxies, ULIRGs), which at the same time may resemble some of the conditions in high-redshift sub-millimeter galaxies \citep[e.g.,][]{Ostriker&Shetty2011}. \cite{HerreraCamus2017} analyzed the [CII] emission in atomic-dominated regions of 31 KINGFISH galaxies to determine the thermal pressure of the neutral gas and related it to $P_{\rm DE}$, obtaining a slope of $1.3$ (dotted blue line). Our results bridge these two extremes; the strong correlation between $\Sigma_{\rm SFR}$ and $P_{\rm DE}$ and its linearity supports the idea of a feedback-regulated scenario, in which star formation feedback acts to restore  balance in the star-forming region of the disk \citep{Sun2020}.

\subsubsection{SFE and Orbital Timescale}
\label{Orbital_timescale_Section}

\begin{figure}
\hspace{-0.6cm}
\begin{tabular}{c}
  \includegraphics[width=8.5cm]{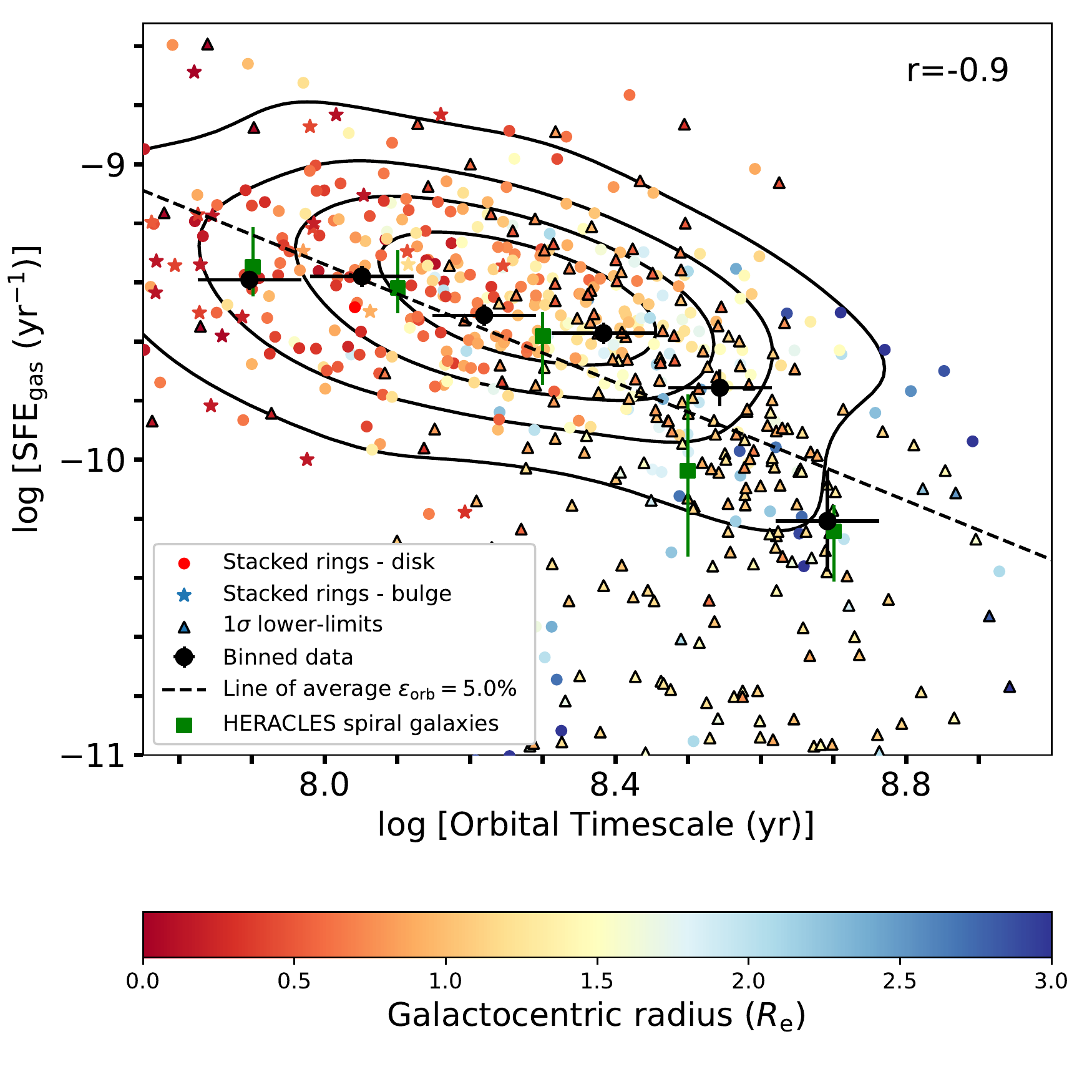}\\
  \hspace{-.3cm}\includegraphics[width=8.5cm]{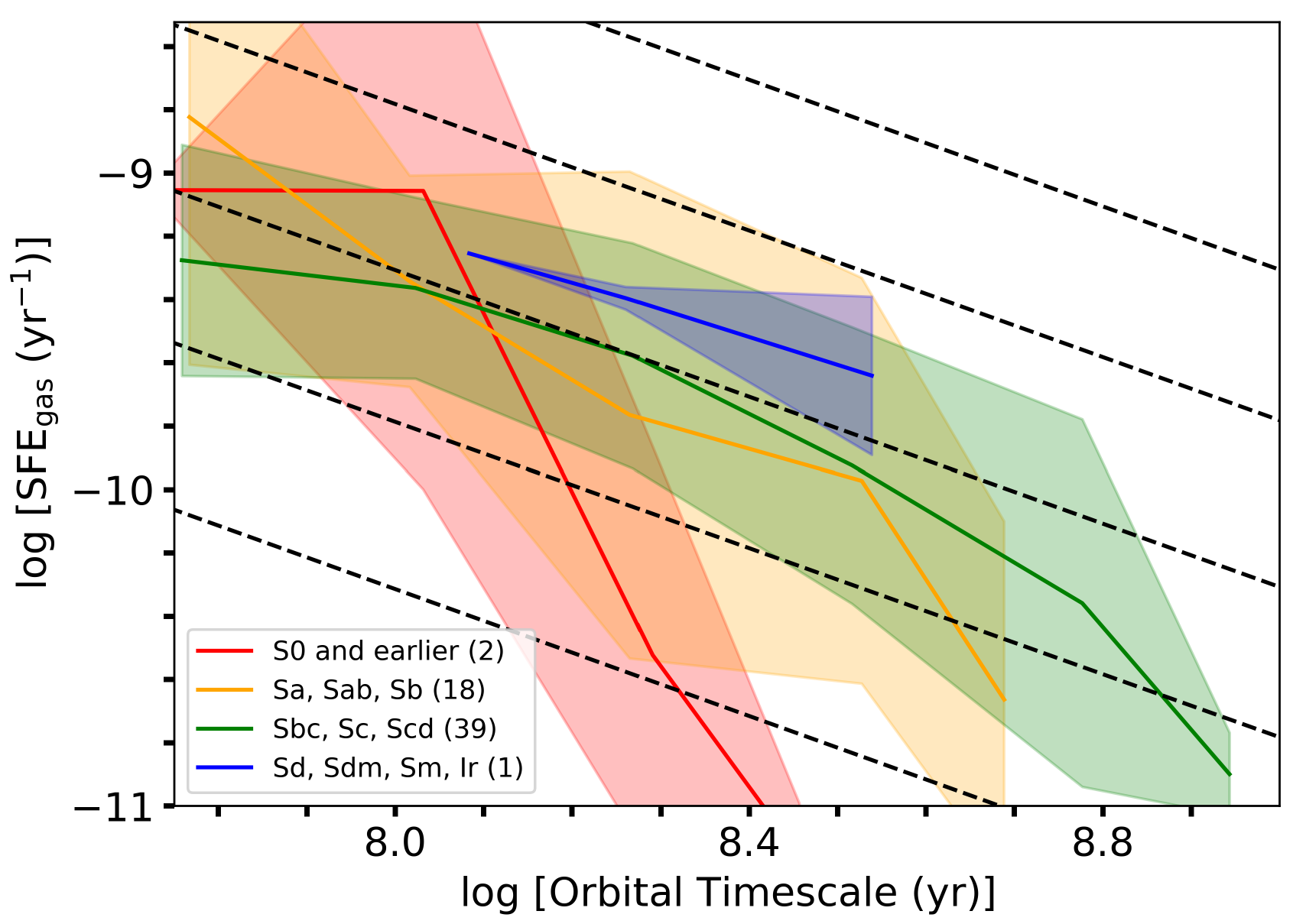}  
  \end{tabular}
  \caption{{\it Top:} SFE$_{\rm gas}$ as a function of the orbital timescale, $\tau_{\rm orb}$. Color coding and symbols are as described in Figure \ref{FIG_3_Leroy2008}. The black-dashed line is the best fit of the binned data and shows 5\% of gas converted into stars per $\tau_{\rm orb}$. The `r' term represents the Pearson correlation coefficient, including the binned annuli, for the relation between the SFE$_{\rm gas}$ and $\tau_{\rm orb}$. {\it Bottom:} SFE$_{\rm gas}$ averaged over $\tau_{\rm orb}$ bins over all galaxies of selected morphological types as in Figure \ref{FIG_1_Leroy2001B}. Black-dashed lines, from top to bottom, represent the 50\%, 17\%, 5\%, 1.7\%, and 0.5\% efficiency of gas converted into stars per $\tau_{\rm orb}$.}
  \label{FIG_7_Leroy2008}
\end{figure}

In the next two sections, we exclude 21 galaxies (out of the 81) since their H$\alpha$ rotation curves \citep[taken from][]{Levy2018} are either too noisy or not well fitted by the universal rotation curve parametric form. The top panel of Figure \ref{FIG_7_Leroy2008} shows SFE$_{\rm gas}$ versus $\tau_{\rm orb}$, the orbital timescale (in units of yr), color-coded by galactocentric radius. When analyzing our efficiencies averaged over orbital timescale bins (black symbols), we note that there is a slightly flattening of the SFE$_{\rm gas}$ at $\log [\tau_{\rm orb}]\sim 7.9-8.1$. We also note that annuli at $\log [\tau_{\rm orb}]\lesssim 8.1$ are usually within the bulge radius in the SDSS $i$-band (reddish-stars symbols). However, the error bars are consistent with SFE$_{\rm gas}$ decreasing as a function of $\tau_{\rm orb}$ including at $\log[\tau_{\rm orb}] < 8.1$. These results are in agreement with what is found in other spatially resolved galaxy samples \citep[e.g.,][]{WongBlitz2002,Leroy2008}. The average gas depletion time for our subsample, $\tau_{\rm dep}=\Sigma_{\rm gas}/\Sigma_{\rm SFR} \approx 2.8^{1.1}_{-1.0}$ Gyr, which agrees farily with the depletion time $\tau_{\rm dep} = 2.2$ Gyr found for HERACLES \citep[not including early-type galaxies;][]{Leroy2013}. \cite{Utomo2017} computed the depletion times for 52 EDGE-CALIFA galaxies using annuli in the region within 0.7 $r_{25}$ (just considering the molecular gas); their average $\tau_{\rm dep} \approx 2.4$ Gyr is in good agreement with our results.

The orbital timescale has a strong correlation with radius, and theoretical arguments expect SFE$_{\rm gas}$ to be closely related to orbital timescale in typical disks \citep{Silk1997,Elmegreen1997,Kennicutt1998}. A correlation between SFE$_{\rm gas}$ and $\tau_{\rm orb}$ is based on the ``Silk-Elmegreen'' relation, which states that $\Sigma_{\rm SFR}=\epsilon_{\rm orb}\,\Sigma_{\rm gas}/\tau_{\rm orb}$, where $\epsilon_{\rm orb}$ is the fraction of the gas converted into stars per orbital time (also called ``orbital efficiency''). Therefore, because ${\rm SFE_{\rm gas}}=\Sigma_{\rm SFR}/\Sigma_{\rm gas}$, SFE$_{\rm gas}$ and $\tau_{\rm orb}$ are related by
\begin{equation}
{\rm SFE_{\rm gas}} = \frac{\epsilon_{\rm orb}}{\tau_{\rm orb}}.
\label{SE_relation}
\end{equation}

It is interesting to analyze the relations between the different timescales since they can give intuition about the physical processes underlying the star formation activity \citep[e.g.,][]{SemenovKravtsovGnedin2017,Colombo2018}. Equation \ref{SE_relation} shows that the timescale to deplete the gas reservoir and the orbital timescale are related through $\epsilon_{\rm orb}$. Although there is large scatter, the median values of $\tau_{\rm orb}$ and $\tau_{\rm dep}$ for our sample are $(2.0^{+0.9}_{-0.7}) \times 10^{8}$ yr and $(2.8^{+1.1}_{-1.0}) \times 10^{9}$ yr, respectively. These values are in good agreement with previous EDGE-CALIFA sample results found by \cite{Colombo2018}, who analyze a more limited subsample of 39 galaxies without the benefit of CO line stacking and more constrained to inclination below $65^{\circ}$, with $\tau_{\rm orb} = (3.2^{+2.0}_{-1.2})\times 10^{8}$ yr and $\tau_{\rm dep} = (2.8^{+2.3}_{-1.2})\times 10^{9}$ yr.  The black-dashed line in the top panel of Figure \ref{FIG_7_Leroy2008} corresponds to the best fit to our binned data (black symbols); our fit excludes lower limits (shown as triangles in the figure), and it shows that $\epsilon_{\rm orb}\approx 5\%$ of the total gas mass is converted to stars per $\tau_{\rm orb}$. This average efficiency is lower but similar to the $\epsilon_{\rm orb}\approx7\%$ of efficiency found by \cite{WongBlitz2002} and \cite{Kennicutt1998}, and the $\epsilon_{\rm orb}\approx6\%$ efficiency for HERACLES \citep{Leroy2008}. Also, this efficiency is the same to the average molecular gas orbital efficiency found by \cite{Colombo2018} for their subsample of EDGE-CALIFA galaxies ($\epsilon_{\rm orb}\approx5\%$). Similar to our results, all of these studies did not find a clear correlation between SFE$_{\rm gas}$ and $\tau_{\rm orb}$ in the inner regions of disks, where the ISM is mostly molecular.

Like \cite{Colombo2018}, however, we find that a constant $\epsilon_{\rm orb}$ is not a good approximation for the data. The efficiency per orbital time depends on the Hubble morphological type, with $\epsilon_{\rm orb}$ increasing from early- to late-types. This is shown in the bottom panel of Figure \ref{FIG_7_Leroy2008}, which shows the data grouped according to the same four morphological classes used in Figure \ref{FIG_1_Leroy2001B}. Our results show that annuli from Sbc, Sc, and Scd galaxies, which are the most numerous in our sample, seem to group around $\epsilon_{\rm orb}\sim5\%$. This value is also representative of the typical $\epsilon_{\rm orb}$ seen for the morphological bins comprised by Sa, Sab, and Sb and Sd, Sdm, Sm, and Ir types in the range $8.0<\log[\tau_{\rm orb}]<8.4$. However, these groups also show $\epsilon_{\rm orb}\lesssim5\%$ in the ranges $\log[\tau_{\rm orb}]<8.0$ and $\log[\tau_{\rm orb}] > 8.4$. However, early-type galaxies (with admittedly limited statistics, 21 annuli in total) show substantially lower $\epsilon_{\rm orb}$, with a median of $\epsilon_{\rm orb}=1.2\%$. These values are in agreement with previous results for EDGE-CALIFA galaxies by \cite{Colombo2018}, even though sample selection and processing were different. They observe a $\epsilon_{\rm orb}\sim 10\%$ for Sbc galaxies (most numerous in their sub-sample), and a systematic decrease in orbital efficiencies from late- to early-type galaxies. 

As concluded in \cite{Colombo2018}, our results support the idea of a non-universal efficiency per orbit for the ``Silk-Elmegreen'' law. Figure \ref{FIG_7_Leroy2008} shows that $\epsilon_{\rm orb}$ depends not just on morphological type, but the behavior also varies with galactocentric radius: at short orbital time scales ($\log[\tau_{\rm orb}]\lesssim8.3$), or small radii ($\log[r/R_{\rm e}]\lesssim1.1-1.3$) the efficiency per unit time SFE$_{\rm gas}$ tends to be constant, and as a consequence the observed $\epsilon_{\rm orb}$ tends to systematically decrease as $\tau_{\rm orb}$ decreases. This is best seen in the top panel in the departure of the binned data (black symbols) from the dashed line of constant $\epsilon_{\rm orb}$. Note that this is also the approximate radius of the molecular disk, the region where molecular gas dominates the gaseous disk (Figure~\ref{FIG_1_Leroy2008_2}). 

Other studies have also reported SFE$_{\rm gas}$ deviations as a function of morphology. \cite{Koyama2019} analyze CO observations of 28 nearby galaxies to compute the C-index $= R90_{petro,r}/R50_{petro,r}$ as an indicator of the bulge dominance in galaxies (where $R90_{petro,r}$ and $R50_{petro,r}$ are the radius containing the 90\% and 50\% of Petrossian flux for SDSS $r$-band photometric data, respectively). Although they do not detect a significant difference in the SFE$_{\rm gas}$ for bulge- and disk-dominated galaxies, they identify some CO-undetected bulge-dominated galaxies with unusual high SFE$_{\rm gas}$s. Their results may reflect the galaxy population during the star formation quenching processes caused by the presence of a bulge component, and they could explain the flattening shown in top panel (mostly dominated by annuli within bulges) and bottom panel (mainly due by early-type and Sb-Scd galaxies annuli) of Figure \ref{FIG_7_Leroy2008}.

\begin{figure*}
\begin{center}
  \includegraphics[width=8.cm]{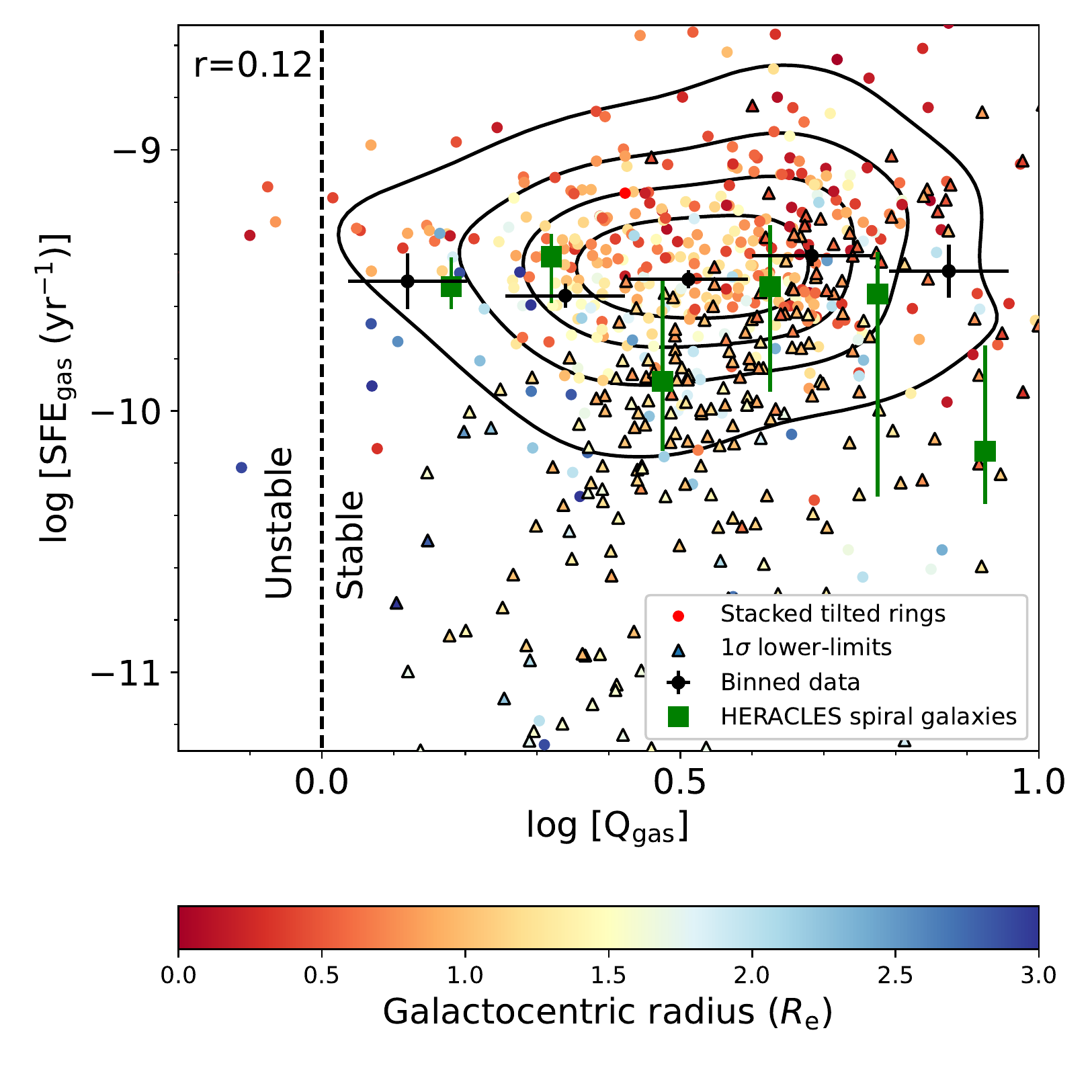}
  \includegraphics[width=8.cm]{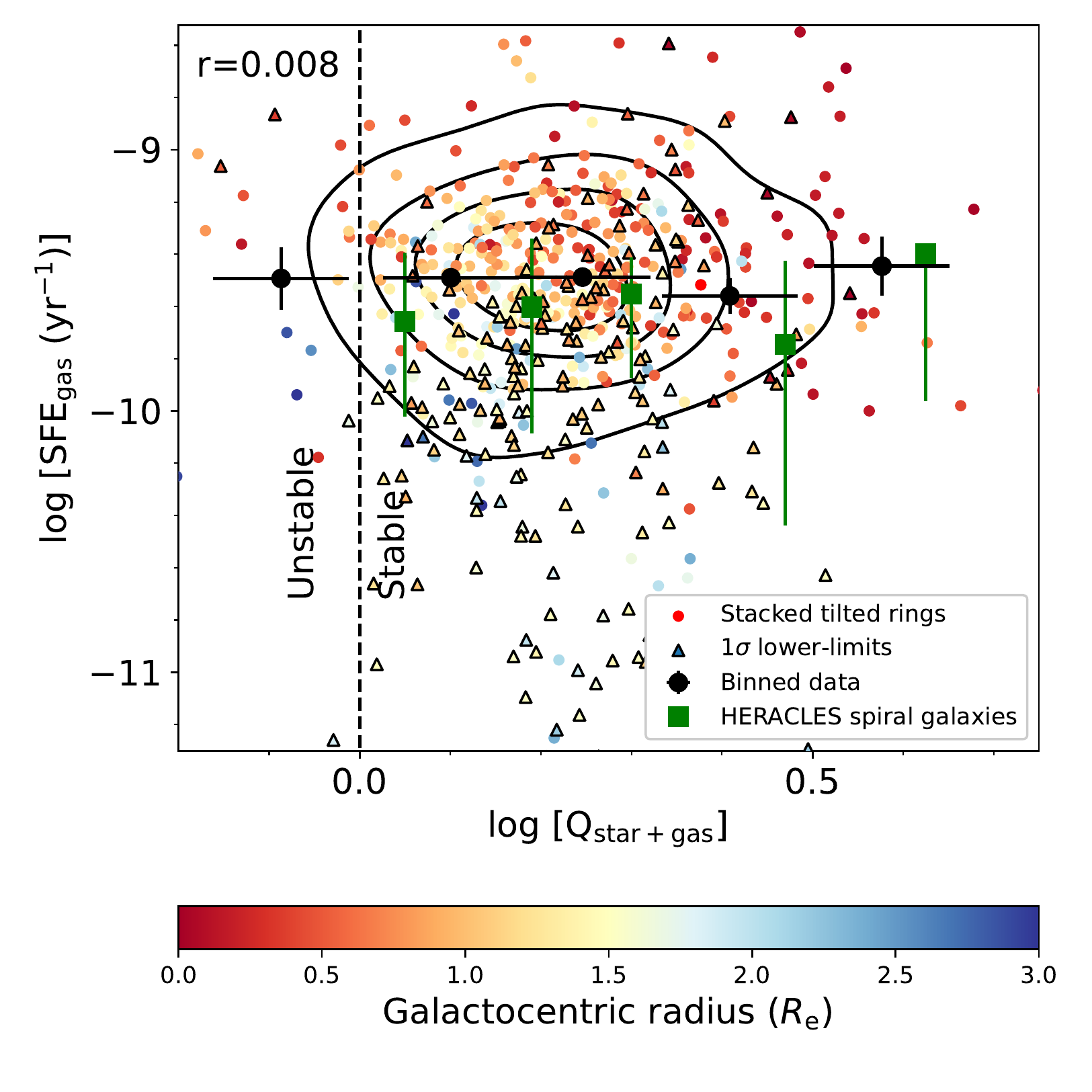}\\
  \vspace{-0.25cm}
  \includegraphics[width=7.8cm]{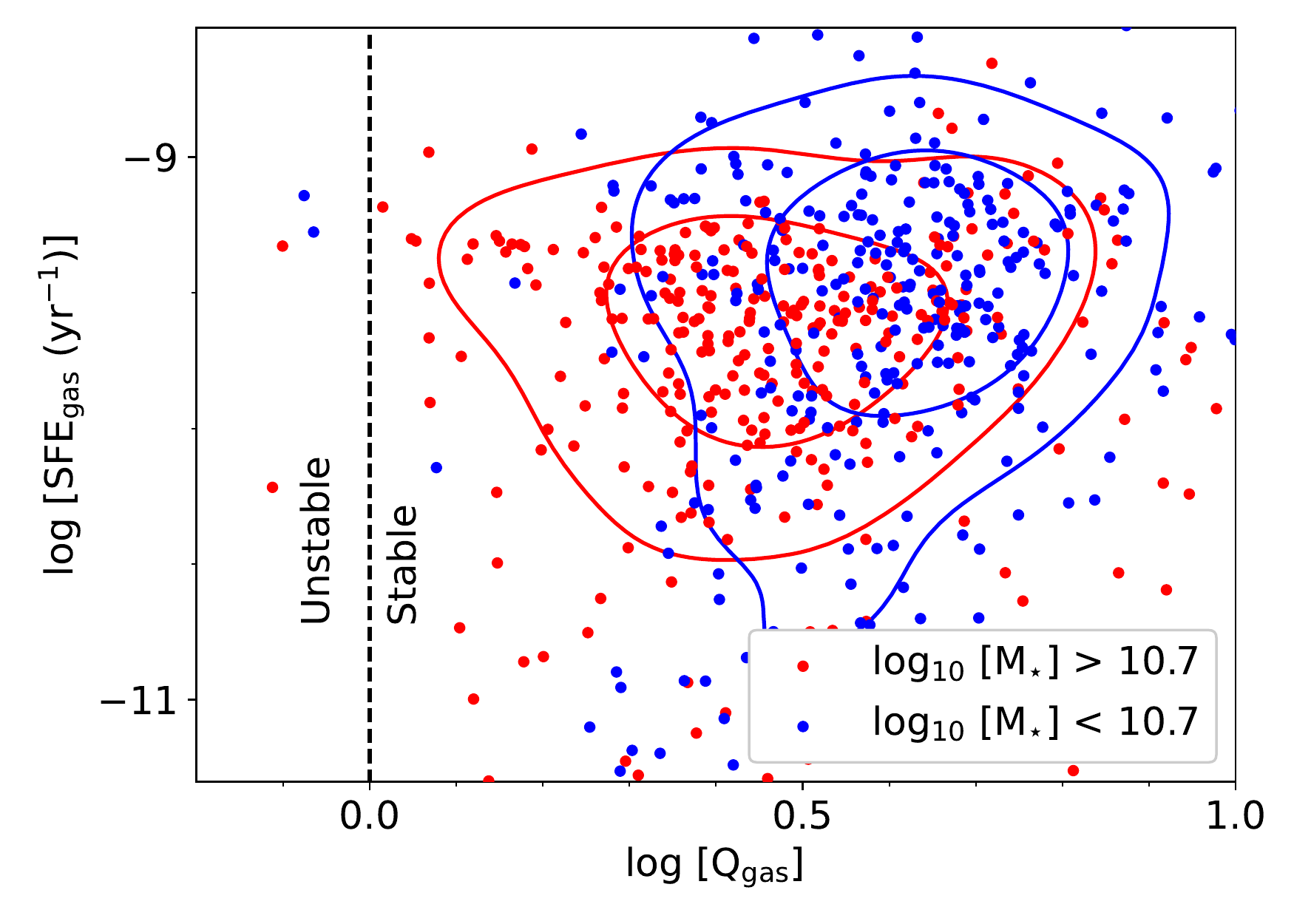}
  \includegraphics[width=7.8cm]{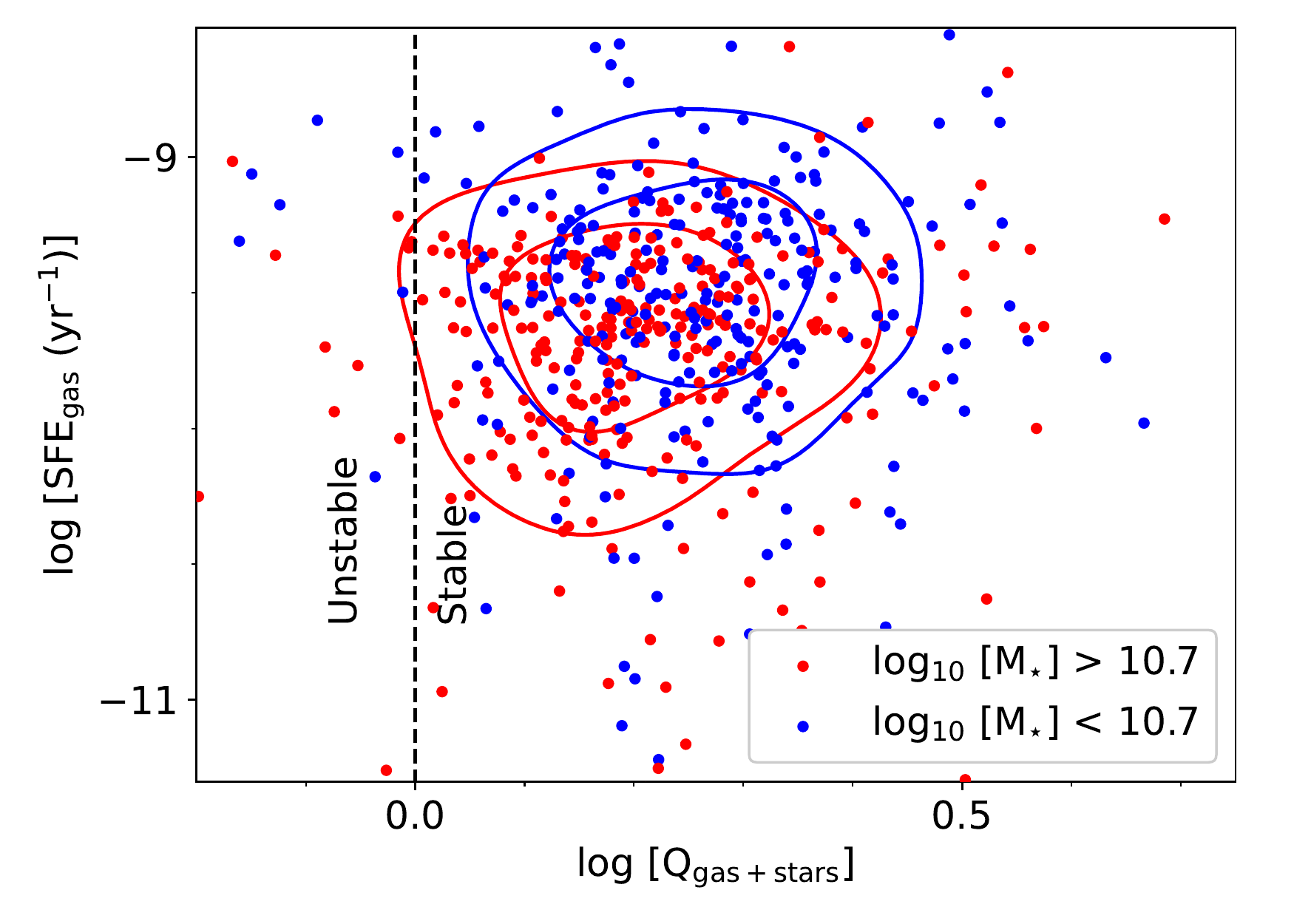}\\
  
  \vspace{-3.15cm}\hspace{-.2cm}\includegraphics[width=9.3cm]{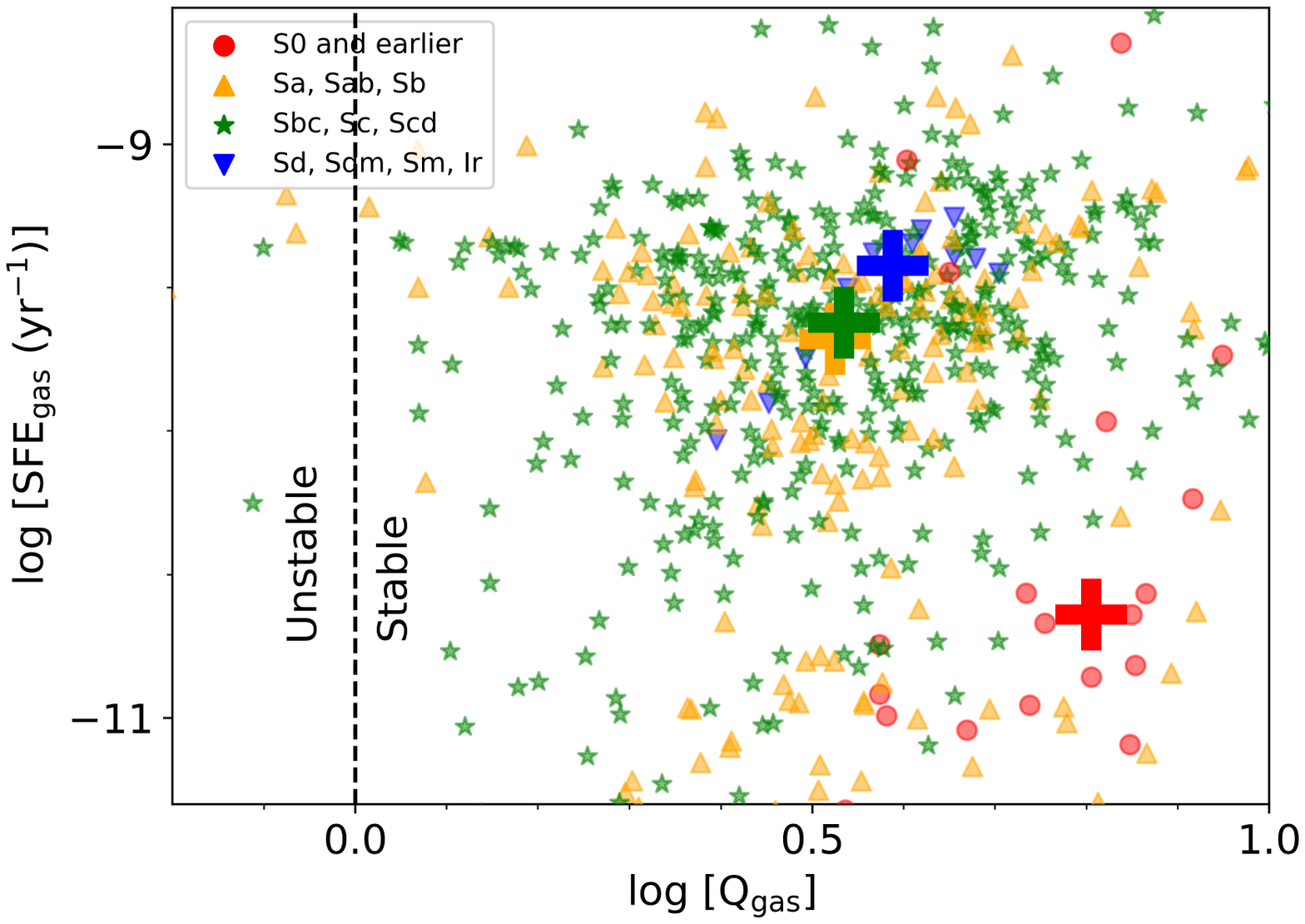}
  \hspace{-1.4cm}\includegraphics[width=9.3cm]{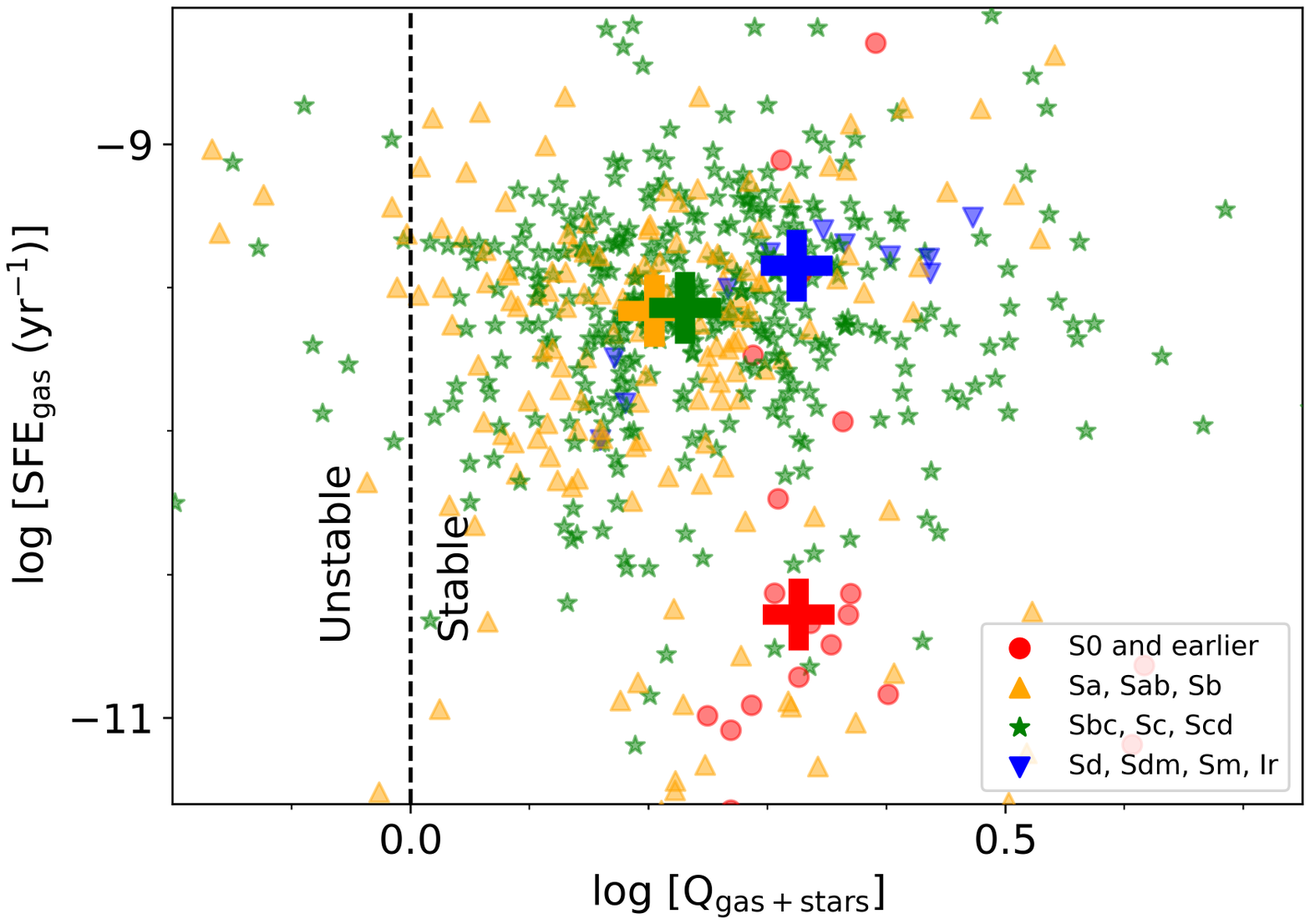} 
\end{center}  
\vspace{-3.3cm}
  \caption{SFE$_{\rm gas}$ as a function of Toomre's gravitational instability Q parameter for two different scenarios. {\it{Left:}} The SFE$_{\rm gas}$ is plotted as a function of the Toomre Q parameter for gas, $Q_{\rm gas}$. {\it{Right:}} The SFE$_{\rm gas}$ is plotted as a function of the Toomre Q parameter redefined by \cite{Rafikov2001} to include the contribution of stars and gas, $Q_{\rm stars+gas}$. Green squares in the upper left and right panels correspond to HERACLES spiral galaxies; black dashed line sets the limit where the gas is unstable or stable. The center left and right panels show the same points included in the upper ones but this time divided into low and high galaxy stellar mass sets; red points correspond to binned annuli which belong to galaxies with $\log_{10}$($M_{\star}$)$>10.7$, while blue points belong to galaxies with stellar masses below this limit. Blue and red contours are the 66\% and 33\% of the points for each mass set. The bottom left and right panels provide information about the morphological type of the host galaxy for a given annulus. The crosses correspond to the center of mass of the $\log_{10}$SFE$_{\rm gas}$ vs center of mass of $\log_{10} Q$ points for each set of morphological types.}
  \label{FIG_12_Leroy2008}
\end{figure*}

\subsubsection{Gravitational instabilities}
\label{Gravitational_instabilities}

The formulation of the Toomre $Q$ gravitational stability parameter \citep[][see Section \ref{Basic_equations} for more details]{Toomre1964} provided a useful tool to quantify the stability of a thin disk  disturbed by axisymmetric perturbations. Some studies have shown that the star formation activity is widespread where the gas disk is $Q$-unstable against large-scale collapse \citep[e.g., ][]{Kennicutt1989,MartinKennicutt2001}. 

First we examine the case where only gas gravity is considered; the top left panel of Figure \ref{FIG_12_Leroy2008} considers this case, showing the SFE$_{\rm gas}$ as a function of both the Toomre instability parameter for a thin disk of gas (x-bottom axis), $Q_{\rm gas}$, and  galactocentric radius (indicated by dot color). The vertical black-dashed line marks the limit where the gas becomes unstable to axisymmetric collapse. The vast majority of our points are in stable (or marginally stable) annuli with an average $Q_{\rm gas} = 3.2$. There is no apparent correlation of SFE$_{\rm gas}$ with $Q_{\rm gas}$ (Pearson correlation coefficient of 0.17), and that is independent of galaxy mass (middle left panel) or type (bottom left panel). In other words, SFE$_{\rm gas}$ does not decrease as stability increases (i.e as $Q_{\rm gas}$ increases). This is in agreement with similar results reported in previous studies. For example, using H$_{\rm I}$ observation for 20 dwarf Irregular galaxies selected from the Local Irregulars That Trace Luminosity Extremes, The H$_{\rm I}$ Nearby Galaxy Survey \citep[LITTLE THINGS; ][]{Hunter2012}, \cite{Elmegreen2015} find that dIrr galaxies are $Q_{\rm gas}$-stable, with a mean $Q_{\rm gas}\sim4$. They also find their galaxies have relatively thick disk, with typical (atomic) gas scale heights of $h_{\rm gas}\sim0.3-1.5$kpc. Consequently, they are more stable than the infinitely thin disks for which the $Q_{\rm gas}=1$ criterion is derived. 

Stars represent the dominant fraction of mass in disks at galactocentric radii with active star formation. Thus, it makes sense to account for their gravity when determining the stability of the ISM in these regions. The top right panel of Figure \ref{FIG_12_Leroy2008} shows the SFE$_{\rm gas}$ as a function of  Toomre's instability parameter modified by \citealt{Rafikov2001} to include the effects of both gas and stars, $Q_{\rm stars+gas}$, again galactocentric radius is indicated by color. As expected, we find that disks become more unstable when stellar gravity is included in addition to gas with a few points appearing in the nominally unstable region for thin disks. The bulk of the annuli, however, are found at around $Q_{\rm stars+gas}\approx1.6$. This is roughly consistent with calculations of $Q$ in other samples \citep{Romeo2020}. There is, however, no correlation of SFE$_{\rm gas}$ with $Q$.

\begin{figure}

\hspace{-0.6cm}
\begin{tabular}{c}
  \includegraphics[width=8.cm]{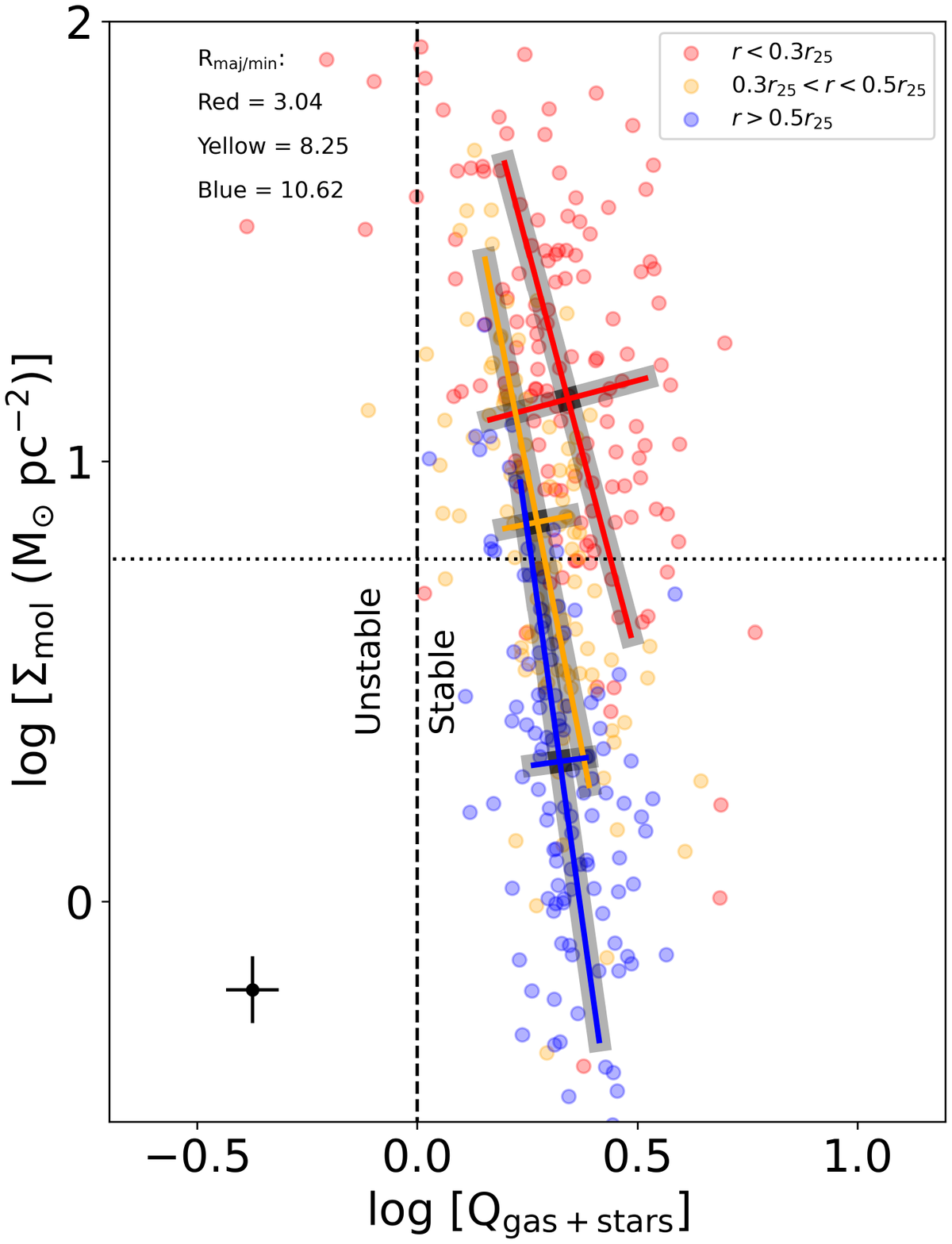}\\
  \includegraphics[width=7.7cm]{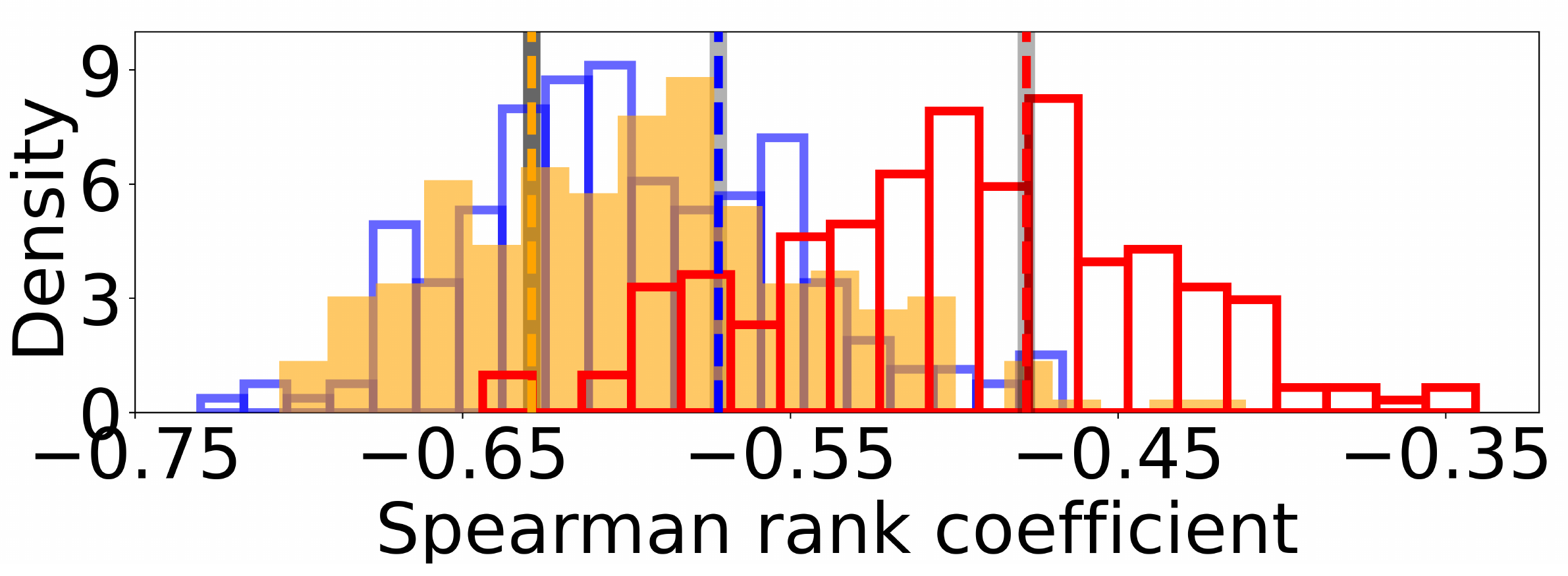}
\end{tabular}
\caption{{\it Top:} Molecular gas surface density, $\Sigma_{\rm mol}$, as a function of Toomre's instability parameter for gas and stars, $Q_{\rm stars+gas}$, for annuli with galactocentric radii within $0.3r_{25}$ (red points), between $0.3r_{25}$ and $0.5r_{25}$ (yellow points), and outside $0.5r_{25}$ (blue points). Each point represents the value of $\Sigma_{\rm mol}$ averaged over a $r/r_{25}$-wide annuli. Points are color-coded by galactocentric radius (in $r_{25}$), as indicated by the color bar on the right side. Solid-lines are PCA major and minor axes for which each of the groups vary most. The axes are normalized to fit the major and minor axes of the elliptical contours that enclose 50\% of the annuli within a given range. The ratio between the major and the minor axes from the PCA, $R_{\rm maj/min}= r_{\rm maj}/r_{\rm min}$, is in the upper left. Typical 1-$\sigma$ error bars are shown in bottom left. The horizontal black-dotted line represents the assumed $\Sigma_{\rm atom} = 6$ M$_{\odot}$ pc$^{-2}$. {\it Bottom:} Distribution of the Spearman rank correlation coefficients for the three $r$ ranges in the top panel after randomizing the $\Sigma_{\rm mol}$ data, per range, in 200 realizations to test for the degree of internal correlation of the axes. The horizontal dashed-red, dashed-yellow, and dashed-blue lines are the Spearman rank coefficients for the actual data, from inner to outer ranges, respectively. This shows that the correlations observed in the top panel are completely consistent with being a result of the definition of $Q_{\rm gas+stars}$ (see discussion in the text), and thus (although tantalizing) are not particularly meaningful.}
\label{Ratio_vs_Q}
\end{figure}

The center panels of Figure \ref{FIG_12_Leroy2008} show the SFE$_{\rm gas}$s versus $Q_{\rm gas}$ and $Q_{\rm stars+gas}$ but this time splitting the points in two groups of different galaxy stellar mass; as in Section \ref{SFE_radius}, we choose $\log_{10}$[$M_{\star}$]$=10.7$ to split the groups. Although the two groups separate in $Q_{\rm gas}$, with annuli from galaxies with $\log_{10}$[$M_{\star}$]$<10.7$ tending to be in general more stable, the separation disappears once the stars are taken into account in the $Q$ calculation. 

In one of the ideas on how stars relate to SFE$_{\rm gas}$, \cite{Dib2017} show that star-formation may be associated with the fastest growing mode of  instabilities. In that case, the relation between SFR and gas in spiral galaxies may be modulated by the stellar mass, which will contribute to the gravitational instability and regulation of star formation \citep[like in the case of NGC 628; ][]{Dib2017}. Also, the $\Sigma_{\rm SFE_{\rm gas}}$-$\Sigma_{\star}$ relation, known as the ``extended Schmidt law'', suggests a critical role for existing stellar populations in ongoing star formation activity, and it may be a manifestation of more complex physics where $\Sigma_\star$ is a proxy for other variables or processes \citep[][]{Shi2011}. Our results may reflect the importance of instabilities in enhancing the SFE$_{\rm gas}$ due to the strong gravitational influence from stars, particularly in galaxies with $\log_{10}$[$M_{\star}$]$>10.7$. But in the aggregate there is no apparent evidence for a trend showing that annuli with more unstable $Q$ have higher star formation efficiencies.

The bottom panels of Figure \ref{FIG_12_Leroy2008} show the same relations as upper panels but this time the data are grouped in four bins by morphological type. In both panels crosses correspond to the "center of mass" for each morphological group. Although annuli in early-type galaxies are more ``Toomre stable", the statistics are very sparse and the Toomre calculation may not apply (since these are not thin disks). Otherwise, we do not find a clear trend between morphology and stability based on the Toomre parameter for stars and gas. Previous studies have reported that $Q_{\rm star+gas}$ increases towards the central parts of spirals. For example, \cite{Leroy2008} found that although molecular gas is the dominant component of the ISM in the central regions, HERACLES galaxies seems to be more stable there than near the H$_{\rm 2}$-to-H$_{\rm I}$ transition. If the type of gravitational instability that $Q$ is sensitive to plays a role in star formation in galaxies, we would expect to see some links between $Q$ and molecular gas abundance. It is therefore interesting to test if there is dependence of the H$_2$-to-H$_{\rm I}$ ratio, $R_{\rm mol}=\Sigma_{\rm mol}$/$\Sigma_{\rm atom}$ on the degree of gravitational instability in EDGE galaxies. Since we assume a constant $\Sigma_{\rm atom}$, however, for us $R_{\rm mol}$ is simply a normalized molecular gas surface density, $\Sigma_{\rm mol}$. We use the typical H$_2$-to-H$_{\rm I}$ transition radius found in  \S\ref{SFE_radius} to split the annuli into three groups: i) annuli at $r<0.3 r_{25}$ ($r < 0.6 R_{\rm e}$; red points) which should be strongly molecular, ii) annuli between $0.3 r_{25}<r$ and $r<0.5 r_{25}$ ($0.6 R_{\rm e} < r < 1.4 R_{\rm e}$; yellow points) which should be around the molecular to atomic transition region, and iii) annuli at $r > 0.5 r_{25}$ ($r > 1.4 R_{\rm e}$; blue points) which should be dominated by atomic gas. The top panel of Figure \ref{Ratio_vs_Q} shows that $\Sigma_{\rm mol}$ has a large scatter and does not seem to depend strongly on $Q_{\rm star+gas}$. Within the each range, however, we find that annuli with smaller galactocentric radii tend to be slightly more stable. 

A suggestive trend emerges when we limit the range of galactocentric radii. We compute a Principal Component Analysis (PCA; \citealt{Pearson1901}) to find the main axis along which the three populations vary most. The top panel of Figure \ref{Ratio_vs_Q} shows the PCA major and minor axis for annuli in the three defined zones. The axes have been normalized to fit the minor of major axes of the elliptical contours that enclose 50\% of the annuli over a given range. The figure suggest that, within a given range, we tend to find more plentiful molecular gas in regions where annuli are more Toomre unstable. A concern, however, is that the axes in this plot have a degree of intrinsic correlation since the computation of $Q_{\rm star+gas}$ includes $\Sigma_{\rm mol}$. Therefore to assert that the correlation we observe is physically meaningful we need to show that it is stronger than that imposed by the mathematics of the computation. We quantify the strength of the correlations using the Spearman rank correlation coefficient, which is a non-parametric measure of the monotonicity of the observed correlations. To investigate the degree to which the axes are internally correlated, we randomize the $\Sigma_{\rm mol}$ data (within each range) and recompute  $Q_{\rm star+gas}$ in 200 realizations, to obtain the distributions of the Spearman rank correlation coefficient for each randomized group. Clearly in the randomized data we would expect only the degree of correlation caused by the mathematical definition of the quantities.  
The bottom panel of Figure \ref{Ratio_vs_Q} shows that the Spearman rank correlation coefficients for the actual data (dashed-red, dashed-yellow, and dashed-blue vertical lines) are consistent with the distributions seen in the randomized histograms. These results suggest that the correlation between $\Sigma_{\rm mol}$ and $Q_{\rm star+gas}$ seen in the top panel of Figure \ref{Ratio_vs_Q} is purely driven by the implementation of equation \ref{Rafikov}, in which $Q_{\rm star+gas}$ depends on $\Sigma_{\rm mol}$.

\section{Summary and conclusions}
\label{S6_Conclusion}

We present a systematic study of the star formation efficiency and its dependence on other physical parameters in 81 galaxies from the EDGE-CALIFA survey. We analyse CO 1-0 datacubes which have ${7}\arcsec$ angular resolution and  20 km s$^{-1}$ channel width, along with H$\alpha$ velocities extracted from the EDGE database, {\tt edge\_pydb} (Wong et al. in prep.). We implement a spectral stacking procedure for CO spectra shifted to the H$\alpha$ velocity to enable detection of faint emission and obtain surface densities averaged over  annuli  of width $0.1r_{25}$  ($\sim {3.5}\arcsec$), and   measure $\Sigma_{\rm mol}$ out to typical galactocentric radii of $r\approx1.2\,r_{25}$ ($r\sim 3\,R_{\rm e}$). We assume a constant \citep{Walter2008}, a Milky-Way constant conversion factor of $\alpha_{\rm CO} = 4.3$ M$_\odot$ $[\rm K \, km \, s^{-1} \, pc^{-2}]^{-1}$, and a constant $\sigma_{\rm g}=11$ km s$^{-1}$ \citep{Leroy2008,Tamburro2009}. We perform a systematic analysis to explore molecular scale lengths and the dependence of the star formation efficiency SFE$_{\rm gas}$=$\Sigma_{\rm SFR}/(\Sigma_{\rm mol}+\Sigma_{\rm atom})$ on various physical parameters. Our main conclusions are as follows:

\begin{enumerate}
    \item We determine the molecular and stellar exponential disk scale lengths, $l_{\rm mol}$ and $l_{\star}$, by fitting the radial $\Sigma_{\rm mol}$ and $\Sigma_{\star}$ profiles, respectively. We also obtain the radii that encloses 50\% of the total molecular mass, $R_{50,\rm mol}$, and stellar mass, $R_{50,\star}$ (see Fig. \ref{FIG_14_LENGTH}). To quantify the relations, we use an OLS linear bisector method to fit all our $3\sigma$ detections beyond $r > 1.5$ kpc. We find that $l_{\rm mol} = [0.86\pm0.07] \times l_{\star}$, $l_{\rm mol} = $ [$0.24 \pm 0.01$]$ \times r_{25}$, and $R_{50,\rm mol } = [0.93\pm 0.05$]$ \times R_{50,\star}$. These results are in agreement with values from the current literature, and indicate that on average the molecular and stellar radial profiles are similar.

	\item We find that on average the SFE$_{\rm gas}$ exhibits a smooth exponential decline as a function of galactocentric radius, without a flattening towards the centers of galaxies seen in some previous studies (see Fig. \ref{FIG_1_Leroy2001B}), in agreement with recent results \citep[e.g.,][]{Sanchez2020_1,Sanchez2020_2}. We note a systematic increase in the average SFE$_{\rm gas}$ from early to late type galaxies. In HI-dominated regions, this conclusion depends strongly on our assumption of a constant HI surface density for the atomic disk. The EDGE-CALIFA survey encompasses a galaxy sample that has not been well represented by prior studies, which includes a larger number of galaxies with a broader range of properties and morphological types. This may explain the differences we observe when we compare our result with previous work.
	
	\item The SFE$_{\rm gas}$ has a clear dependence on $\Sigma_{\star}$ (see Fig. \ref{FIG_3_Leroy2008}), a relation that holds for both the atomic-dominated and the molecular-dominated regimes. The SFE$_{\rm gas}$ has a comparatively flatter dependence on $\Sigma_{\rm gas}$ for high values of the gas surface density. This suggests that the stellar component has a strong effect on setting the gravitational conditions to enhance the star formation activity, not just converting the gas from HI to H$_2$. However, statistical tests, which are beyond the scope of this work, may be required to demonstrate that this secondary relation is not induced by errors \citep{Sanchez2021}.
	
	\item There is a clear relationship between SFE$_{\rm gas}$ and the dynamical equilibrium pressure, $P_{\rm DE}$, particularly in the innermost regions of galactic disks. Moreover, we find a strong correlation between $\Sigma_{\rm SFR}$ and $P_{\rm DE}$. We identify a transition at $\log[P_{\rm DE}/k ({\rm K \, cm^{-3}})] \sim$ 3.7, above which we find a best-linear-fit slope of $1.11\pm0.15$. Our results are in good agreement with the current literature and support a self-regulated scenario in which the star formation acts to restore the pressure balance in active star-forming regions.
	
	\item We find a power-law decrease of SFE$_{\rm gas}$ as a function of orbital time $\tau_{\rm orb}$ (see Fig. \ref{FIG_7_Leroy2008}). The average $\tau_{\rm orb}$ within $0.7 r_{25}$ for our galaxies is $2.6\pm0.2$ Gyr, with a typical efficiency for converting gas into stars of $\sim5\%$ per orbit. Note, however, that there are systematic trends in this efficiency. In particular, we note that there is a flattening of the SFE$_{\rm gas}$ for $\log [\tau_{\rm orb} ({\rm yr})]\sim 7.9-8.1$ which may reflect  star formation quenching  due to the presence of a bulge component. Although our methodology is different, our findings support the conclusion that the star formation efficiency per orbital time is a function of morphology \citep{Colombo2018}. 
    
    \item Finally, under the assumption of a constant velocity dispersion for the gas, we do not find clear correlations between the SFE$_{\rm gas}$ and $Q_{\rm gas}$ or $Q_{\rm stars+gas}$. It is possible that larger samples of galaxies may be required to confidently rule out any trends. Our typical annulus has $Q_{\rm stars+gas}\sim1.6$, independent of galaxy mass or morphological type. The range of $\Sigma_{\rm mol}$ is very broad, and we do not find any meaningful trends. 

\end{enumerate}

Future VLA HI and ALMA CO data may improve the spatial coverage and sensitivity, allowing us remove some limitations and extend this analysis to fainter sources (e.g., earlier galaxy types), contributing to a more extensive and representative sample of the local universe.

\section{acknowledgments}

V. Villanueva acknowledges support from the scholarship ANID-FULBRIGHT BIO 2016 - 56160020 and funding from NRAO Student Observing Support (SOS) - SOSPA7-014. A. D. Bolatto, S. Vogel, R. C. Levy, and V. Villanueva, acknowledge partial support from NSF-AST1615960. J.B-B acknowledges support from the grant IA-100420 (DGAPA-PAPIIT, UNAM) and funding from the CONACYT grant CF19-39578. R.H.-C. acknowledges support from the Max Planck Society under the Partner Group project "The Baryon Cycle in Galaxies" between the Max Planck for Extraterrestrial Physics and the Universidad de Concepci{\'o}n. Support for CARMA construction was derived from the Gordon and Betty Moore Foundation, the Kenneth T. and Eileen L. Norris Foundation, the James S. McDonnell Foundation, the Associates of the California Institute of Technology, the University of Chicago, the states of California, Illinois, and Maryland, and the NSF. CARMA development and operations were supported by the NSF under a cooperative agreement and by the CARMA partner universities. This research is based on observations collected at the Centro Astron\'{o}mico Hispano-Alem\'{a}n (CAHA) at Calar Alto, operated jointly by the Max-Planck Institut f\"{u}r Astronomie (MPA) and the Instituto de Astrofisica de Andalucia (CSIC). M. Rubio acknowledge support from ANID(CHILE) Fondecyt grant No 1190684 and partial support from ANID project Basal AFB-170002. AST-1616199 for Illinois (TW/YC/YL) and AST-1616924 for Berkeley (LB/DU). This research has made use of NASA’s Astrophysics Data System.

\software{ Astropy \citep{AstropyCollaboration2018}, MatPlotLib \citep{Hunter2007},
NumPy \citep{Harris2020}, SciPy \citep{2020SciPy-NMeth}, seaborn \citep{Waskom2021}, Scikit-learn \citep{scikit-learn}.}

\bibliography{main}{}
\bibliographystyle{aasjournal}



\end{document}